\documentclass[superscriptaddress,aps,prl,english,floatfix,twocolumn,10pt]{revtex4-2}
\usepackage{graphicx}
\usepackage{float}
\usepackage{natbib}
\usepackage{amssymb}
\usepackage{amsmath}
\usepackage{mathtools}
\usepackage[utf8]{inputenc}
\usepackage{braket}
\usepackage{graphicx}
\usepackage{subfigure}
\usepackage{pgfplots}
\usepackage{csquotes}
\usepackage{hhline}
\usepackage{amssymb}
\usepackage{xr}
\usepackage{dcolumn}
\usepackage{tabularx}
\usepackage[colorlinks=true,linkcolor=blue,citecolor=blue,urlcolor=blue]{hyperref}
\usepackage{longtable}
\usepackage{braket}
\usepackage{float}
\usepackage{listings}
\usepackage{color}
\usepackage[export]{adjustbox}
\definecolor{mygreen}{rgb}{0,0.6,0}
\definecolor{mygray}{rgb}{0.5,0.5,0.5}
\definecolor{mymauve}{rgb}{0.58,0,0.82}
\newcolumntype{C}{>{\centering\arraybackslash}X}

\begin{document}
	
	\title{\textbf{The fate of Wannier-Stark localization and skin effect in periodically driven non-Hermitian quasiperiodic lattices}}
	\author{Aditi Chakrabarty}
	\email{aditichakrabarty030@gmail.com}
	\affiliation{Department of Physics and Astronomy, National Institute of Technology, Rourkela, Odisha-769008, India}
	\author{Sanjoy Datta}
	\email{dattas@nitrkl.ac.in}
	\affiliation{Department of Physics and Astronomy, National Institute of Technology, Rourkela, Odisha-769008, India}
	\date{\today}
	
	\begin{abstract}
		The eigenstates of one-dimensional Hermitian and non-Hermitian tight-binding systems (in the presence/absence of quasiperiodic potential) and an external electric field undergo complete localization with equally spaced eigenenergies, known as the Wannier-Stark (WS) localization.
		In this work, we demonstrate that when the electric field is slowly modulated with 
		time, new non-trivial phases with multiple mobility edges emerge in place of WS localized phase,
		which persists up to a certain strength of the non-Hermiticity.
	    On the other hand, for a large driving frequency, we retreive the usual sharp delocalization-localization transition to the usual (no WS) localized phase, similar to the static non-Hermitian Aubry-Andr\'e-Harper type without any electric field.
	    This vanishing of WS localization can be attributed solely to the time-periodic drive and occurs irrespective of the non-Hermiticity.
	    Interestingly, under the open boundary condition (OBC), we find that contrary 
	    to the undriven systems where an external electric field destroys the SE completely, the SE appears in certain regime of the parameter space when the electric field is temporally driven.
	    This appearance of SE is closely related to the absence of extended unitarity. 
	    In addition, in the presence of the drive, the skin states are found to be \emph{multifractal},
	    contrary to its usual nature in such non-Hermitian systems.  An in-depth understanding about the behavior of the states in the driven system is established from the long-time dynamics of an initial excitation.
		
	\end{abstract}
	
	\maketitle
	\section{Introduction}\label{Sec:Introduction}
	The quest for exotic and novel quantum phases of matter has led to investigations of systems with time-periodic drive \cite{Hanggi,Stockmann}. Such systems exhibit intriguing features in contrast to the static counterparts. For example, a time-periodic drive reduces the tunneling rate of particles in optical lattices \cite{Lignier,Ivanov,Chen}, and introduces unusual dynamics in kicked rotors \cite{Bitter}. Besides, periodically driven cold-atomic systems have been used for precise measurement of gravitational acceleration \cite{Poli}, and realizing anomalous topological phases \cite{Karen}. Parallelly, recent years have also witnessed rich features in non-Hermitian driven 
	systems in the context of the spectral, transport and topological properties
	\cite{Valle,Hockendorf,Banerjee,Wu,Liu_2022,Sengupta}. Such a time-periodic drive	in non-Hermitian systems has also been useful in stabilizing its dynamics \cite{Gong,Timms}. Furthermore, several non-Hermitian 
	time-periodic systems have been implemented in experiments on electric circuit \cite{Tsampikos}, ultracold atoms \cite{Jiaming} and acoustic lattice \cite{Chong}.\\
	\indent
	The non-Hermitian systems have also been extensively studied to investigate 
	electronic properties and phases, especially in one-dimensional (1D) systems, wherein it was demonstrated that akin to its static Hermitian analogue \cite{Wannier,Wannier_1962,Fukuyama,Emin,Holthaus}, the phenomenon of localization (first developed in Anderson's seminal work on random disordered models \cite{Anderson}) also emerges in disorder-free lattices upon application of a non-zero external uniform electric field \cite{MingLiu}, well known in the literature as the Wannier-Stark (WS) localization. This complete localization of all the electronic eigenstates due to the electric field has also been demonstrated in 1D Hermitian quasicrystals \cite{Carpena,Carpena_2,Salazar} that otherwise demonstrate a  delocalization-localization (DL) phase transition \cite{Aubry,Harper,Sokoloff} at a finite strength of quasiperiodic potential. In addition, the WS localization is reported in non-Hermitian systems with quasiperiodic potential \cite{Zeng}.
	Such an unique localization typically appears in the form of discrete equally spaced energy levels in the localized phase. \\
	\indent
	The WS localization is often interlinked with the emergence of Bloch oscillation, where the dynamics of the electrons undergo a periodic oscillation with time, in both Hermitian \cite{Popescu,Prates} as well as non-Hermitian systems \cite{Song_2024,Song_2024_arXiv}.
	Furthermore, in the last few years, the investigations on the effect of the interplay of non-Hermiticity and the time-periodic drive has become a central theme for researchers working in this field in the context of localization \cite{Valle} and mobility edges \cite{Zhou_2021}.
	Moreover, a special class of non-Hermitian Hamiltonian \cite{HatanoNelson1996,HatanoNelson1998} has garnered widespread attention in the scientific community in recent times, where an unidirectionality in fermionic hopping leads to an extensive number of bulk eigenstates being exponentially localized towards one of the edges of the system under the open boundary condition (OBC), a phenomenon known as the skin effect (SE) \cite{Lee,Lee_2019,Jiang,Sato}. However, the influence of a time-periodic electric field on the DL transition, WS localization, and the SE in such non-Hermitian systems remains largely unanswered.\\
	\indent
	In one of the past works, it was demonstrated that in a purely tight binding Hermitian Hamiltonian with a sinusoidal electric field, the states are delocalized, except at some critical ratios of the strength of the electric field and driving frequency \cite{Kenkre}. However, the nature of these so-called delocalized states were not analyzed in details. 
	Motivated by the lack of such understanding, in this work, we thoroughly investigate the effect of a continuously modulated external electric field in non-Hermitian quasiperiodic lattices with asymmetric hopping. We find that unique localization traits in the system can be harnessed using the electric field in different regimes of the drive.
	Similar to Ref.~\cite{Kenkre}, we find that in stark contrast to the static scenario, the time-periodic drive destroys the WS localization completely irrespective of the degree of non-Hermiticity. However, we clearly demonstrate that the states are not completely delocalized. Under the periodic boundary condition (PBC), we rather unveil that the time-periodic electric field enhances the delocalized regime, and generates multiple mobility edges when the frequency of the drive is moderately small, unlike the generic undriven counterparts, wherein the mobility edges appear only with  modified quasiperiodic potentials \cite{Cao,Longhi_2020,He}. The mobility edges persist only upto a critical value of the non-Hermiticity, beyond which we retreive the usual DL transition. The sharp DL transition emerges irrespective of the degree of non-Hermiticity, which is also captured analytically.
	We verify the absence of WS localization in the high driving frequency limit as well. Unlike the WS ladders with a $\delta$-type energy level statistics, in the localized phase we clearly find the distribution to approach the usual Poissonian limit when the electric field is time-dependent.\\
	\indent
	In addition, we demonstrate that the SE appears in the non-Hermitian systems with a time-dependent electric field, unlike the static counterparts where the SE vanishes completely. Moreover, similar to the finding in our recent work \cite{Chakrabarty}, under the OBC, we emphasize that this appearance of SE is crucially dependent upon the absence of extended unitarity. Interestingly, we reveal that the skin states are multifractal ($\emph{extended}$, yet $\emph{non-ergodic}$) in nature, which can be attributed to the periodicity in the external electric field. Furthermore, we have studied the wave-packet dynamics in the non-trivial phases with mobility edges, and identified the superdiffusive nature of electron transport. We also verify the absence of Bloch oscillations concommitant with the vanishing of WS ladder-like structure. The evolution of the wave-packet under the OBC reflects  
	the multifractal nature of the skin states after a stroboscopic period.\\
	\indent The birds-eye view of the rest of this article is as follows: In Sec.~\ref{Sec:Model}, we introduce the non-Hermitian quasiperiodic Hamiltonian with a time-dependent electric field. The understanding of such time-dependent Floquet systems are built in Sec.~\ref{Sec:Floquet_approach}. This is followed by the estimation of the critical threshold of quasiperiodic potential for DL transition explained analytically and numerically in Secs.~\ref{Sec:Effective_Hamiltonian} and \ref{Sec:Numerical_techniques} respectively.
    A brief discussion on the energy level statistics is presented in Sec.~\ref{Sec:Level statistics}.
    Our crucial findings on the existence of multiple mobility edges, absence of WS localization and the unique nature of SE are elucidated in Secs.~\ref{Sec:Transition},\ref{Sec:WS_localization}~and \ref{Sec:SE} respectively.
    A complete understanding of the long-time dynamics under both PBC and OBC in the driven system has been presented in Sec.~\ref{Sec:Dynamics}. Finally, the crucial highlights of this work are outlined in Sec.~\ref{Sec:Conclusions}.
	
	\section{Non-Hermitian Hamiltonian with a time-modulated electric field}\label{Sec:Model}
\begin{figure}[t]
		\begin{center}
			\begin{tabular}{p{\linewidth}c}
				\centering
				\includegraphics[width=0.300\textwidth,height=0.235\textwidth]{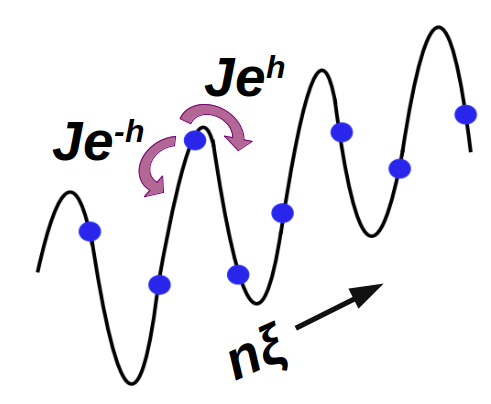}
				\caption{Schematic diagram of the underlying time-independent Hamiltonian discussed in Eq.~(\ref{Eq:Hamiltonian}), showing the directions of fermionic hopping and the electric field, where the fermions are influenced by an underlying quasiperiodic potential.}
				\vspace{-0.9cm}
				\label{Fig:Fig1}
			\end{tabular}
		\end{center}
	\end{figure}

	The time-dependent non-interacting Hamiltonian considered in this work is an amalgamation of the paradigmatic Aubry-Andr\'e-Harper (AAH) quasiperiodic model \cite{Aubry,Harper}, an uniform external electric field \cite{Wannier,Wannier_1962} and non-reciprocity in the hopping amplitudes \cite{HatanoNelson1996}, and can be defined as \cite{Kenkre, Zhou_2021},
	\begin{eqnarray}
		\mathcal{H}(t)= \displaystyle\sum_{n} (Je^{h} c^\dag_{n+1} c_{n}
		+Je^{-h} c^\dag_{n} c_{n+1})~~~~~~~~~~~\nonumber \\
		~~~~~~~~+\sum_{n} V \text{cos} (2\pi\alpha n+\phi) c^\dag_{n} c_{n} + \sum_{n} n \xi\text{cos}(\omega t)c^\dag_{n} c_{n}.~~~~~
		\label{Eq:Hamiltonian}
	\end{eqnarray}
	$c^\dag_{n}$ and $c_{n}$ represent the usual creation and annihilation operators in the second quantization notation respectively.
	$n$ indicates the site index of a lattice with $N$ sites.
	The size of the lattice is $L=Na$, where $a$ is the translational vector (considered to be 1 in arb. units).
	The first term in the Hamiltonian constitutes the asymmetricity in the fermionic hopping ($J$) towards the left and right directions, incorporated using the parameter $h$.
	The second term describes the quasiperiodicity in the potential with strength $V$, brought in by $\alpha$ (set as $(\sqrt{5}-1)/2$ throughout this work). $\phi$ is an arbitrary phase in the quasiperiodic potential, and is set to $0$ unless specified. The third part of the Hamiltonian consists of an electric field gradient ($\xi$) which is responsible for the localization of all the eigenstates in the form of WS ladders in static systems with or without the quasiperiodic potential. 
	It is important to note that the cosine modulation in the drive retains the time-reversal symmetry in the system, similar to the Hatano-Nelson Hamltonian. A schematic of the time-independent model
	described here is depicted in Fig.~\ref{Fig:Fig1}.\\
	\indent
	An experimental realization of the static Hamiltonian has been recently proposed in a 1D optical lattice by Li $et.~al.$ \cite{Li_24}, where two counter-propagating laser beams generate the 
	potential whose depth can be tuned by an efficient control of the laser intensities, and a gravitational or magnetic field gradient (which can be made to be cosine-modulated in our case) 
	can be used to create the linear potential gradient. Such an oscillating electric field can be thought of as an electric dipole that varies with time. The non-reciprocity can be controlled with 
	the help of dissipative channels. Our proposed Hamiltonian can therefore be implemented in a 
	similar optical lattice set-up.
	
	\section{Understanding time-dependent systems using the Floquet approach}\label{Sec:Floquet_approach}
	\indent	To find stationary solutions of the Hamiltonian described in Eq.~(\ref{Eq:Hamiltonian}), which is time-periodic nature with periodicity $T$, i.e., $\mathcal{H}(t)=\mathcal{H}(t+T)$, the Floquet theory has been utilised where the Floquet operator for an entire stroboscopic period is defined as,
	\begin{eqnarray}
		U(T,T_0)=\mathcal{T}e^{-(i/\hbar)\int_{T_0}^{T}\mathcal{H}(t)dt}=e^{-i\mathcal{H_F}(T-T_o)/\hbar},
		\label{Eq:Floquet_operator}
	\end{eqnarray}
	where $\mathcal{T}$ is the well-known time-ordering operator, and $\mathcal{H_F}$ is an effective Floquet Hamiltonian. Without loss of generality, we set the initial time $T_0=0$ since it does not affect the properties of the Hamiltonian.
	The reduced Planck's constant is set as $\hbar=1$ throughout this work.
		
	\indent The Floquet modes provide a complete basis for the Hilbert space which can be obtained on exact diagonalization of the Floquet operator given as,
	\begin{eqnarray}
		U(T,0)=\displaystyle \sum_n E_n \ket{\psi_{nR}}\bra{\psi_{nL}}, \text{and} ~E_n=e^{-i\epsilon_n^FT} .
		\label{Eq:Floquet_quasienergies}
	\end{eqnarray}
	Since the Floquet propagator $U(T,0)$ is non-unitary when the constitutent Hamiltonians at each time interval is non-Hermitian in nature, we extrinsically conserve the unitarity using the QR decomposition method during the evolution with time.
	In Eq.~(\ref{Eq:Floquet_quasienergies}), $\ket{\psi_{nL}}$ and  $\ket{\psi_{nR}}$ are the left and right eigenvectors and $\epsilon_n^F$ is the obtained Floquet quasienergy, which plays the role of ``energy'' in driven systems.
	However, such quasienergies are not uniquely determined, unlike the normal energies, and are typically obtained modulo $\hbar\omega$ within the first quasienergy Brillouin zone.

	\begin{figure*}[]
		\centering
		\includegraphics[width=0.320\textwidth,height=0.28\textwidth]  
		{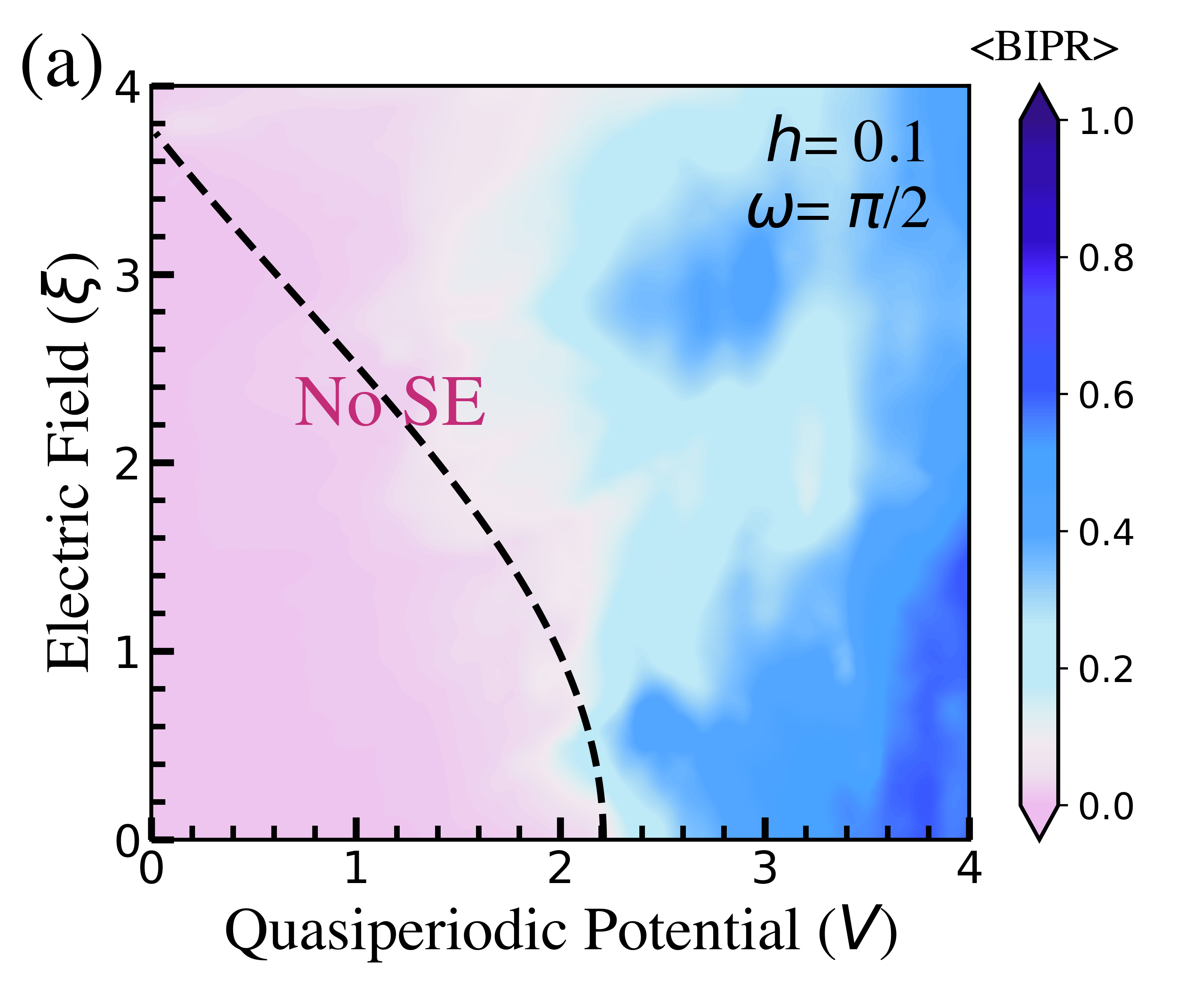}
		\includegraphics[width=0.320\textwidth,height=0.28\textwidth]{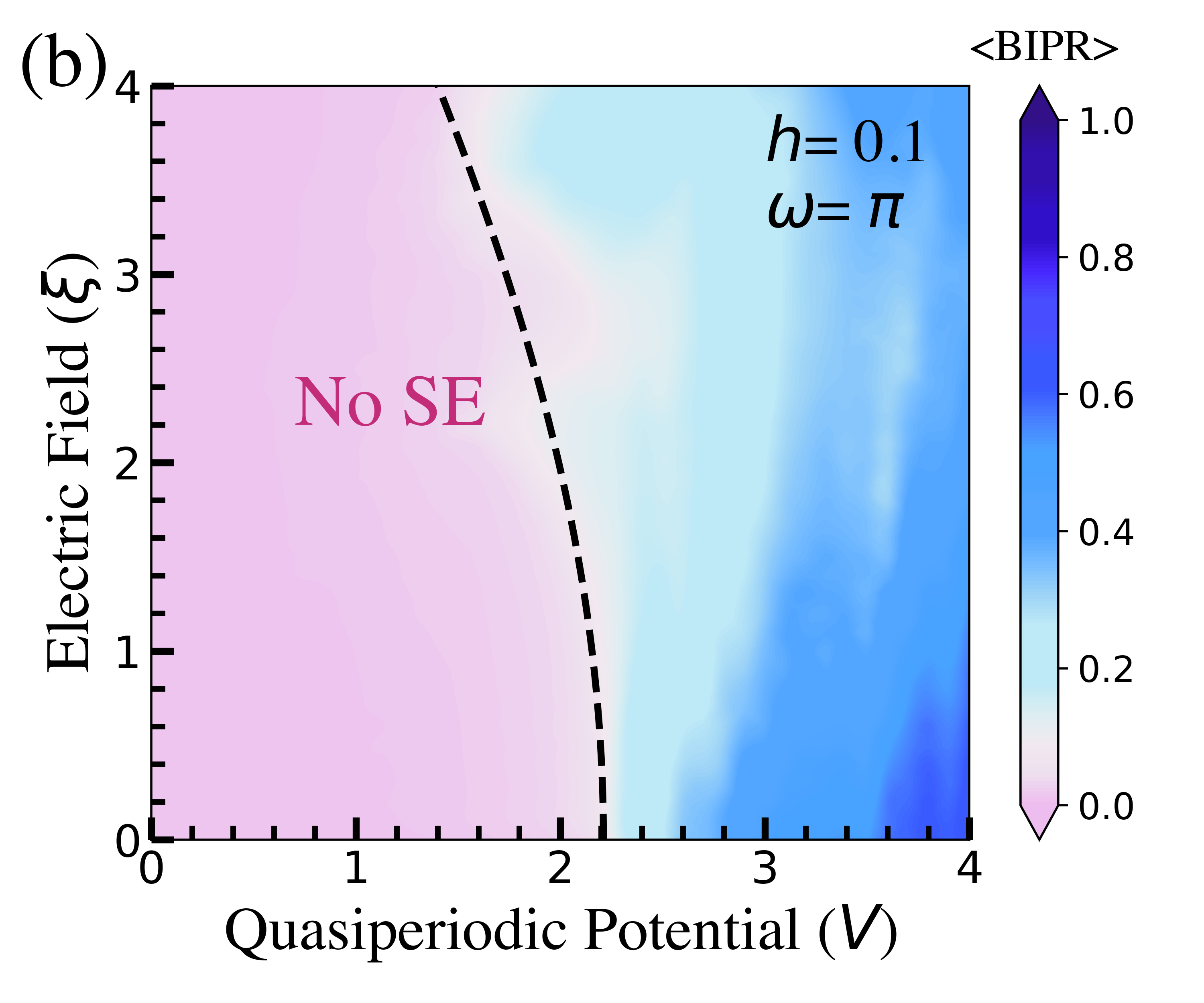}
		\includegraphics[width=0.320\textwidth,height=0.28\textwidth]{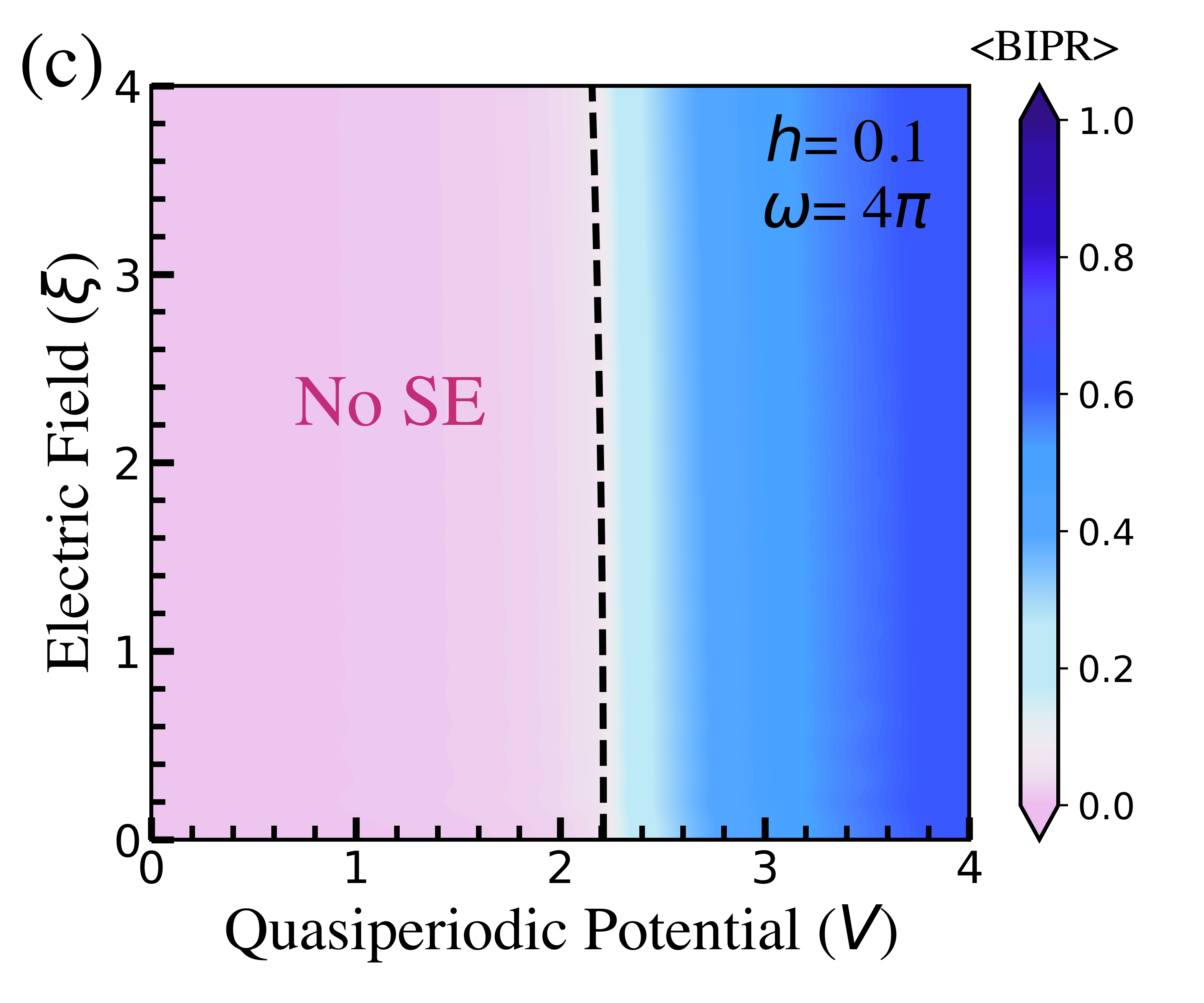}
		\caption{Projection of $\left\langle \text{BIPR} \right\rangle$ under the PBC in the $\xi-V$ parameter space of the non-Hermitian driven system with $h=0.1$ at (a) $\omega=\pi/2$, (b) $\omega=\pi$, and (c) $\omega=4\pi$. The numerically obtained DL transition is shown as a transition from the pink to blue color.
		The analytically predicted DL transition from Eq.~(\ref{Eq:Critical_point}) is illustrated by the black dotted line. The absence of SE (under OBC) in the delocalized regime is also indicated.}
		\label{Fig:Fig_2}
	\end{figure*}
	
	\section{Phase analysis under the drive}\label{Sec:Analysis}
	\subsection{Metal-insulator phase transition from the effective Hamiltonian}\label{Sec:Effective_Hamiltonian}
	In Floquet systems, an approximate analytical understanding can be obtained in the limit of large 
	frequency by averaging the time-dependent Hamiltonian that results in an effective Floquet 
	$\emph{time-independent}$ Hamiltonian $\mathcal{H}_{F}$~\cite{Goldman,Bukov,Eckardt} as, 
	\begin{eqnarray}
		\mathcal{H}_{F}=\frac{1}{T}\int_{0}^{T}\mathcal{H}(t) dt
		\label{Eq:High_frequency_Hamiltonian}
	\end{eqnarray}	
	Following Refs.~\cite{Eckardt_2017,Zhou_2021}, wherein the authors have used a rotated Hamiltonian observed in a comoving frame of reference, we can write,	
	\begin{eqnarray}
		\mathcal{H}_{F}= \displaystyle\sum_{n} (Je^h\mathcal{J}\Big( \frac{\xi}{\omega} \Big) c^\dag_{n+1} c_{n}
		+Je^{-h}\mathcal{J}\Big( \frac{\xi}{\omega} \Big) c^\dag_{n} c_{n+1})\nonumber \\
		+\sum_{n} V \text{cos} (2\pi\alpha n+\phi) c^\dag_{n} c_{n},~~~~~~~~~~~~~~~
		\label{Eq:High_frequency}
	\end{eqnarray}
	where the time-dependent terms in the electric field have been absorbed into the Bessel function of first kind. Interestingly, $\mathcal{H}_{F}$ turns out to be the static non-Hermitian analogue of 
	the AAH Hamiltonian with rescaled hopping amplitudes in terms of $\xi/\omega$.
	It was previously reported that the DL transition for static non-Hermitian AAH-type systems occurs at $V_c=2~\text{max}\{Je^h,Je^{-h}\}$ \cite{Jiang}.
	It is therefore expected that for our Hamiltonian given in Eq.~(\ref{Eq:High_frequency}) in the regime of a large frequency drive in the electric field, the DL transition will be at the critical value of the quasiperiodic potential given as,
	\begin{eqnarray}
		V_c=2~\text{max}\{J'e^h,J'e^{-h}\}~~\text{where}~~J'=J\mathcal{J}\Big( \frac{\xi}{\omega} \Big).~~~~~~~~~
		\label{Eq:Critical_point}
	\end{eqnarray}
	This expected critical value at high frequency will be verified in the subsequent discussions.
	
	\subsection{Numerical identifiers of the phases: ~IPR,NPR, $\eta$ and the mean fractal dimension $D_2$}\label{Sec:Numerical_techniques}
	In order to obtain an idea of the localization characteristics of the driven system, we use the estimate of the physical quantity, i.e., the Inverse Participation Ratio (IPR) \cite{Mirlin,Wessel}.
	In case of a non-Hermitian system where $\ket{\psi_{nL}}$ and $\ket{\psi_{nR}}$ are in general bi-orthonormal \cite{Brody}, the bidirectional-IPR(BIPR) \cite{Wang_2019} has been recently used and is given for an eigenstate `$j$' as,
	\begin{eqnarray}
		BIPR_j=\frac{\displaystyle\sum_{n=1}^{N} |\psi_{nL}^{j}\psi_{nR}^{j}|^2}{\Big(\displaystyle\sum_{n=1}^N |{\psi_{nL}^j\psi_{nR}^{j}}|\Big)^2}  ~~~~~.
		\label{Eq:BIPR}
	\end{eqnarray}
	The BIPR allows values in between 0 and 1, where the two limits pertain to the situations when all the states are either completely delocalized or localized respectively, and in the thermodynamic limit.
	Complementary to the behavior of BIPR, there exists a quantity frequently termed in literature as the Normalized Participation Ratio \cite{Xiao}, which we re-write for the non-Hermitian system as,
	\begin{eqnarray}
		BNPR_j=\Bigg[N\frac{\displaystyle\sum_{n=1}^{N} |\psi_{nL}^{j}\psi_{nR}^{j}|^2}{\Big(\displaystyle\sum_{n=1}^N |{\psi_{nL}^j\psi_{nR}^{j}}|\Big)^2}\Bigg]^{-1}  ~~~~~.
		\label{Eq:BNPR}
	\end{eqnarray}
	Such an indicator, in contrast, attains a finite value for the delocalized states and vanishes when all the states are localized. In addition, an intermediate phase exhibit finite values in both of these indicators. To concretely capture various phases, the authors of Ref.~\cite{DasSharma,Mishra} have recently introduced an useful technique given by the quantity $\eta$ defined as,
		\begin{eqnarray}
		\eta=\log_{10}[\left\langle \text{BIPR} \right\rangle \times \left\langle \text{BNPR} \right\rangle].
		\label{Eq:Eta}
	\end{eqnarray}
	In the above definition, $\left\langle.\right\rangle$ signifies the average of that quantity over all the eigenstates. From Eq.~(\ref{Eq:Eta}), it is evident that if either $\left\langle \text{BIPR} \right\rangle$ or $\left\langle \text{BNPR} \right\rangle$ are vansishing (i.e., $\mathcal{O}(L^{-1})$), $\eta$ $\lesssim -\text{log}_{10}(L)$. On the other hand, when any one among $\left\langle \text{BIPR} \right\rangle$  and $\left\langle \text{BNPR} \right\rangle$ becomes finite (i.e.,$\mathcal{O}(1)$), $-\text{log}_{10}(L)\lesssim\eta\lesssim-1$. However, the single-particle intermediate phase(s) may be either a combination of the delocalized and localized states, or might indicate the existence of a mobility edge in the system \cite{DasSharma}, and these indicators fail to capture this difference. In order to circumvent the issue of determining the intermediate phase accurately, we have plotted the quasienergy resolved phase diagram that clearly illustrates the nature of these intermediate phase(s).

	\indent In addition, the delocalized states can be either completely ergodic or multifractal in nature. Therefore, in order to comprehend the nature of the delocalized phase in detail, we estimate the mean fractal dimension $D_2$, which is given as \cite{Deng,Joana},
	\begin{eqnarray}
		D_2=-\frac{1}{N}\displaystyle\sum_{j=1}^N \frac{\text{ln} ~BIPR_j}{\text{ln}~N}.
		\label{Eq:D_2}
	\end{eqnarray}
	The ergodic states have $D_2\sim1$, whereas the localized ones are characterized by $D_2\sim0$. $0<D_2<1$ implies the existence of the multifractal states.
	
	\subsection{Level statistics in the localized phase}\label{Sec:Level statistics}
	Since the Hamiltonian considered in our work possesses an external electric field that is anticipated
	to induce WS localization, it is important to understand the nature of the energy levels, including their separation.
	In this regard, an important quantity that shows a clear distinction in the eigen energy spectrum in the localized regime is the spectral statistics of such a Hamiltonian.
	A dimensionless quantity that characterises the level statistics has been defined in Ref.~\cite{Huse} for Hermitian systems with random on-site disorder as,
	\begin{eqnarray}
		r_n=\frac{\text{min}\{\delta_n,\delta_{n-1}\}}{\text{max}\{\delta_n,\delta_{n-1}\}},
		\label{Eq:r_statistics}
	\end{eqnarray}
	where $r_n$ lies in between $0$ and $1$, $\delta_n$ being the separation between consecutive energy levels, i.e., $\delta_n=E_{n+1}-E_n$.	
	It is well known that the localized eigenstates that are close in energy lie far away and do not exhibit level repulsion, following a Poissonian statistics with a probability distribution $P(r)=2/(1+r)^2$.
	In the thermodynamic limit, $\langle r \rangle \sim 0.386$ over many realizations of the disorder.
	This measure also works well for quasiperiodic systems where $\langle r \rangle$ is estimated by taking realizations over $\phi$ lying in between $[0,2\pi)$ \cite{Vincent}.
	Moreover, in systems with electric field, the equi-spaced WS energy ladders should naturally lead to $\langle r \rangle \sim 1$, with a $\delta$-type probability distribution with its peak value at $r\sim1$.
	It is important to note that this measure works for real eigenvalues that are obtained in Hermitian systems. The reason why such a measure can be used in our analysis to investigate the spectral behavior of the localized eigenstates where the electric field is modulated with time will be discussed in details in Sec.~\ref{Sec:WS_localization}.
	
	\section{Results and discussions}\label{Sec:Results}
	An investigation on Hermitian quasiperiodic systems has revealed that a very tiny strength in the electric field localizes all the states \cite{Carpena_2}.
	We have verified the same for a non-Hermitian system with asymmetric hopping in the entire parameter space of $\xi$ and $V$ in Figs.~\ref{Fig:Fig_S1}(a-b) in Appendix \ref{App:Undriven}.
	In addition, from the past literature, it is known that the WS ladders under PBC and SE under OBC occur in time-independent systems in presence of the electric field.
	Therefore, in this work, we scrutinize different phases by applying a periodic drive to the electric field in a non-Hermitian quasicrystal (as given in Eq.~(\ref{Eq:Hamiltonian})) in different regimes of the driving frequency.
	In this work, we use the Trotter time step $\Delta t=0.001$, which retains the invariance of the Hamiltonian at two instants of time such that Eq.~(\ref{Eq:Floquet_operator}) can be used to estimate the properties of the time-dependent Hamiltonian.
	We use PBC for a system with $J=1$ and $L=144$, unless specifically stated.

	\begin{figure}[]
		\begin{tabular}{p{\linewidth}c}
			\includegraphics[width=0.2450\textwidth,height=0.228\textwidth]  
			{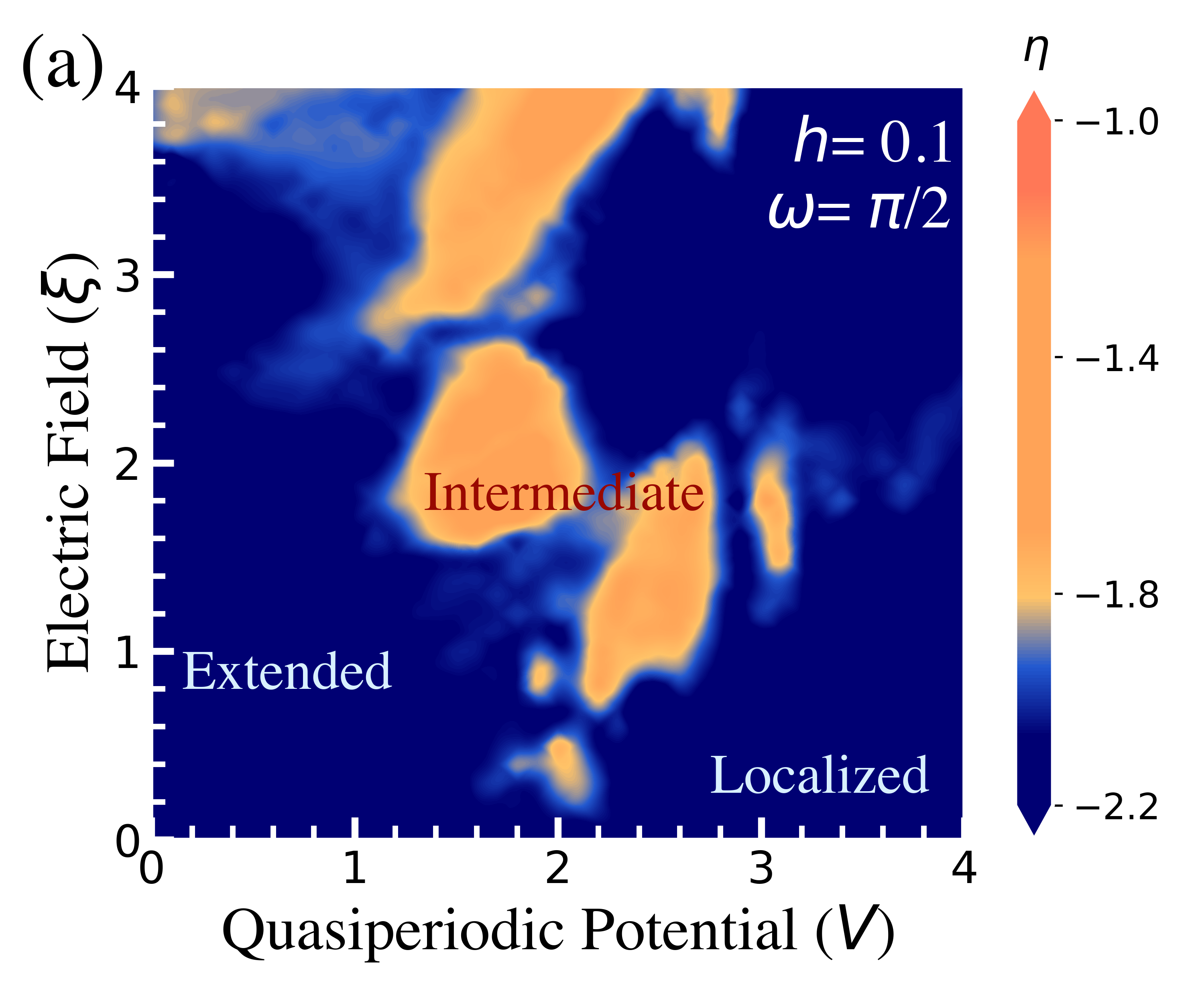}\hspace{-0.1cm}
			\includegraphics[width=0.2450\textwidth,height=0.225\textwidth]{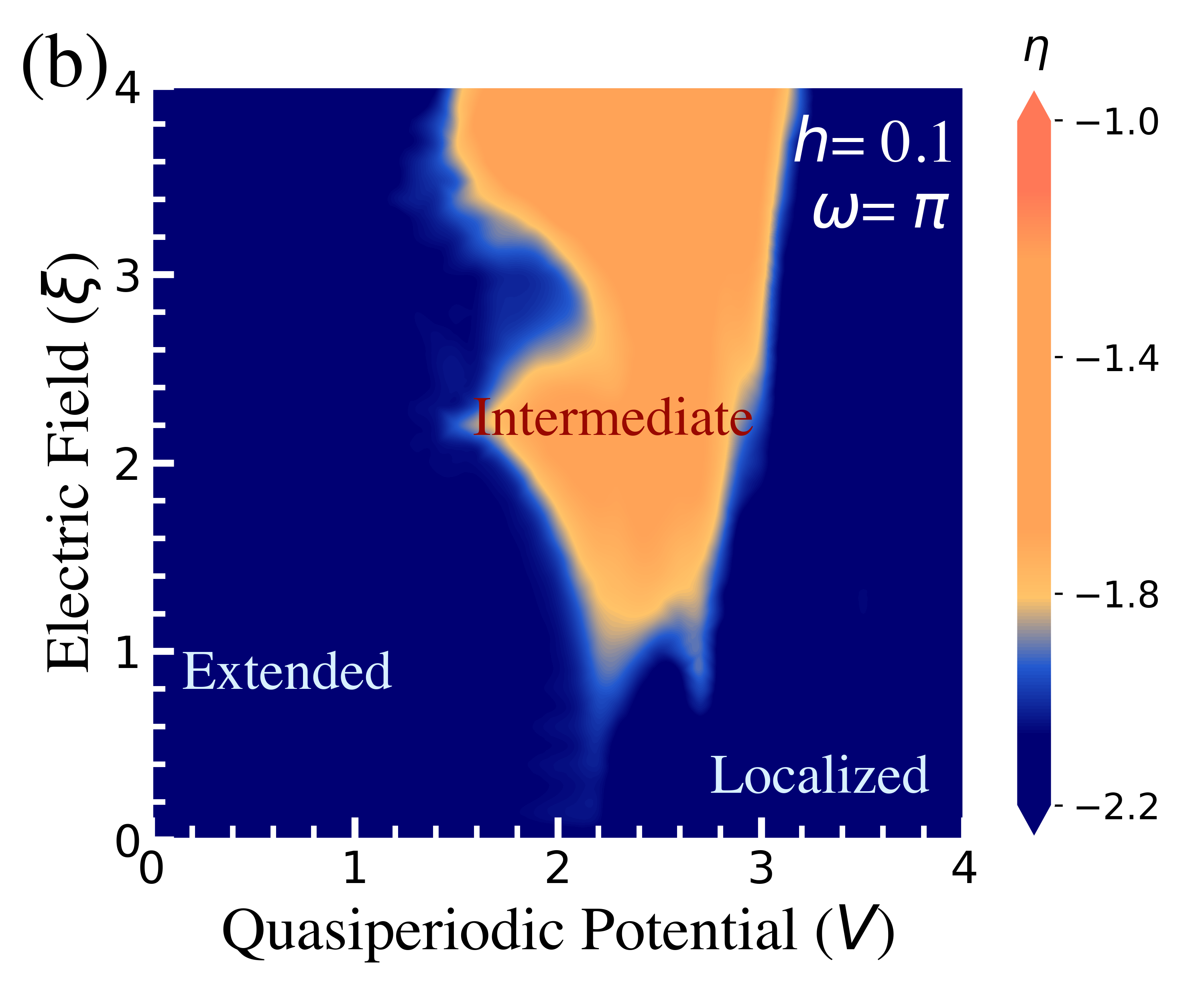}\\
			\hspace{-0.0cm}\includegraphics[width=0.224\textwidth,height=0.228\textwidth]{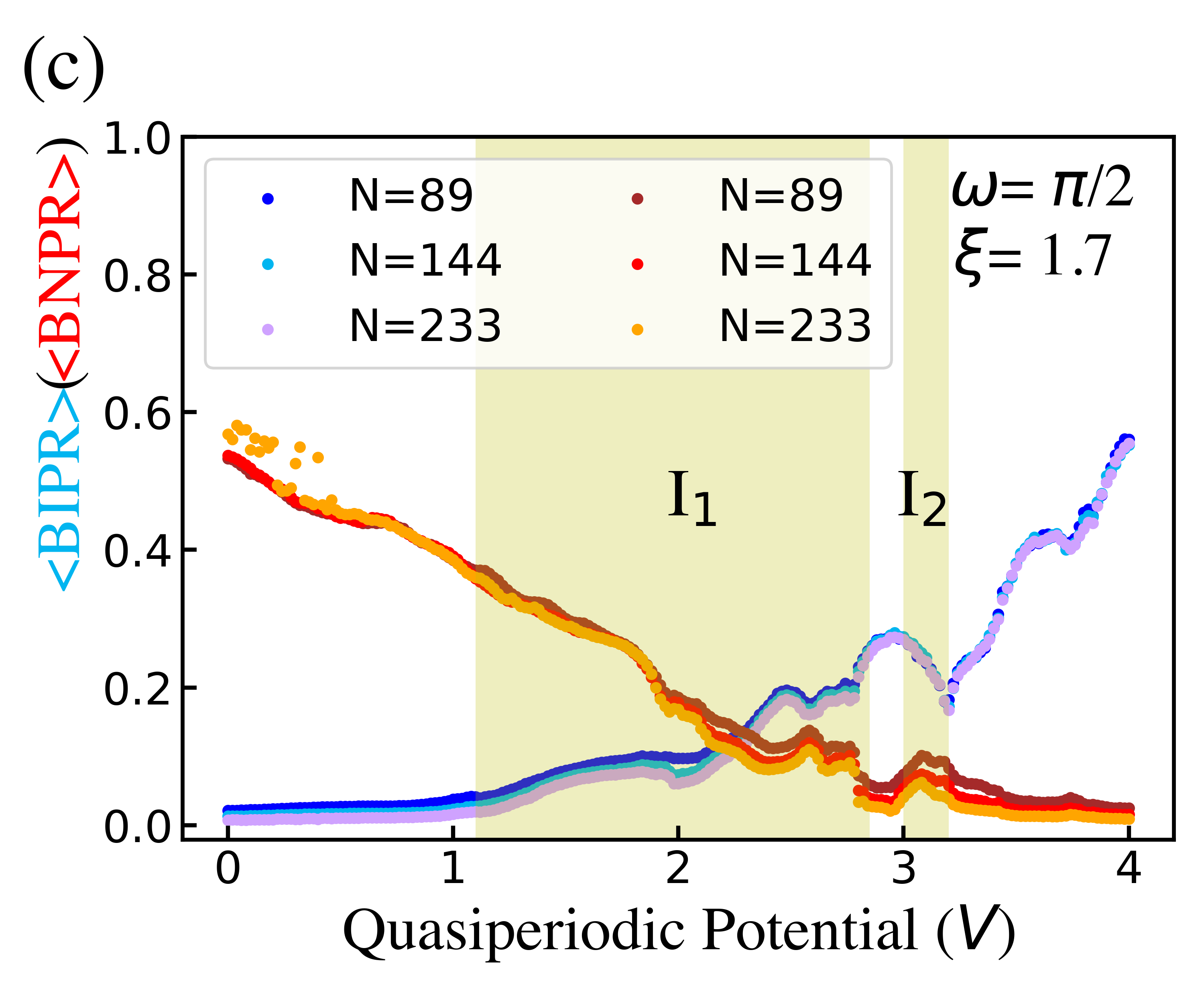}\hspace{0.1cm}
			\includegraphics[width=0.224\textwidth,height=0.232\textwidth]{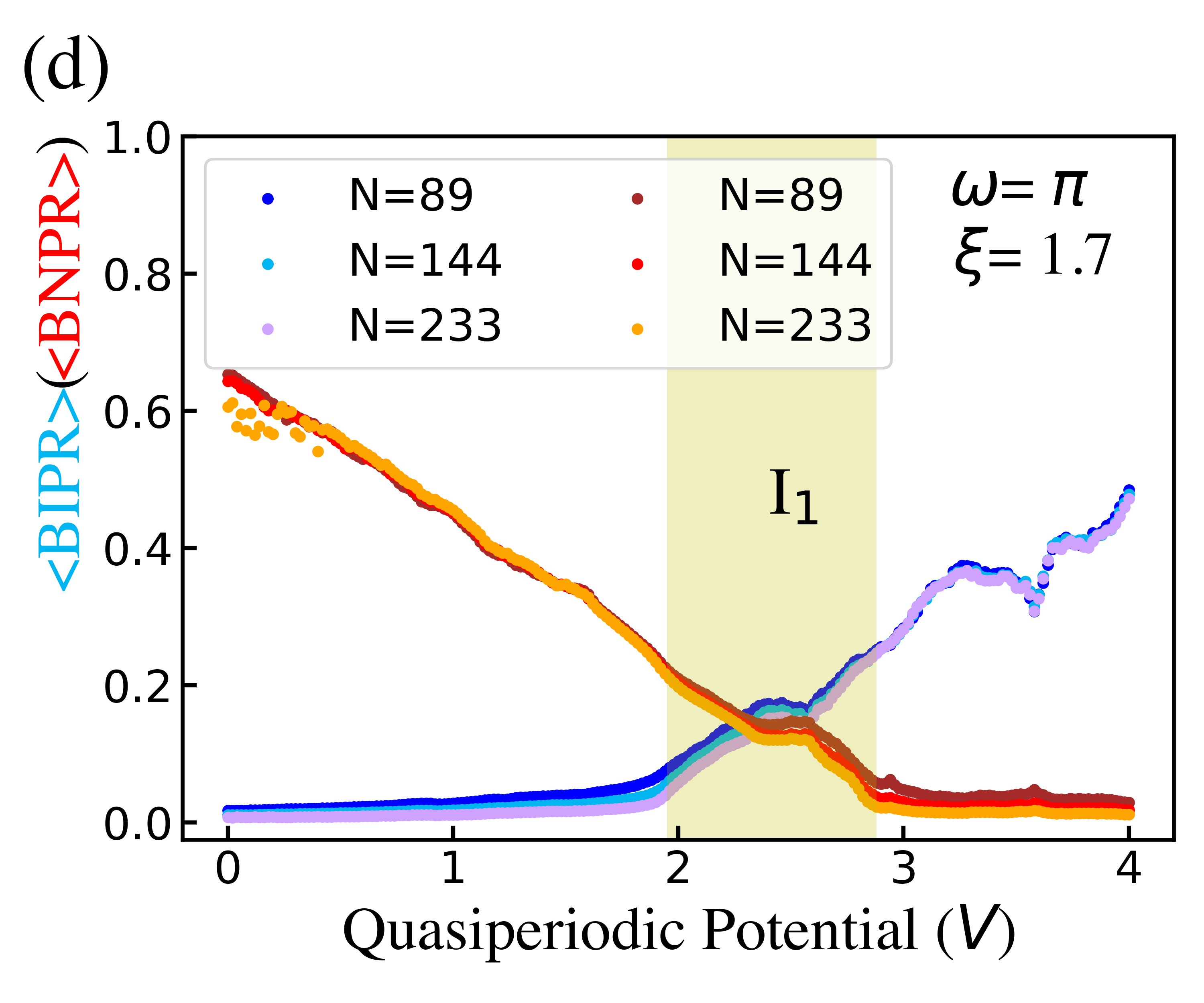}
			\caption{The phase diagram of $\eta$ at (a) $\omega=\pi/2$ and (b) $\omega=\pi$, for different strengths of the electric field ($\xi$) and quasiperiodic potential ($V$) in the time-dependent non-Hermitian system ($h=0.1$) under PBC. The lower panels illustrate the combined behavior of $\left\langle \text{BIPR} \right\rangle$ and $\left\langle \text{BNPR} \right\rangle$ for different system sizes at $\xi=1.7$ and (c) $\omega=\pi/2$, (d) $\omega=\pi$ in the intermediate regimes corresponding to the upper panel.}
			\label{Fig:Fig_3}
		\end{tabular}
	\end{figure}

	\subsection{The metal-insulator transition and the existence of mobility edges}\label{Sec:Transition}
	
\begin{figure}[]
		\begin{tabular}{p{\linewidth}c}
			\centering
			\includegraphics[width=0.249\textwidth,height=0.228\textwidth]  			{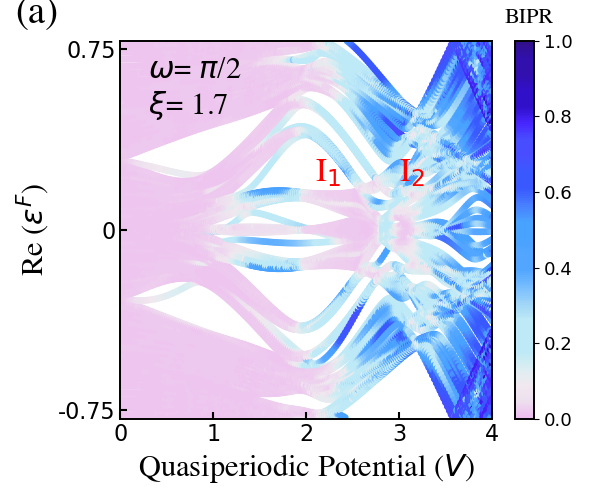}\hspace{-0.3cm}
			\includegraphics[width=0.245\textwidth,height=0.222\textwidth]{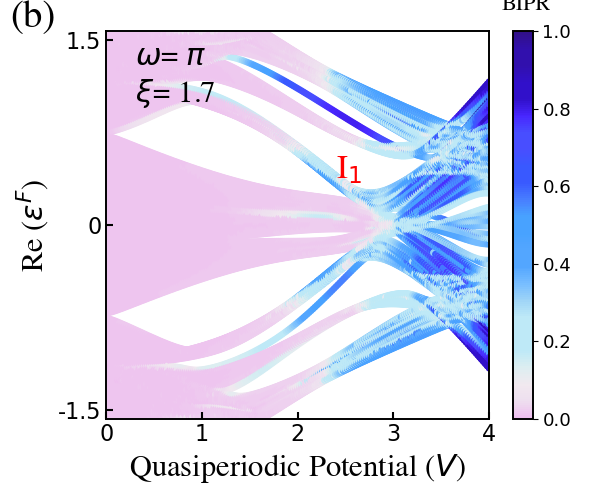}\hspace{-0.3cm}
			\caption{The existence of multiple mobility edges for the driven system demonstrated using the $BIPR$ for the real parts of the Floquet quasienergy and varying strengths of the quasiperiodic potential at $\xi=1.7$, where (a) $\omega=\pi/2$, and (b) $\omega=\pi$. The other parameters are same as in Fig.~\ref{Fig:Fig_3}.}
			\label{Fig:Fig_4}
		\end{tabular}
	\end{figure}

\begin{figure}[b]
			\includegraphics[width=0.245\textwidth,height=0.215\textwidth]{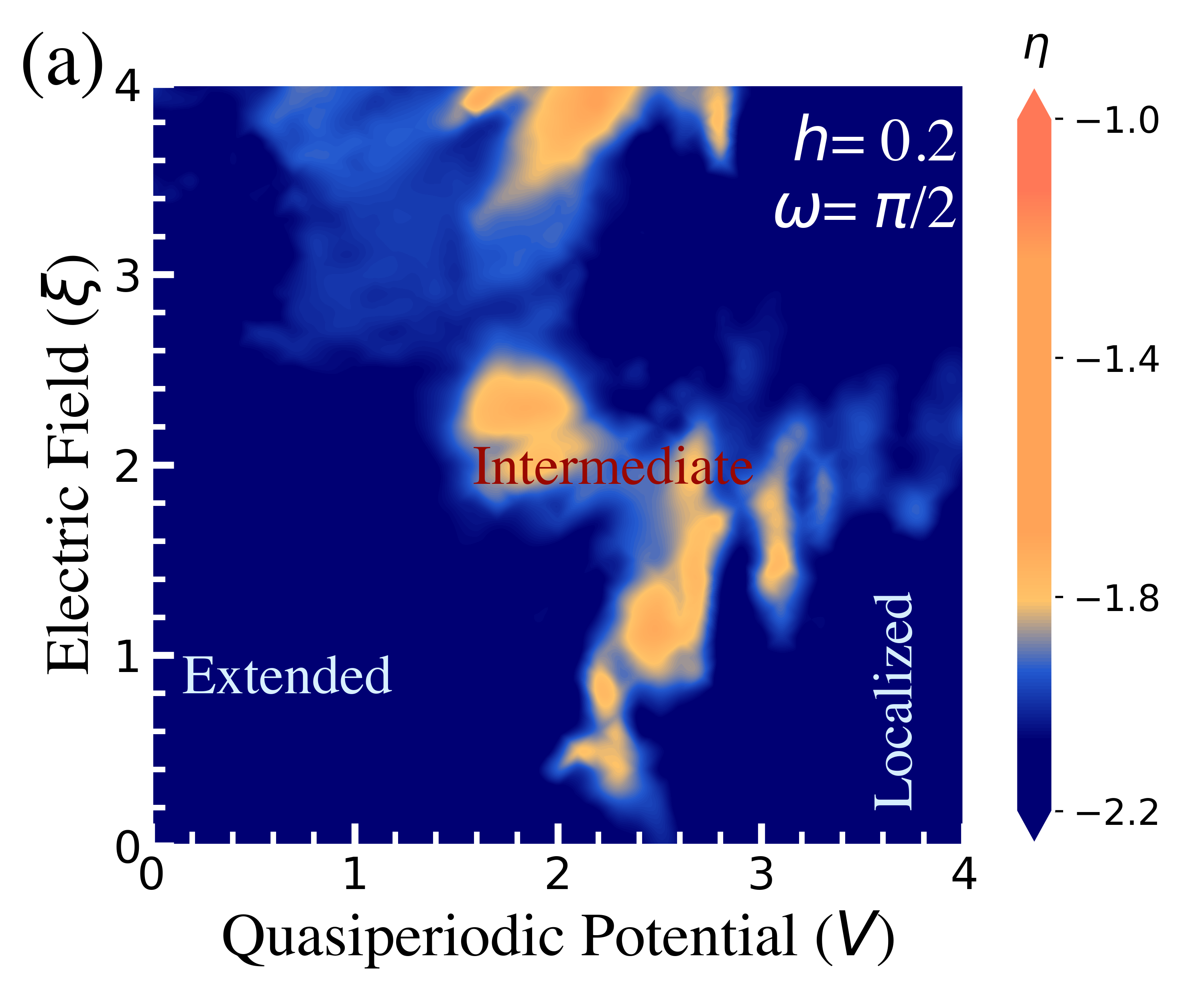}\hspace{-0.25cm}
			\includegraphics[width=0.245\textwidth,height=0.215\textwidth]{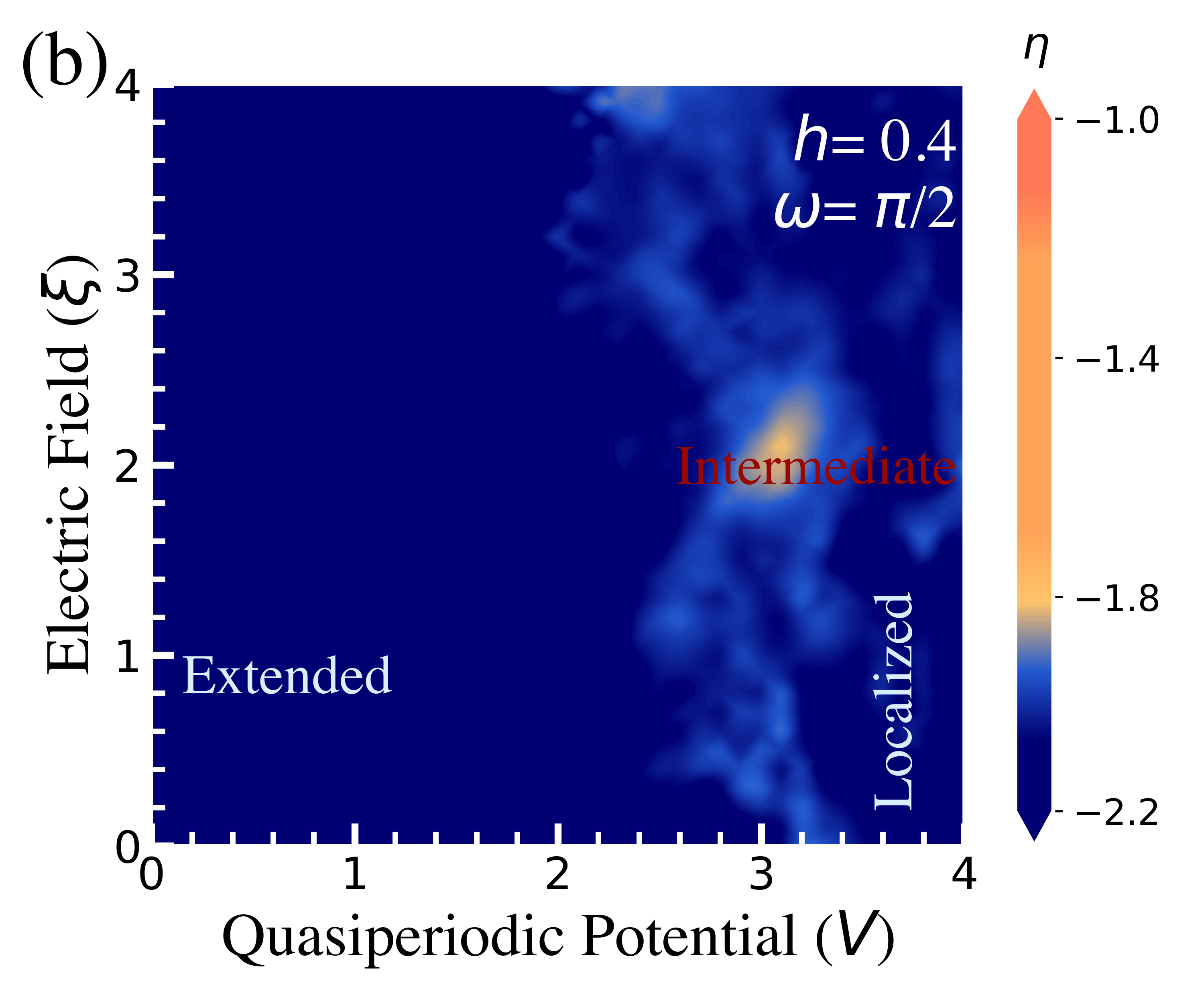}\\
			\includegraphics[width=0.245\textwidth,height=0.215\textwidth]{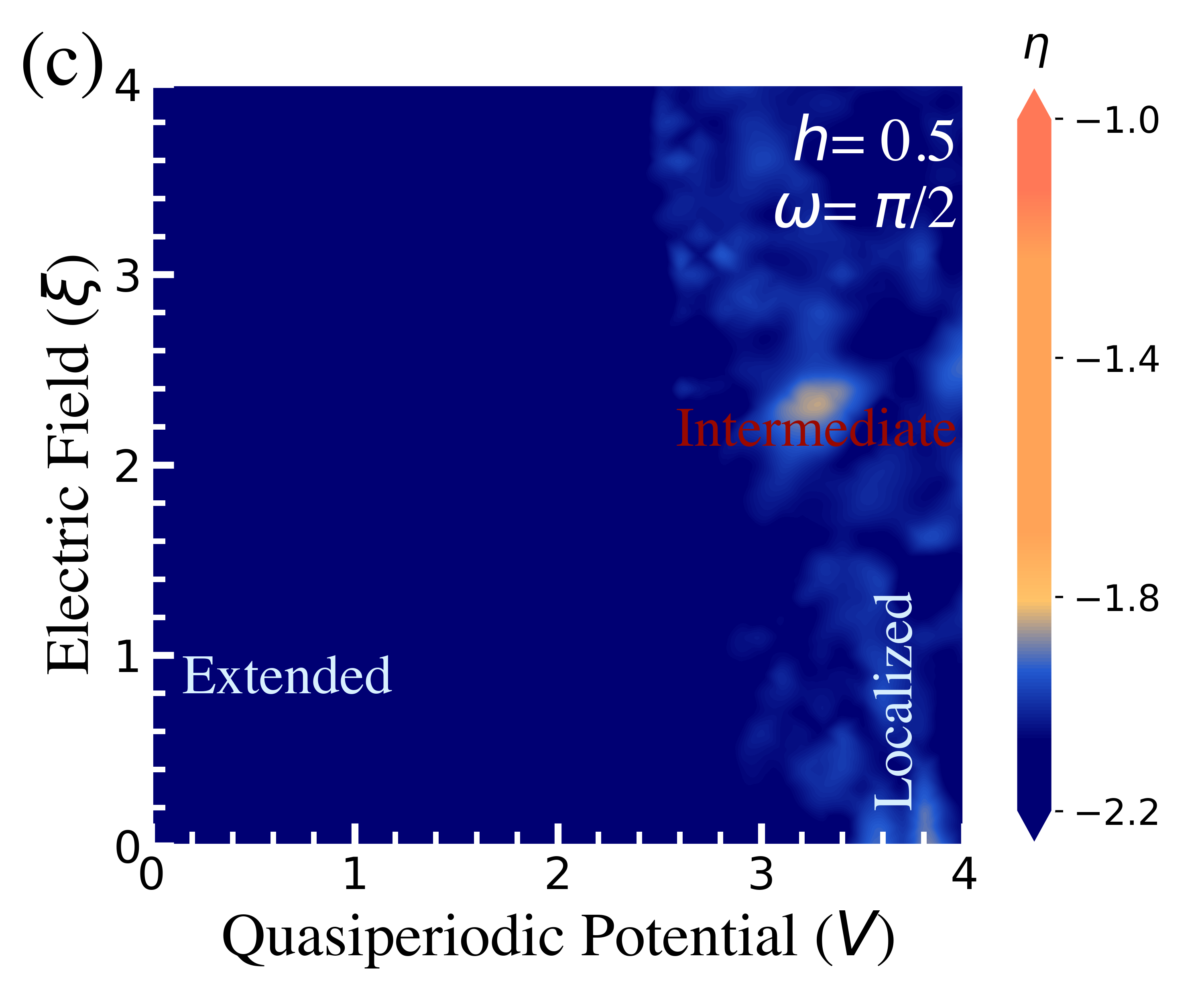}\hspace{-0.25cm}
			\includegraphics[width=0.245\textwidth,height=0.215\textwidth]{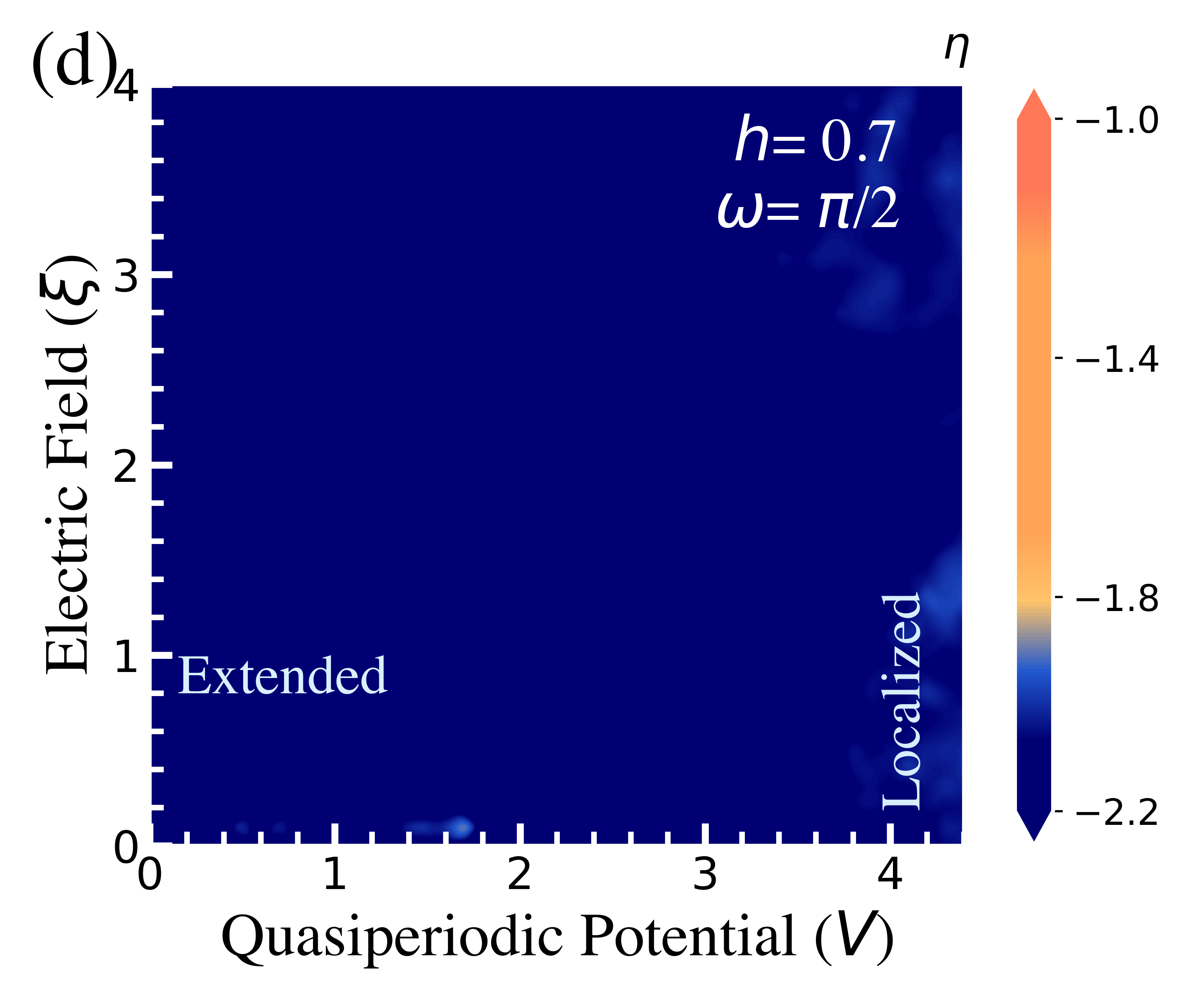}
			\caption{Phase diagram demonstrating the extent of intermediate regime at a low driving frequency ($\omega=\pi/2$) and (a) $h=0.2$, (b) $h=0.4$, (c) $h=0.5$, and (d) $h=0.7$ under the PBC.}
			\label{Fig:Fig_5}
	\end{figure}

	The indicator of $\left\langle \text{BIPR} \right\rangle$ is often used to characterize the metallic/insulating phases of a system as discussed in Sec.~\ref{Sec:Numerical_techniques}. 
	In Figs.~\ref{Fig:Fig_2}(a-c), we present the results of the $\left\langle \text{BIPR} \right\rangle$ projected over the entire parameter space of electric field and the quasiperiodic potential for low, moderate and high driving frequencies.
	It is quite evident that the inclusion of a time-periodic drive in the electric field enhances the delocalized regime to a significantly great extent as compared to its undriven counterpart (the results of the undriven analogue is demonstrated in Appendix \ref{App:Undriven}).
	In addition, interestingly, the boundary of the DL transition approaches the static quasiperiodic Hamiltonian with rescaled hopping amplitudes as determined in Eq.~(\ref{Eq:Critical_point}), separating completely delocalized and localized states when the frequency is very high (Fig.~\ref{Fig:Fig_2}(c)).\\
	\indent
	To determine the exact nature of the eigenstates at $\omega=\pi/2$ and $\omega=\pi$, we have plotted the phase diagram of $\eta$ in Figs.~\ref{Fig:Fig_3}(a,b), where the orange regime suggests a phase which can contain either a mixed spectrum of delocalized and localized states or an energy dependent mobility edge. Next, we select a certain strength of the electric field ($\xi=1.7$) for the two driving conditions in Figs.~\ref{Fig:Fig_3}(a,b) and select different system sizes in ascending order which eliminates the possibility of finite-size effects and confirms that the single-particle intermediate phase persists even for larger system sizes (achieved within our computational limit).
	The different intermediate phases are labelled as $I_1, I_2, \cdots$. We further probe the qualitative nature of the intermediate phase by plotting the quasienergy resolved values of BIPR across the entire quasiperiodic potential regime in Figs.~\ref{Fig:Fig_4}(a-b). It is clear that the intermediate phases in Figs.~\ref{Fig:Fig_3}(c,d) are indeed mobility edges. In the low frequency limit (Fig.~\ref{Fig:Fig_4}(a)), one of the mobility edge (in the region labelled by $I_1$) confirms a delocalized-localized-delocalized separation between the energy eigenstates. However, the mobility edges (labelled by $I_2$ regime) exhibits a localized-delocalized-localized behavior of the eigenstates. Unlike the case for low drive frequency, when the frequency of the drive is moderate (Fig.~\ref{Fig:Fig_4}(b)), there exists a single intermediate phase ($I_1$)  with localized-delocalized-localized behavior of the eigenstates. It is important to note that the mobility edge depends crucially upon the electric field strength, as evident from Figs.~\ref{Fig:Fig_3}(a-b).
	To further understand whether such intermediate regimes at low driving frequency ($\omega = \pi/2$)
	appears irrespective of the strength of non-Hermiticity ($h$), we plot the phase diagram of $\eta$ in Figs.~\ref{Fig:Fig_5}(a-d) with increasing values of $h$. We find that the extent of the intermediate regime narrows down till $h=0.5$. The intermediate regime absolutely vanishes when $h=0.7$ (Fig.~\ref{Fig:Fig_5}(d)), separating completely delocalized and localized eigenstates. The threshold
	value of $h$ upto which the mobility edge persists, decreases with the increase in the driving
	frequency. This is evident from Fig.~\ref{Fig:Fig_2}(c). For brevity, we have not presented the 
	results for $\omega = \pi$.
	In addition, these findings are also verified in a Hermitian driven system (illustrated in Appendix \ref{App:Hermitian}), which indicates that the non-trivial intermediate phases are generated solely due to the time-periodic nature of the electric field.
	
	\begin{figure}[]
		\begin{tabular}{p{\linewidth}c}
			\centering
			\includegraphics[width=0.245\textwidth,height=0.215\textwidth]{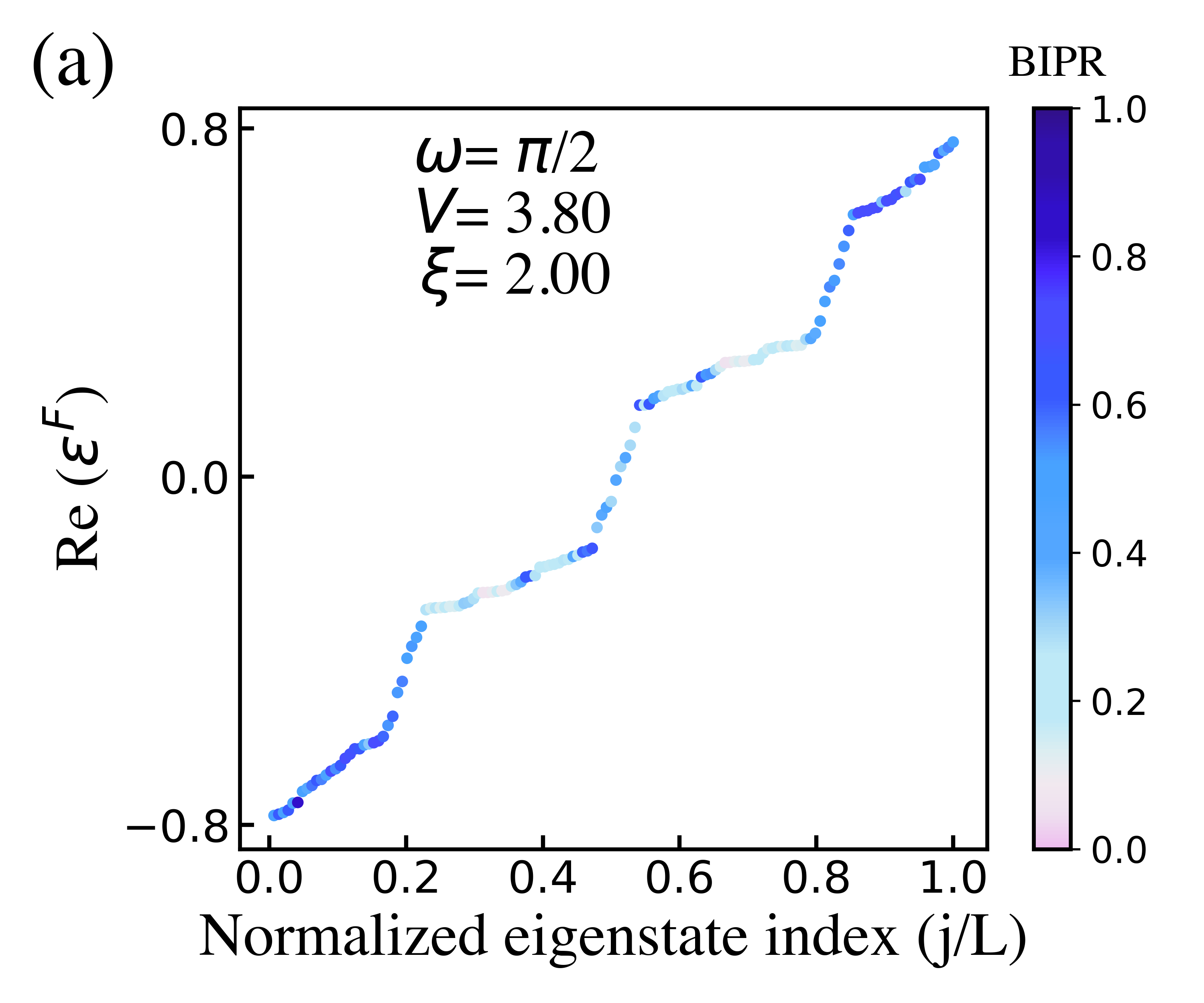}\hspace{-0.25cm}
			\includegraphics[width=0.245\textwidth,height=0.215\textwidth]{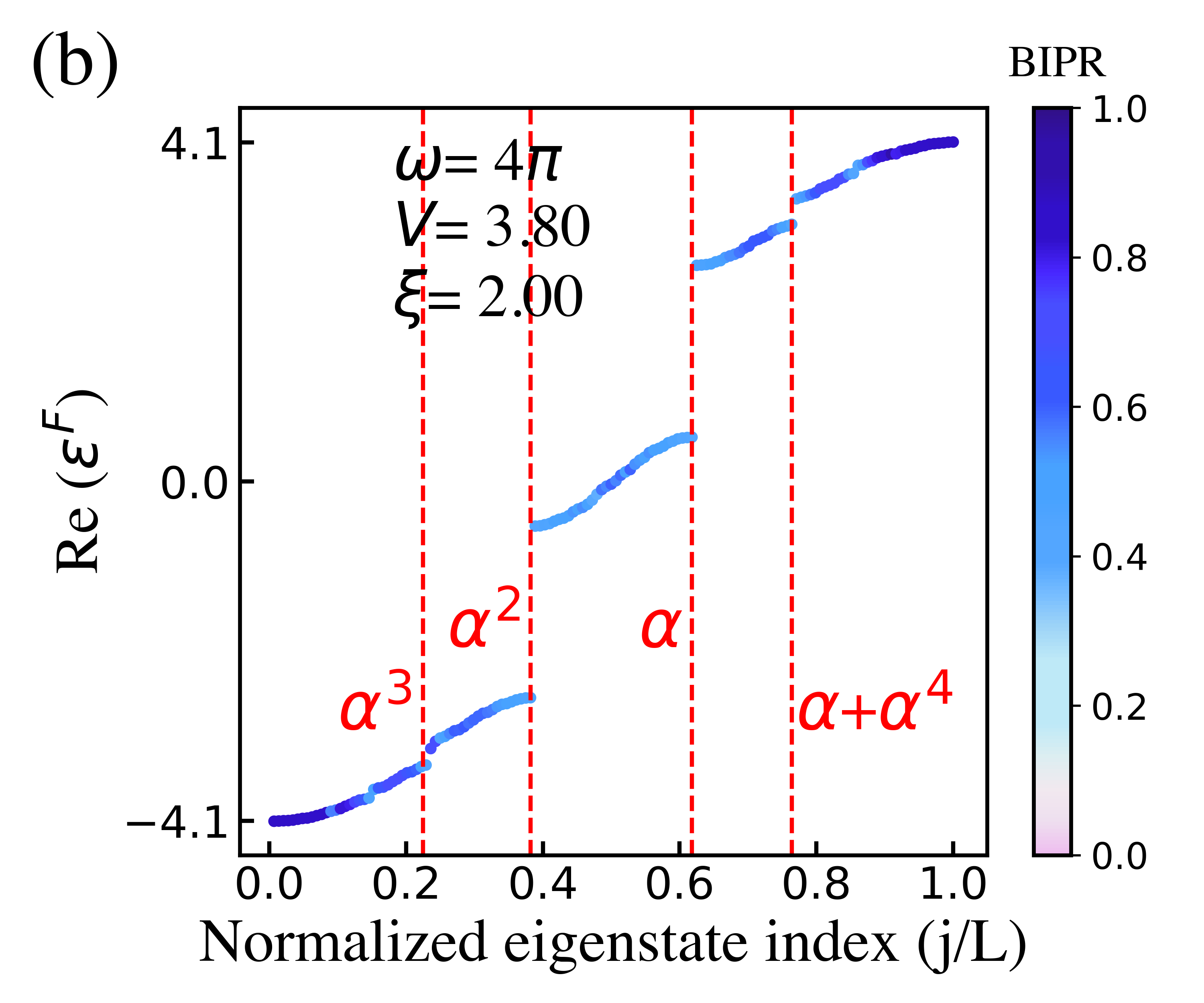}
			\caption{The real Floquet quasienergies as a function of the normalized eigenstate index in the localized regime ($V=3.80$ and $\xi=2.0$) of the periodically driven system under the PBC at (a) $\omega=\pi/2$ and (b) $\omega=4\pi$, denoting the absence of WS ladders. The values of $\alpha^z+\alpha^{z'}$ where the band gaps appear are demonstrated clearly.}
			\vspace{-0.2cm}
			\label{Fig:Fig_6}
		\end{tabular}
	\end{figure}
	
	\begin{figure}[b]
		\begin{tabular}{p{\linewidth}c}
			\centering
			\includegraphics[width=0.245\textwidth,height=0.215\textwidth]{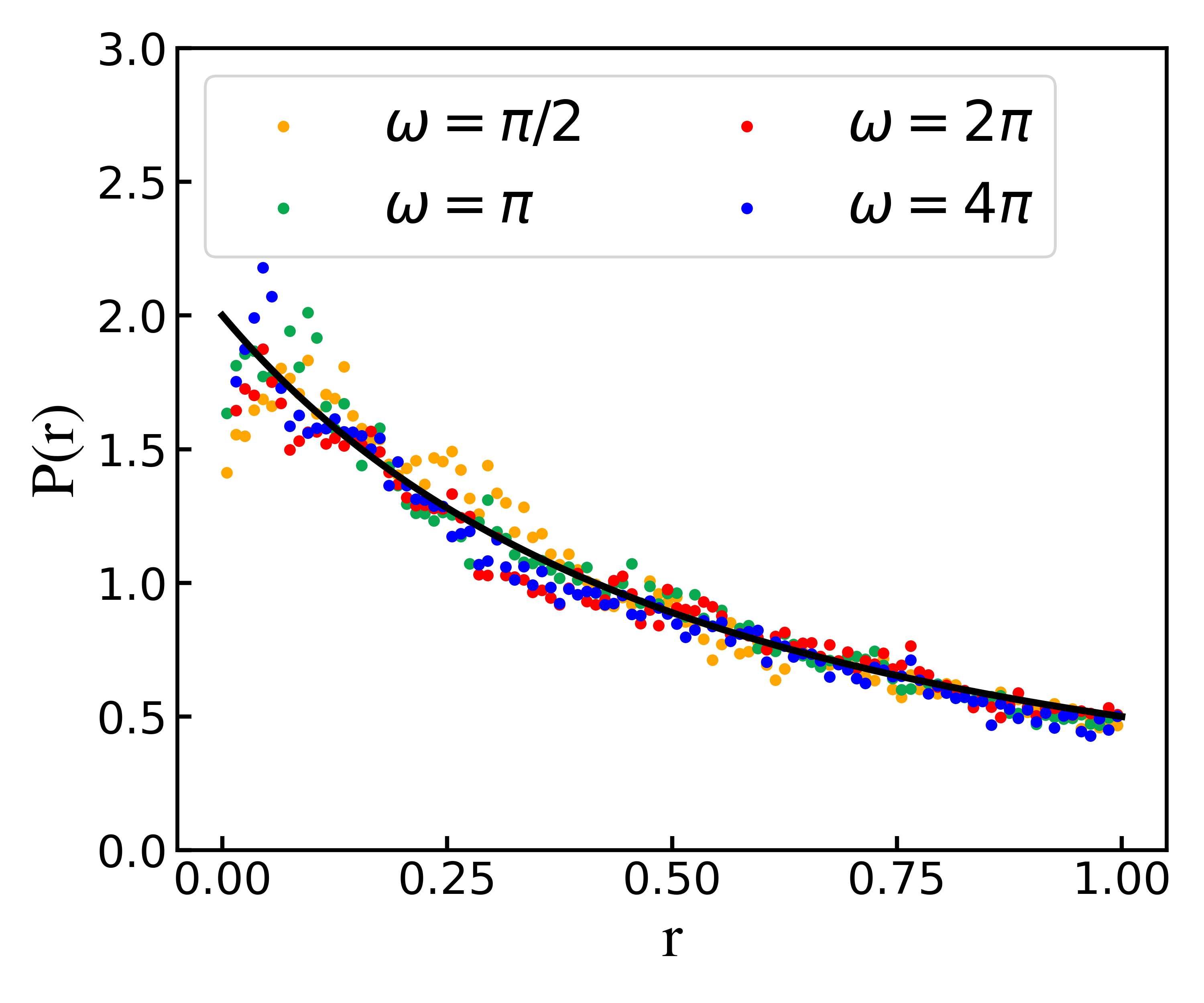}\hspace{-0.25cm}
			\caption{The probability $P(r)$ as a function of $r$ at different frequencies of the drive: $\omega=\pi/2$ (in orange), $\omega=\pi$ (in green), $\omega=2\pi$ (in red) and $\omega=4\pi$ (in blue). The data are obatined for 500 realizations in $\phi$. The distribution of the level statistics in the Poissonian limit, i.e., $P(r)=2/(1+r)^2$ is shown in the black solid line. To ensure that the states are completely localized, we have considered $V=3.8$ and $\xi=2.0$ (blue regime of the phase diagram in Fig.~\ref{Fig:Fig_2}).}
			\vspace{-0.2cm}
			\label{Fig:Fig_7}
		\end{tabular}
	\end{figure}
	
	\begin{figure}[t]
		\begin{tabular}{p{\linewidth}c}
			\centering
			\includegraphics[width=0.245\textwidth,height=0.215\textwidth]{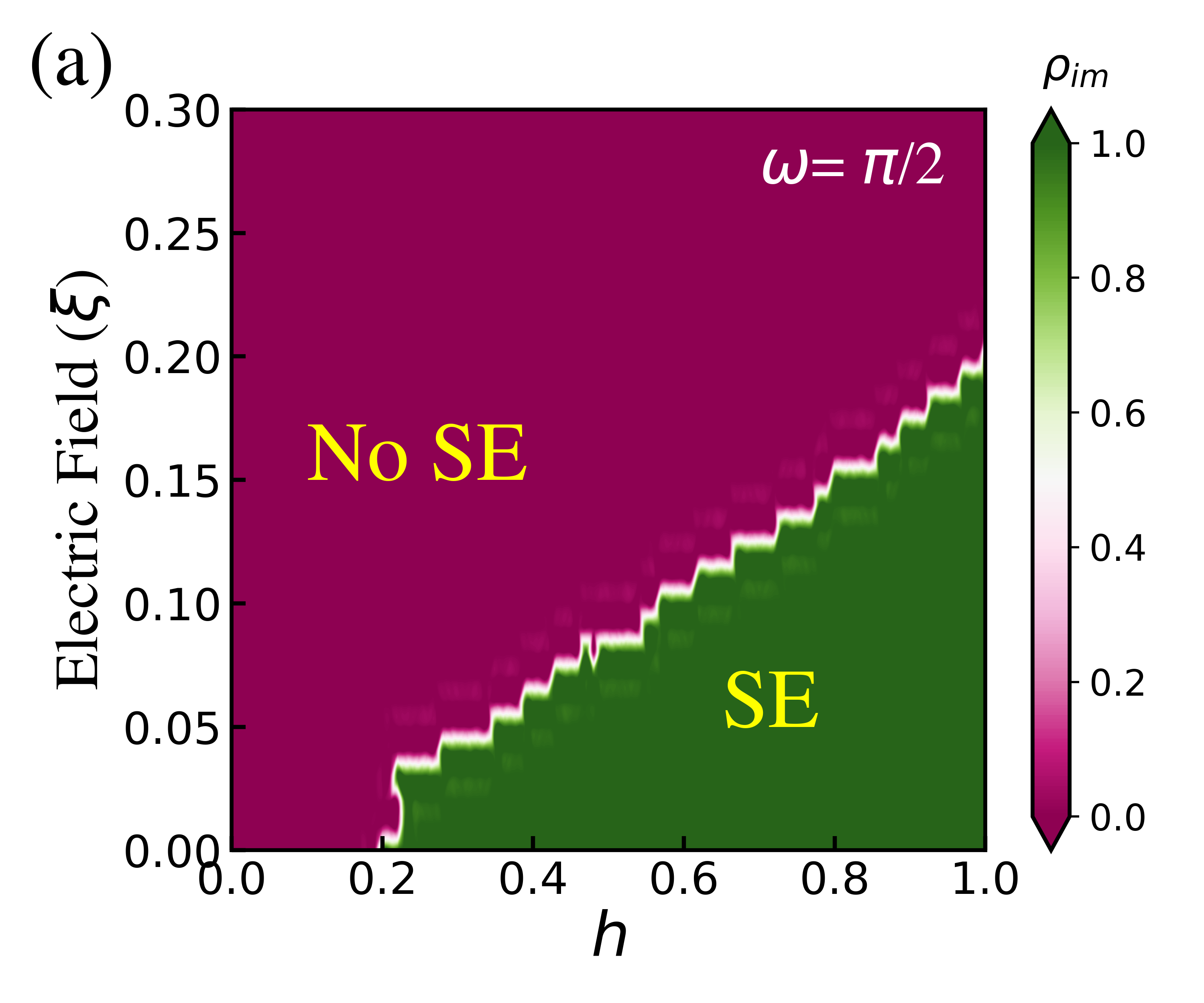}\hspace{-0.25cm}
			\includegraphics[width=0.245\textwidth,height=0.215\textwidth]{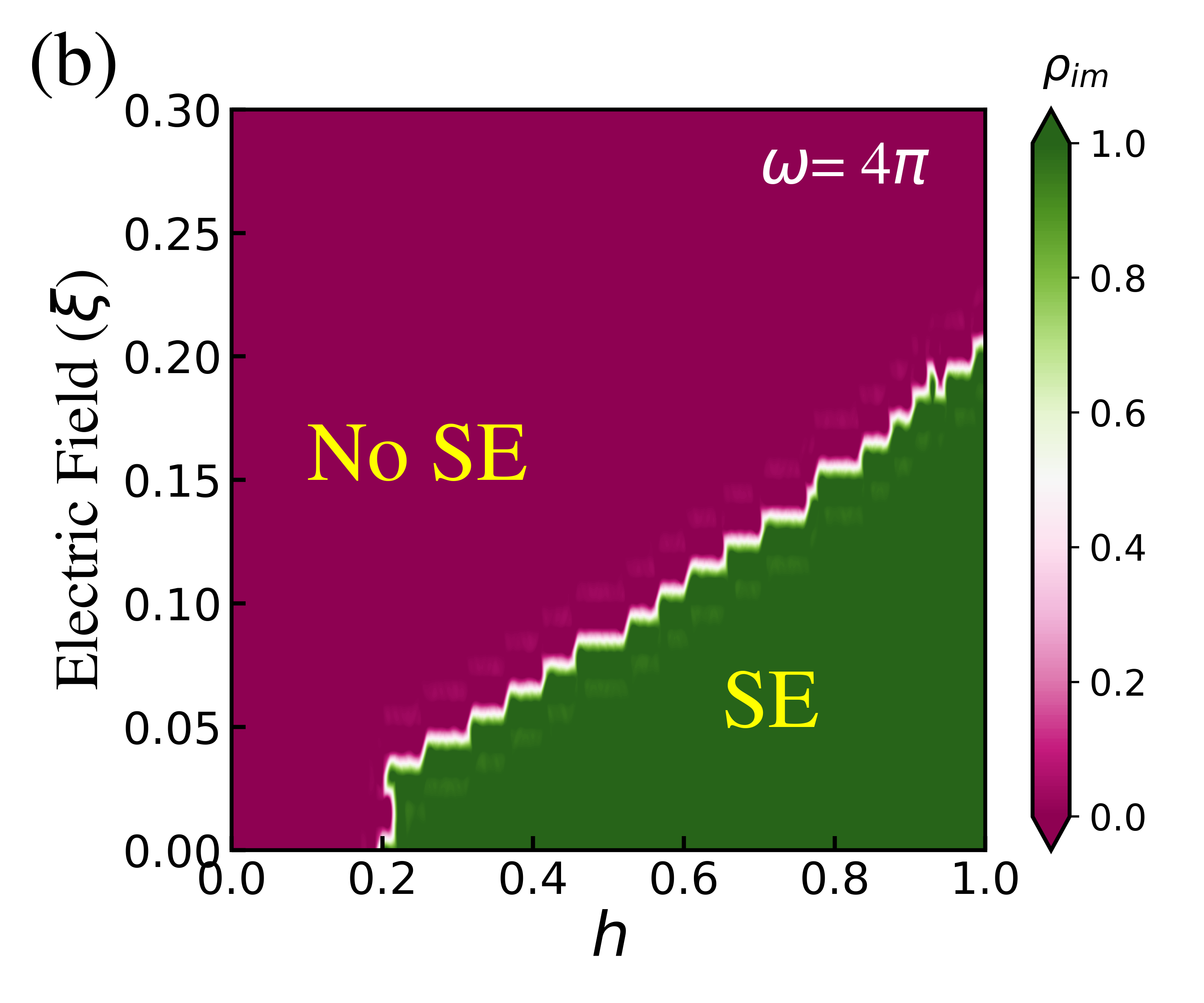}\\
			\includegraphics[width=0.242\textwidth,height=0.200\textwidth]{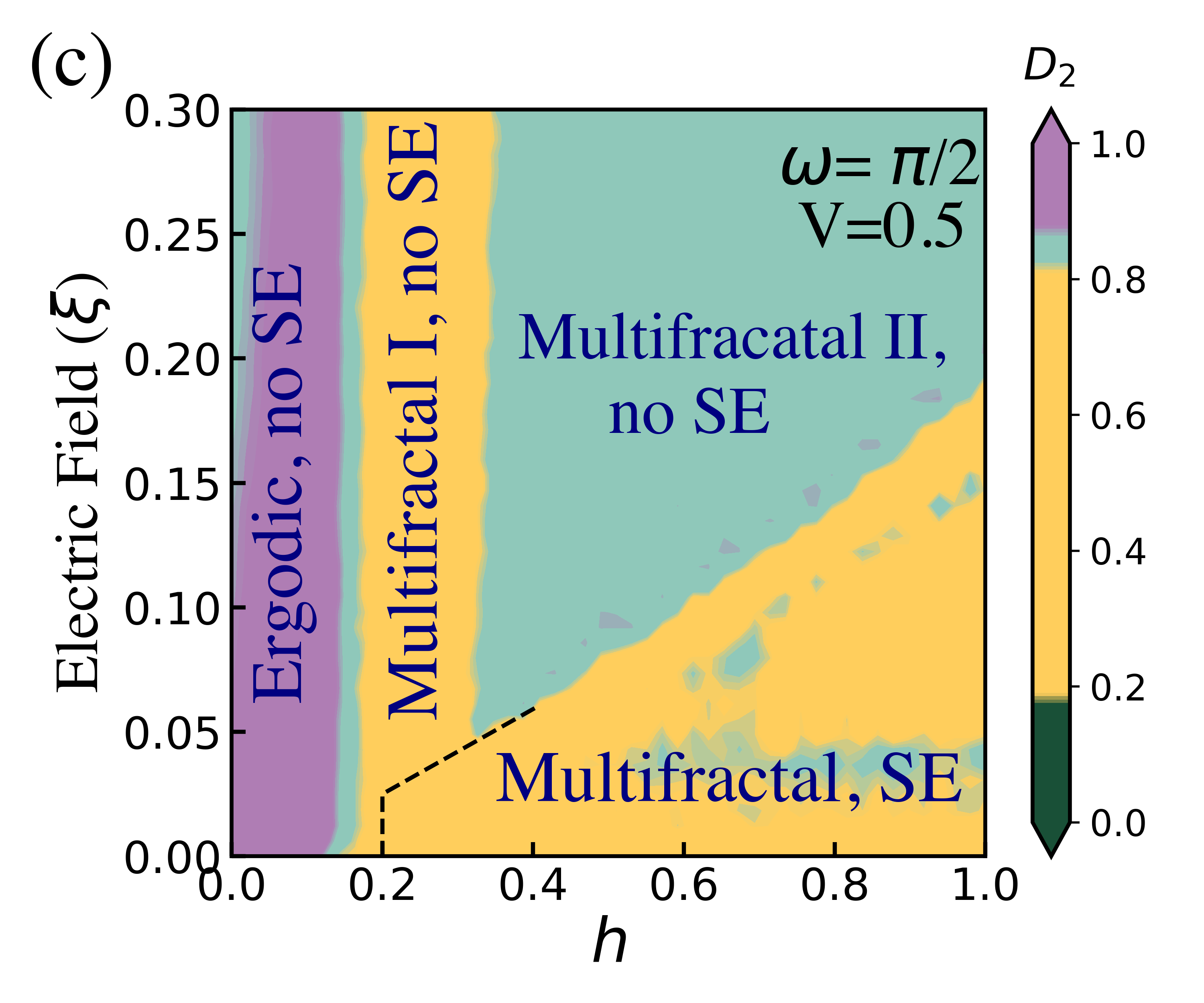}\hspace{-0.20cm}
			\includegraphics[width=0.242\textwidth,height=0.200\textwidth]{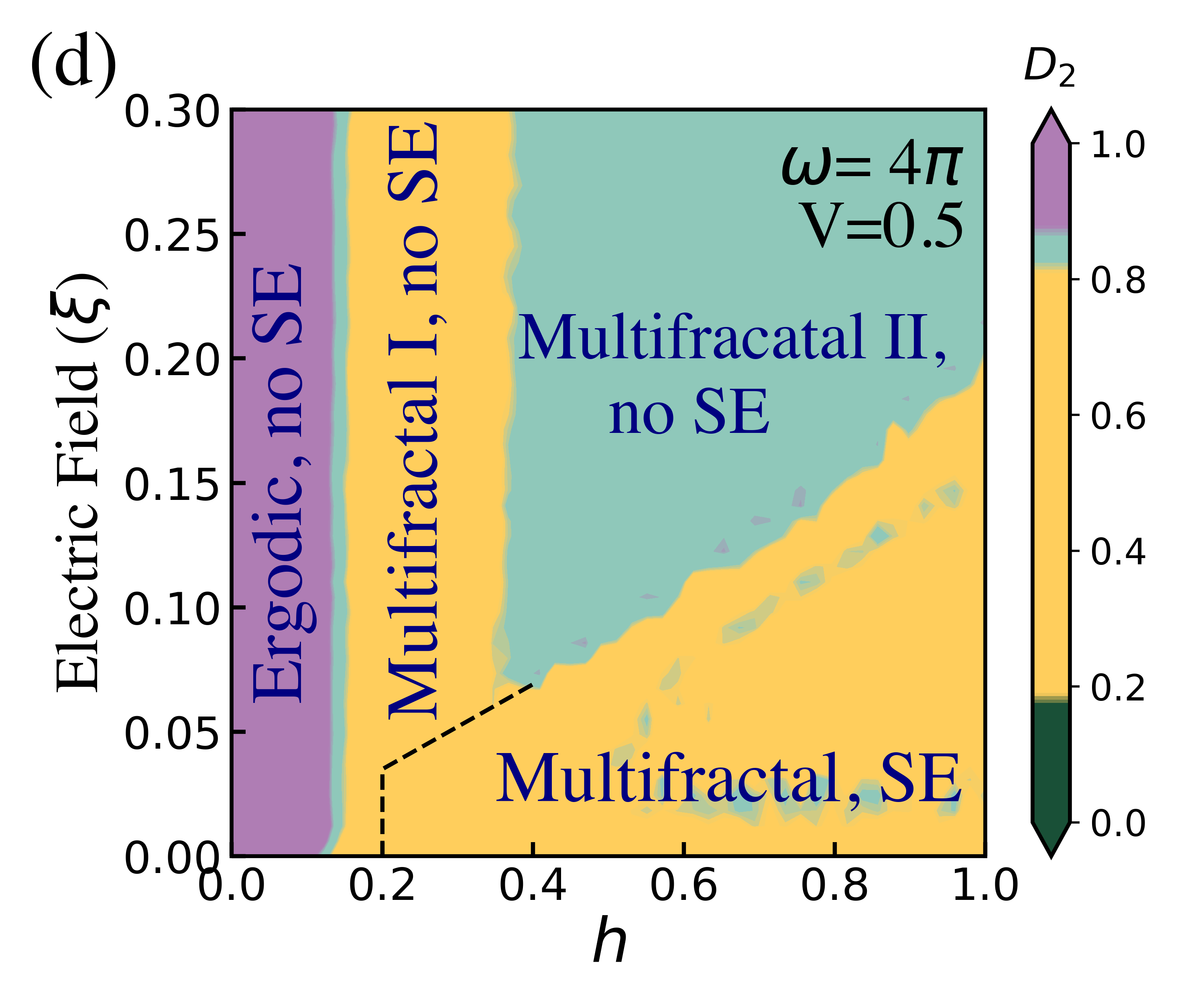}
			\caption{(a)-(b) Phase diagram demonstrating the demarcation of the regime where the SE under OBC is either present or absent in the parameter space of $\xi$ and $h$ using the indicator $\rho_{im}$ at (a) $\omega=\pi/2$ and (b) $\omega=4\pi$. (c)-(d) The phase diagram of mean fractal dimension $D_2$ at driving frequencies (c) $\omega=\pi/2$, and (d) $\omega=4\pi$ corresponding to (a) and (b) respectively. The different phases characterized by $D_2$ are labelled for clarity. The black dotted line is a guide to the eye to separate the SE-no SE phases. The quasiperiodic potential is set at $V=0.5$ (in the delocalized regime of Fig.~\ref{Fig:Fig_2}).}
			\vspace{-0.2cm}
			\label{Fig:Fig_8}
		\end{tabular}
	\end{figure}

	\indent	
	The importance of these results are two-fold.
	In the first place, it can now be inferred that the drive in the electric field suppresses the localization behavior, typically observed due to even a tiny electric field \cite{Wannier_1962}.
	Next, our results suggest that an external time periodic drive generates mobility edge(s) in 
	pure AAH type quasicrystals. Interestingly, however, the mobility edges are absent for high frequencies in the drive as is clearly evident from Fig.~\ref{Fig:Fig_2}(c). 

	\subsection{The Wannier-Stark localization and the level statistics in the localized phase}\label{Sec:WS_localization}
	
	It is well known that the presence of a time-independent  electric field gives rise to WS localization, wherein the states are separated by equal spacing in energies.
	The equi-spaced energy levels in a WS ladder are of the form, $E_n=E_0+ne{\xi}a$, where $E_0$ is the energy of the lower band, $n$ being a positive integer.
	The WS ladders for an undriven system in the localized regime of a non-Hermitian system is shown for clarity in Fig.~\ref{Fig:Fig_S1}(d) in Appendix \ref{App:Undriven}.
	However, unlike the scenario of a static electric field, in the presence of the cosine modulated drive, the WS ladders disappear for the frequencies considered in Figs.~\ref{Fig:Fig_6}(a,b). This 
	conclusion is equally valid for $\omega = \pi$. 
	In addition, surprisingly, when the driving frequency is sufficiently high, the energy bands in the localized phase are splitted into sub-bands separated by an energy gap, quite similar to the delocalized regime of the undriven AAH systems, as illustrated in Fig.~\ref{Fig:Fig_6}(b).
	These bandgaps appear at $\alpha^z$ or $\alpha^z$+$\alpha^{z'}$ (as marked in Fig.~\ref{Fig:Fig_6}(b)), where $z,z'\in\mathbb{Z}$ as discussed in Appendix \ref{App:Undriven}.
	Moreover, the Floquet quasienergies of such non-Hermitian driven systems are real in the localized regime due to the existence of time-reversal symmetry in the system. This has been explicilty shown in Appendix~\ref{App:Floquet_energies} for clarity.\\
	\indent To identify the distribution of the energy levels in the localized regime, we use the measure of $r_n$ as described in Eq.~(\ref{Eq:r_statistics}).
	It is possible to use this measure even in the non-Hermitian system considered in our case, since the quasienergies turn out to be real in the localized phase as discussed.
	We find that at all frequencies of the drive, $\langle r \rangle \simeq 0.38$, i.e., the energy levels are uncorrelated and belong to the Poissonian distribution. This is verified in Fig.~\ref{Fig:Fig_7}, which shows that the probability distribution in the driven non-Hermitian system that we have considered agrees quite well with $P(r)=2/(1+r)^2$.
	This is in stark contrast to the WS energy level distribution which shows a $\delta$-type probability distribution with $\langle r \rangle \simeq 1$, as discussed in Sec.~\ref{Sec:Level statistics}.
	These results indicate that a time-periodic drive in the electric field can alter the nature of localization.
		
	\begin{figure}[]
		\begin{tabular}{p{\linewidth}c}
			\includegraphics[width=0.2450\textwidth,height=0.228\textwidth]  
			{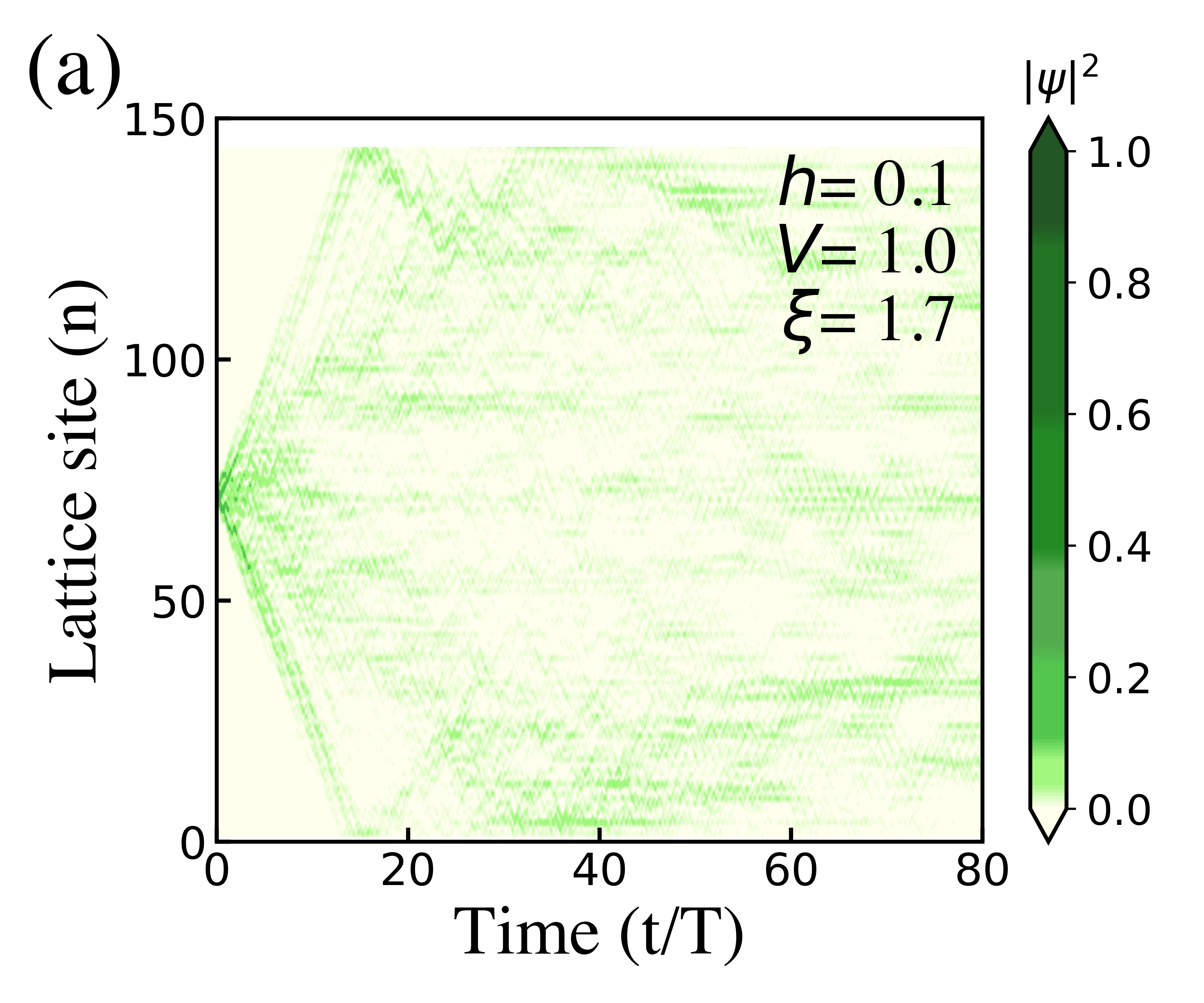}\hspace{-0.1cm}
			\includegraphics[width=0.2450\textwidth,height=0.228\textwidth]{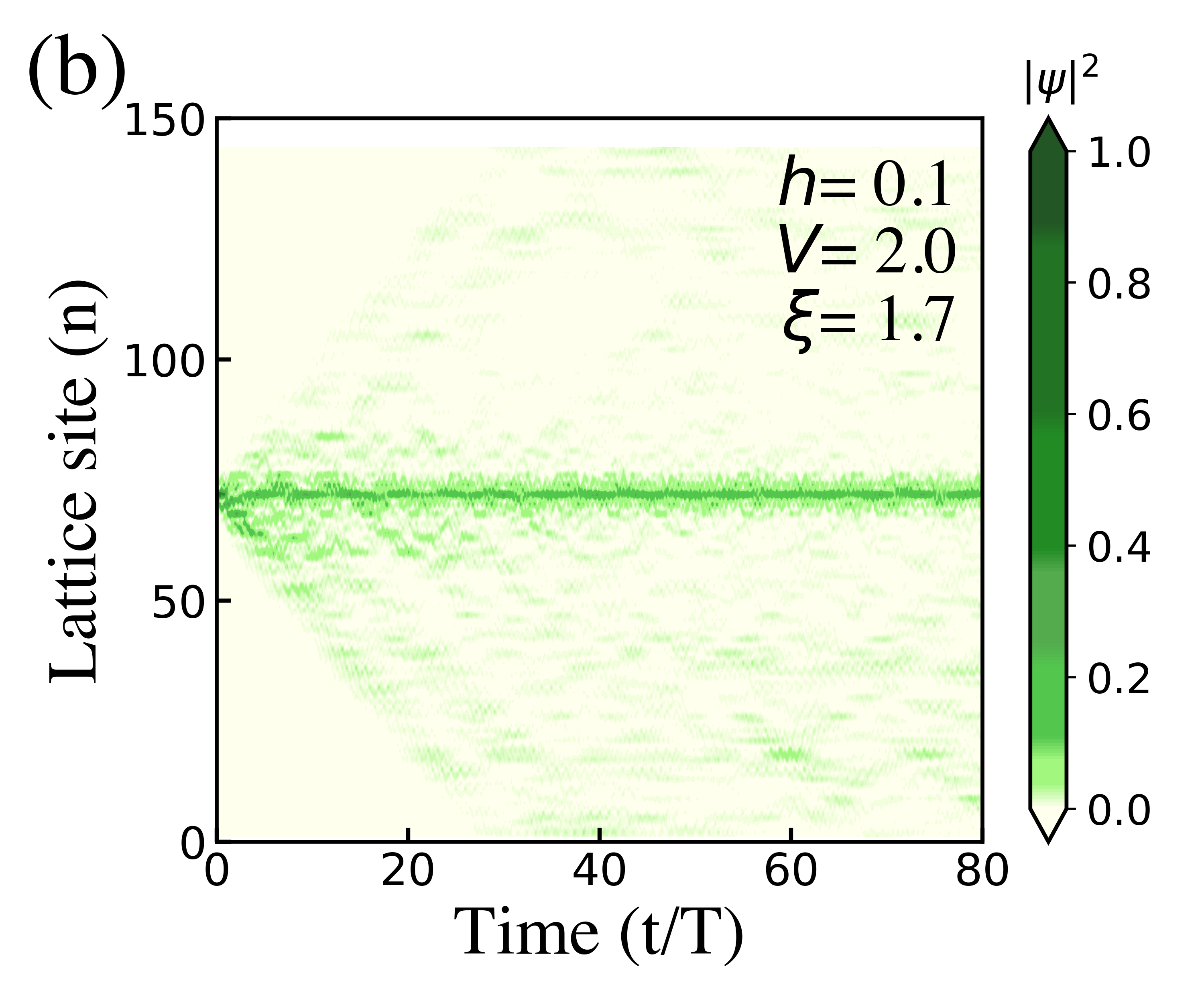}\\
			\includegraphics[width=0.245\textwidth,height=0.228\textwidth]{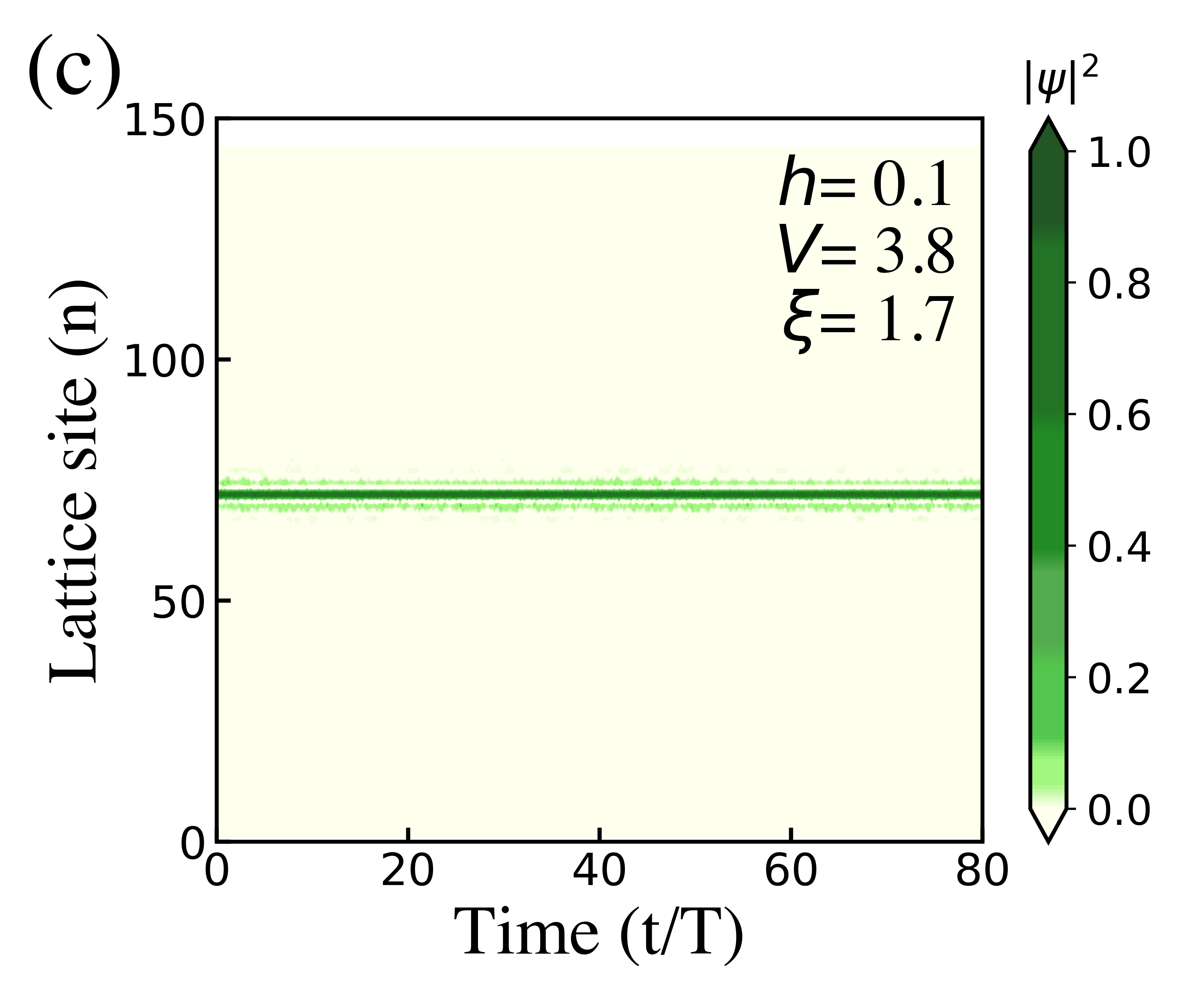}\hspace{-0.1cm}
			\includegraphics[width=0.22\textwidth,height=0.224\textwidth]{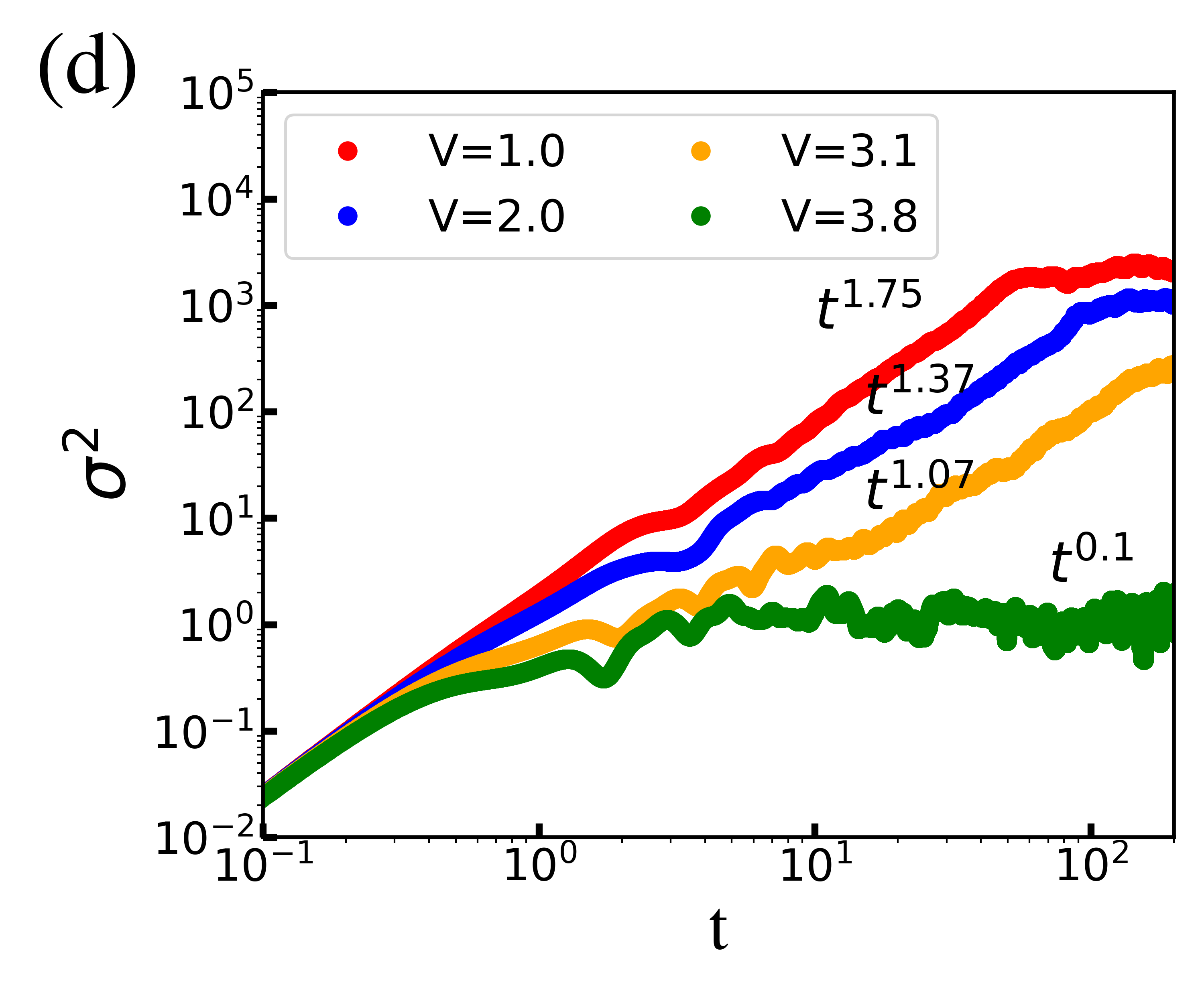}
			\caption{Amplitude of the time-evolved wavepacket ($|\psi|^2$) as a function of all lattice sites and at long times (scaled with the driving period $T$) at $\omega=\pi/2$. The electric field strength is set at $\xi=1.7$. Other parameters are $h=0.1$ and (a) $V=1.0$, (b) $V=2.0$ and  (c) $V=3.8$ and corresponding to the delocalized, intermediate ($I_1$) and localized regime in Figs.~\ref{Fig:Fig_3}(a,c) respectively. The time-evolution in the intermediate regime $I_2$ is qualitatively similar to that shown in (b). (d) MSD as a function of time in double log-scale in the four regimes as mentioned.}
			\label{Fig:Fig_9}
		\end{tabular}
	\end{figure}

	\begin{figure}[]
		\begin{tabular}{p{\linewidth}c}
			\includegraphics[width=0.2450\textwidth,height=0.228\textwidth]{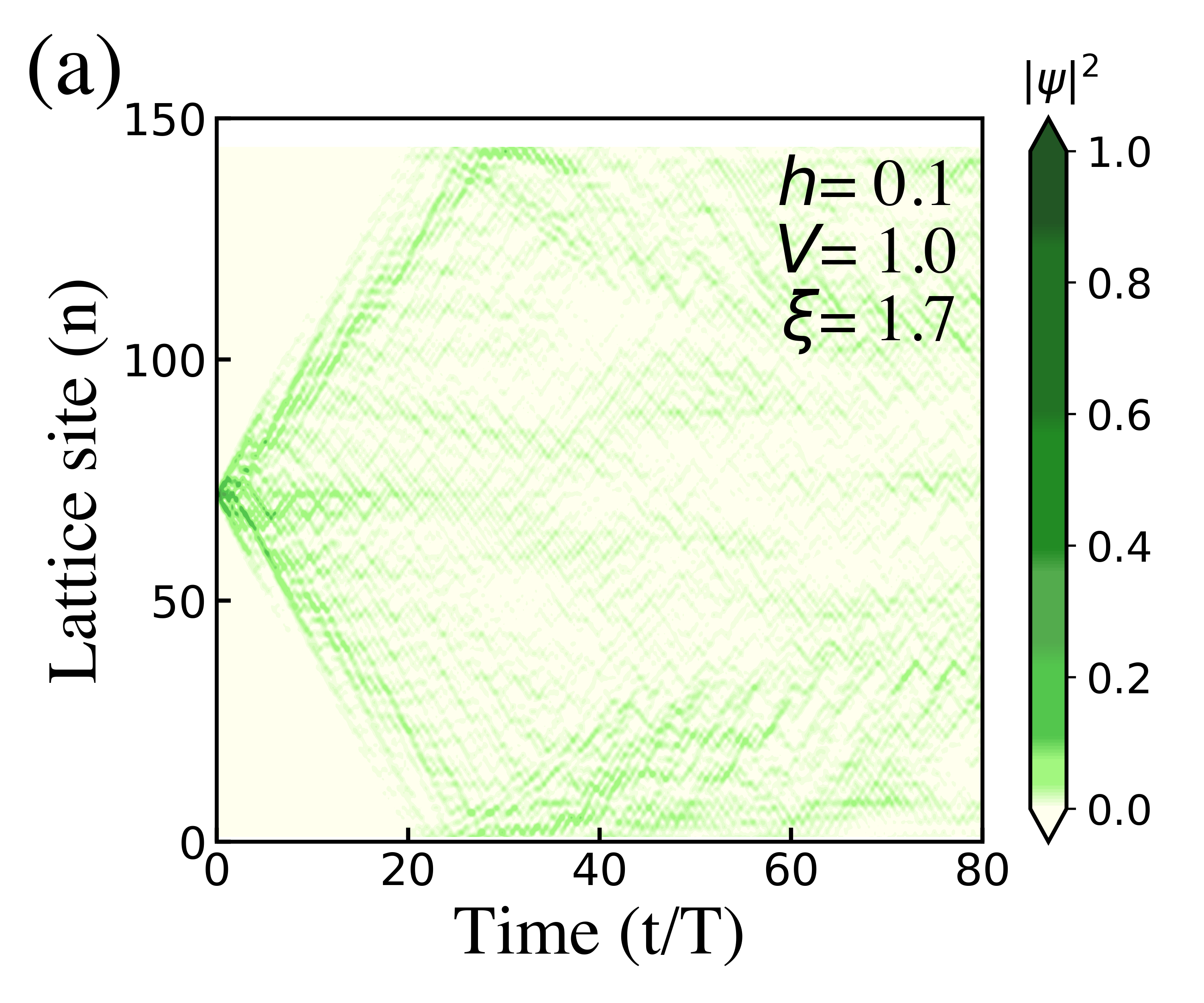}\hspace{-0.1cm}
			\includegraphics[width=0.2450\textwidth,height=0.228\textwidth]{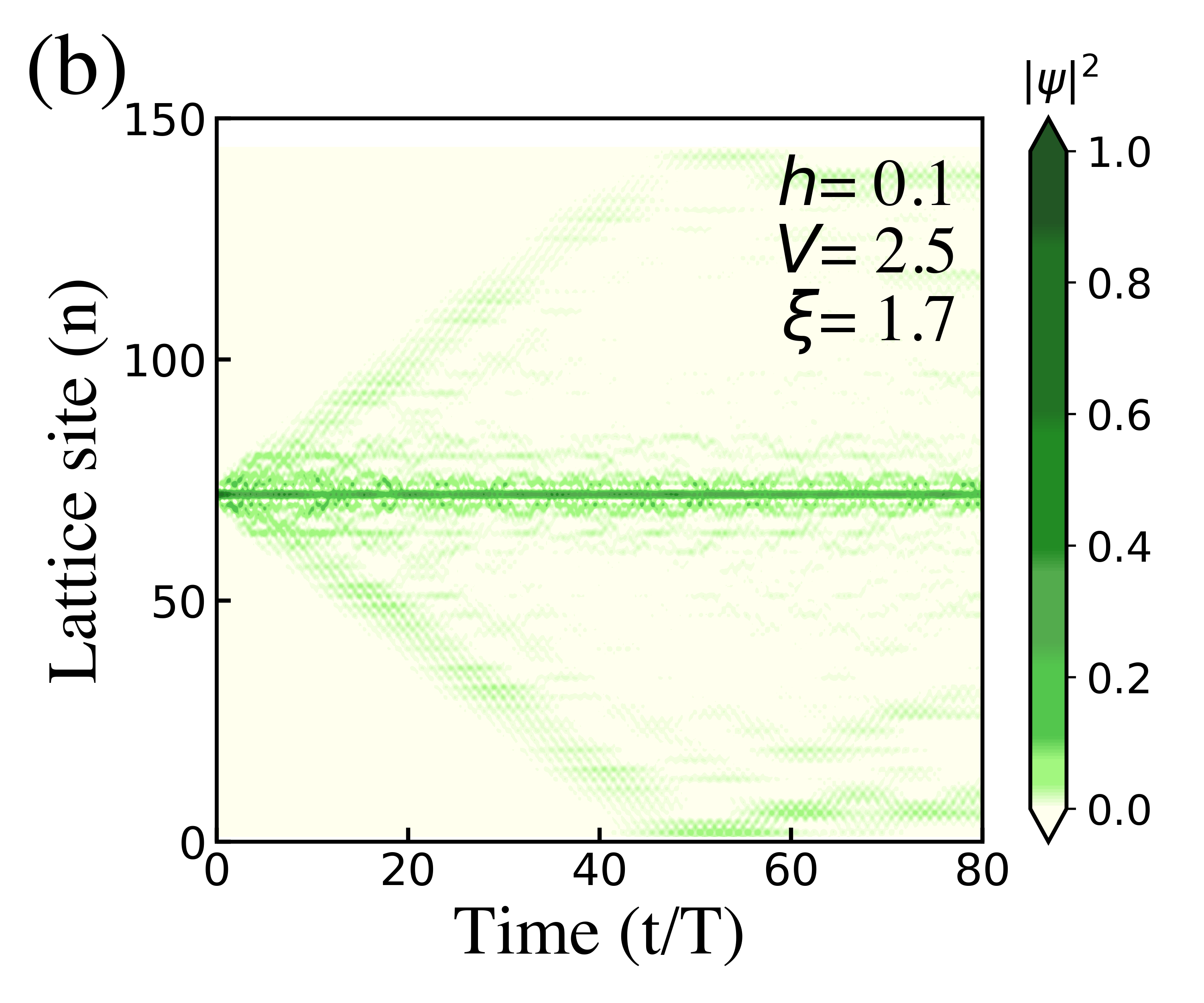}\\
			\includegraphics[width=0.245\textwidth,height=0.228\textwidth]{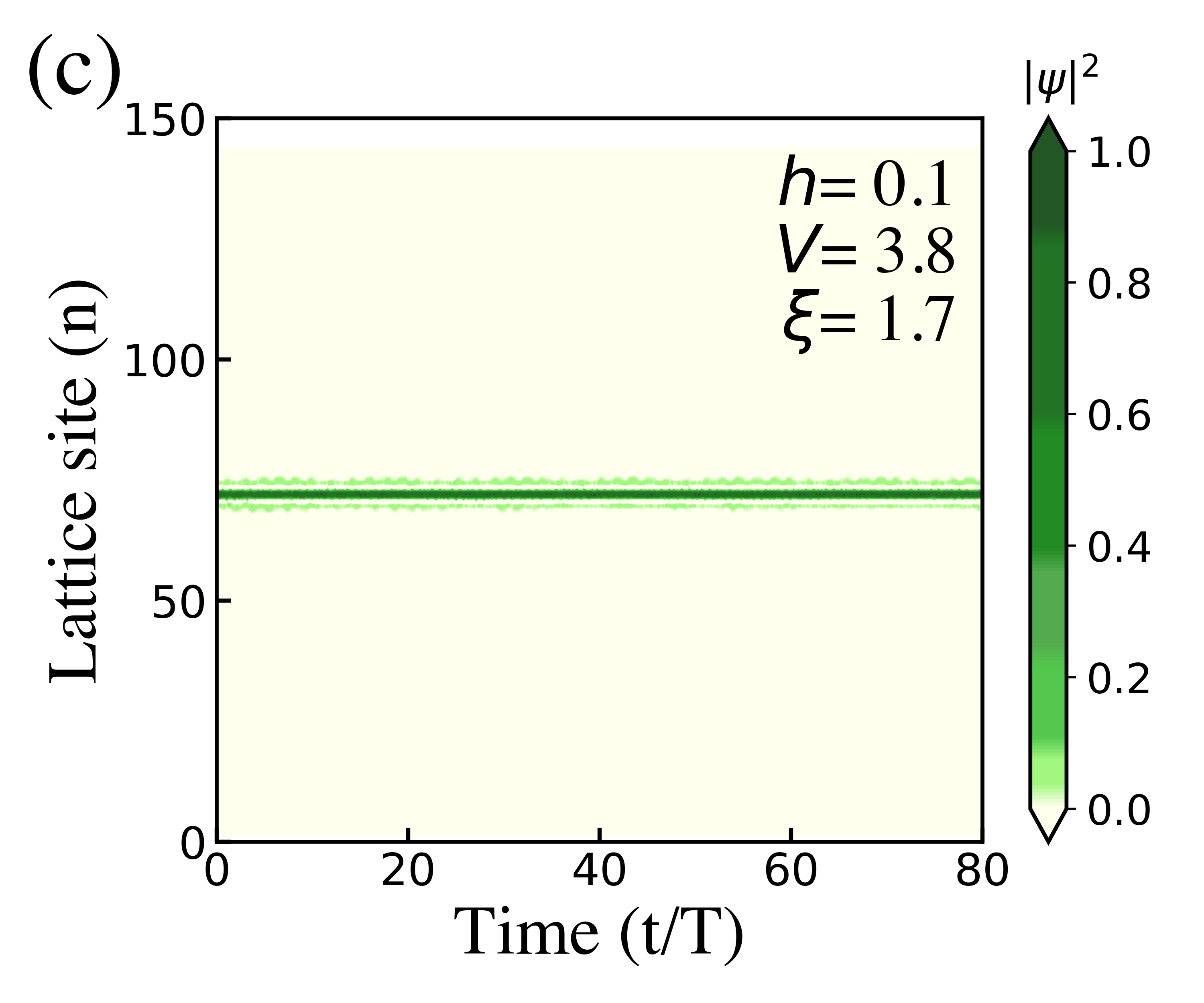}\hspace{-0.1cm}
			\includegraphics[width=0.22\textwidth,height=0.224\textwidth]{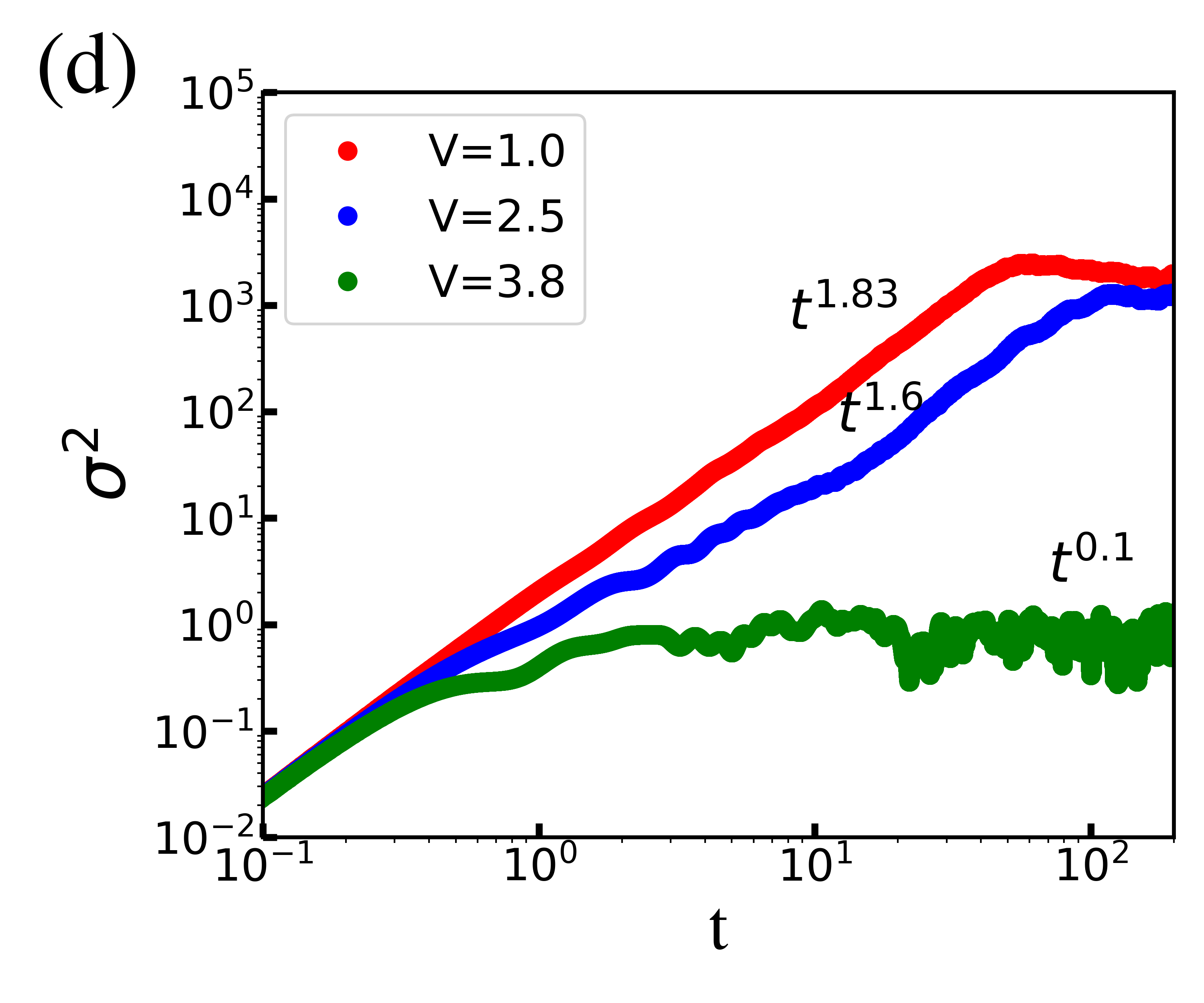}
			\caption{$|\psi|^2$ plotted for all lattice sites and at long times ($t/T$), similar to Fig.~\ref{Fig:Fig_9} at $\omega=\pi$. The strength of the quasiperiodic potential are chosen as (a) $V=1.0$, (b) $V=2.5$ and (c) $V=3.8$ and (d) $V=3.8$ such that the excitation at $t=0$ lies in the delocalized, $I_1$ and localized regime as manifested in Figs.~\ref{Fig:Fig_3}(b,d) respectively. (d) MSD as a function of time in these three phases as discussed.}
			\vspace{-1cm}
			\label{Fig:Fig_10}
		\end{tabular}
	\end{figure}
	
	\subsection{Skin effect under the OBC}\label{Sec:SE}

	The undriven counterpart of the non-Hermitian systems  with asymmetric hopping amplitudes considered in our work are well known to exhibit SE towards one end of the lattice possessing greater magnitude in directionality \cite{Sato}, in the absence of an electric field. Therefore, one of the natural motives is the investigation of SE in such driven counterparts. First, we verify that in the delocalized regime ($V=0.5$) of the undriven
	non-Hermitian AAH Hamiltonian with an uniform electric field, under the open boundaries the SE remains absent for all values of $\xi~\text{and}~h\neq0$, as shown in Fig.~\ref{Fig:Fig_S2}. This is in striking contrast to the undriven non-Hermitian system with asymmetric hopping amplitudes, where the SE occurs for any infinitesimal asymmetric hopping amplitude ($h$).\\
	\indent
	However, when the electric field is driven, as considered in our work, surprisingly, we find that below a critical value of the degree of non-Hermiticity, the states are completely ergodic, in contrast to the undriven non-Hermitian counterparts. In addition, we find that the SE appears with $\xi\neq0$ and beyond this critical value of $h$.
	This presence/absence of SE is intricately related to the emergence of real Floquet quasienergies (termed as $``extended~unitarity"$) and has also been found in our recent work where $\xi=0$ and $h$ is driven \cite{Chakrabarty}.
	The existence/disappearance of SE in the driven systems is demonstrated and verified in Fig.~\ref{Fig:Fig_S3} in Appendix \ref{App:SE_driven}.
	We have verified that in the presence of drive, our quasienergy spectrum is either completely real or completely imaginary.
	Therefore, in order to complete our understanding on the existence of SE at different parameters of $\xi$ and $h$, we use the measure $\rho_{im}=N_{im}/N$, which counts the fraction of imaginary Floquet quasienergies.
	It is therefore expected that the SE appears only when $\rho_{im}=1$.
	We plot a phase diagram at both low and high driving frequencies for a wide parameter space in $\xi-h$ at $V=0.5$ (corresponding to the delocalized regime under PBC) as shown in Fig.~\ref{Fig:Fig_8} to demonstrate the regime where the SE persists (as shown in green).
	As is clearly evident, for both low and high frequency in the drive, the SE remains absent below $h\sim0.2$, irrespective of the strength of the electric field.\\
\indent In addition, to fully determine the nature of these skin states, we employ the multifractal analysis and estimate the mean fractal dimension $D_2$ as discussed in Sec.~\ref{Sec:Numerical_techniques}. Surprisingly, from Figs.~\ref{Fig:Fig_8}(c-d) we find that under the OBC, the phase diagram is fragmented into a small regime (at low $h$) where the states are truly ergodic and do not demonstrate the SE. For most other parameters of $h$ and $\xi$, the states are multifractal in nature, either with or without the SE. It is important to note that the states that exhibit the SE are indeed multifractal, irrespective of the frequency of the drive. This is in stark contrast to the scenario that is generally observed in the absence of a time-periodic drive, where the bulk states under PBC become skin (localized) modes when the boundaries are open. The existence of multifractal states and its dynamics over long time is elaborately discussed in Sec.~\ref{Sec:Dynamics}.

	\begin{figure*}[]
			\includegraphics[width=0.250\textwidth,height=0.25\textwidth]{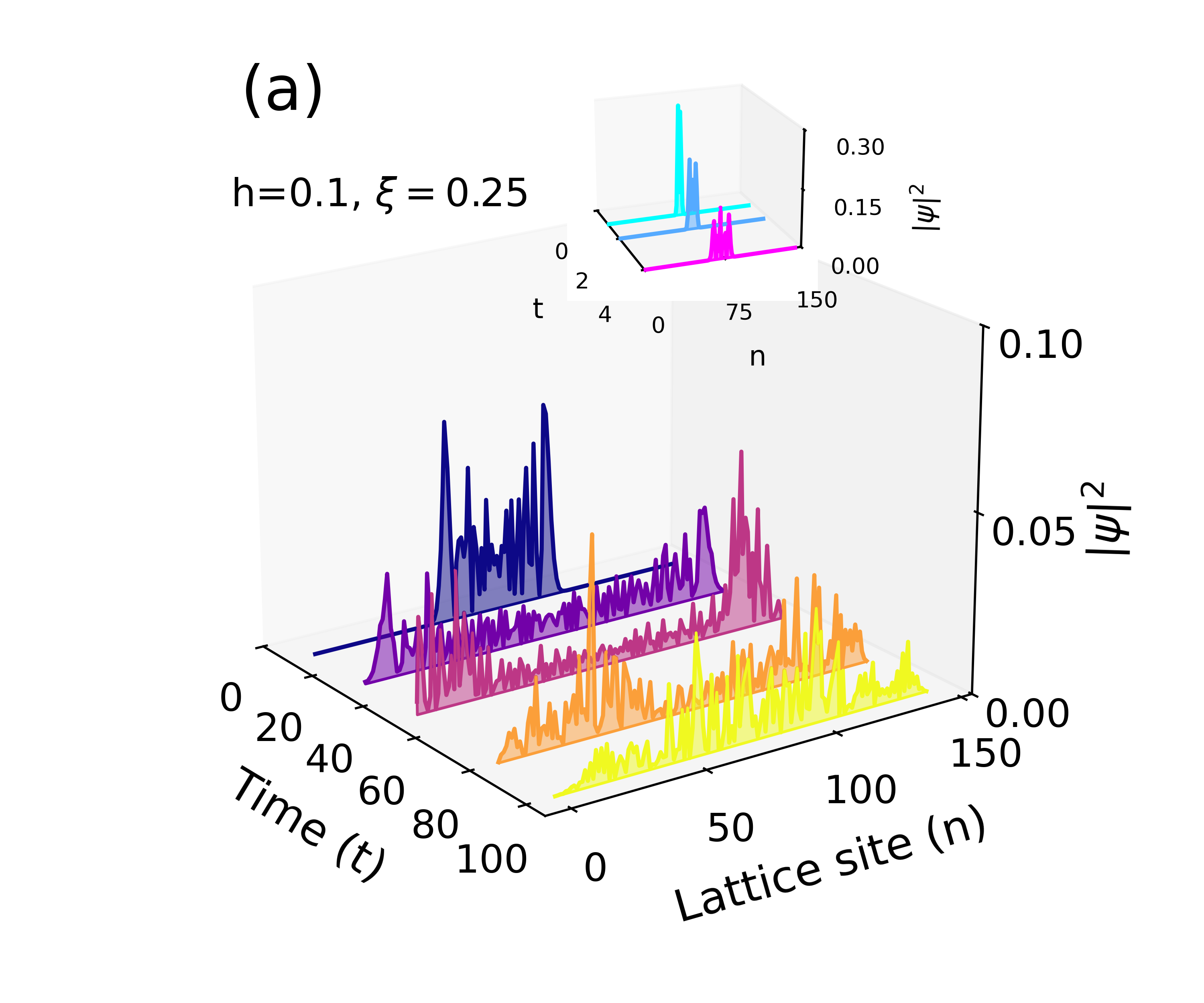}\hspace{-0.1cm}
			\includegraphics[width=0.250\textwidth,height=0.25\textwidth]{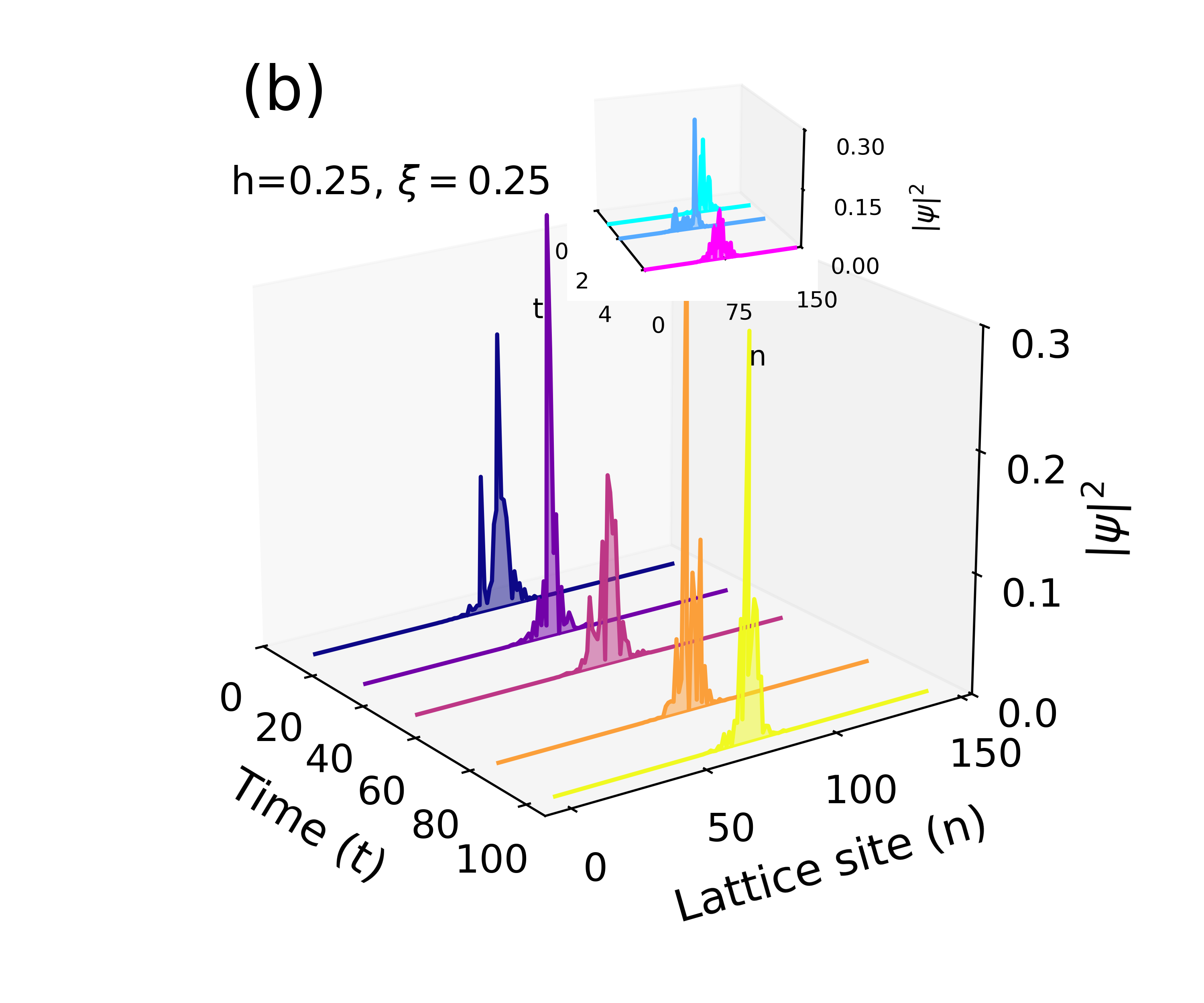}\hspace{-0.1cm}
			\includegraphics[width=0.25\textwidth,height=0.25\textwidth]{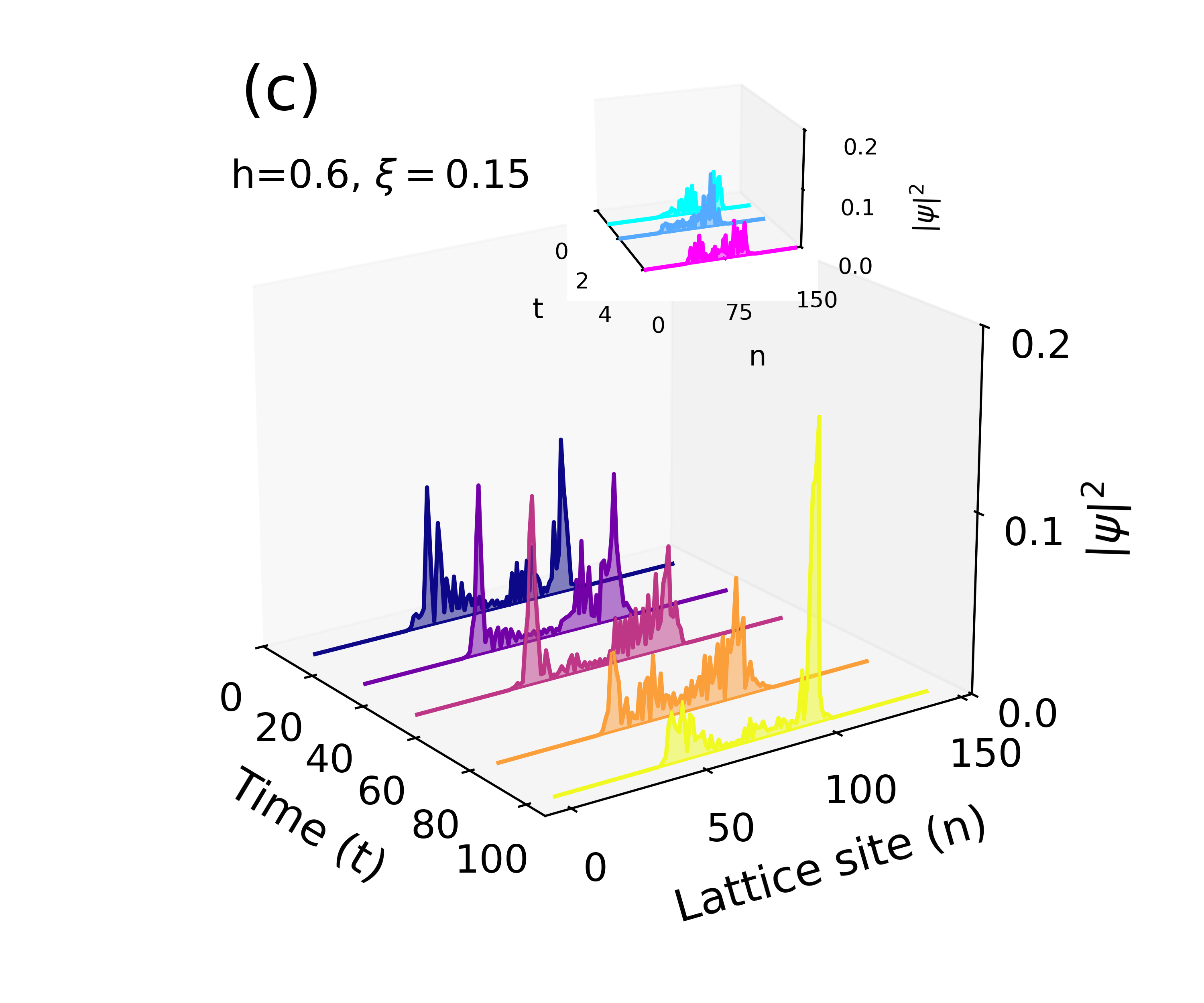}\hspace{-0.1cm}
			\includegraphics[width=0.25\textwidth,height=0.25\textwidth]{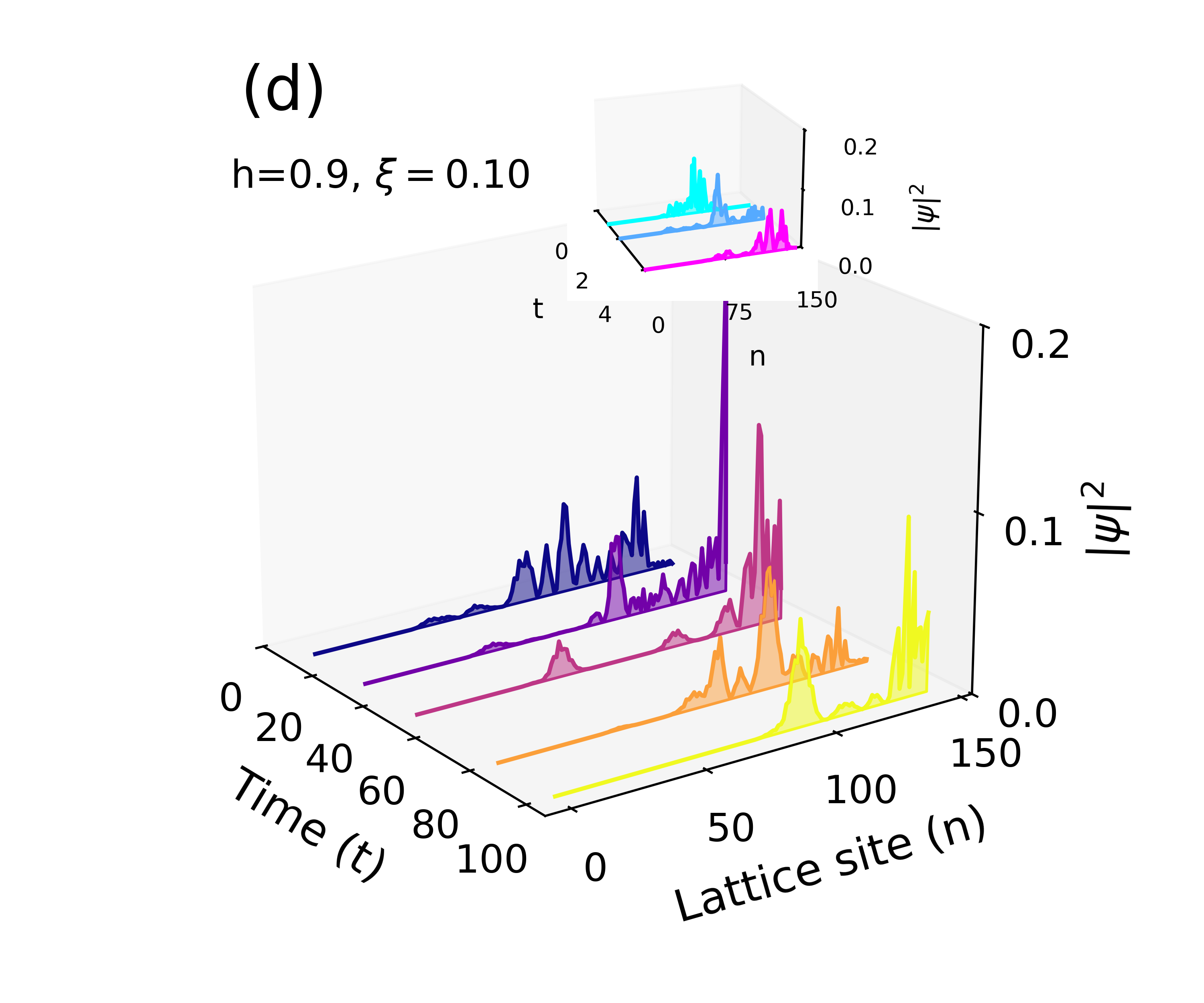}\\
			\hspace{-0.2cm}\includegraphics[width=0.232\textwidth,height=0.22\textwidth]{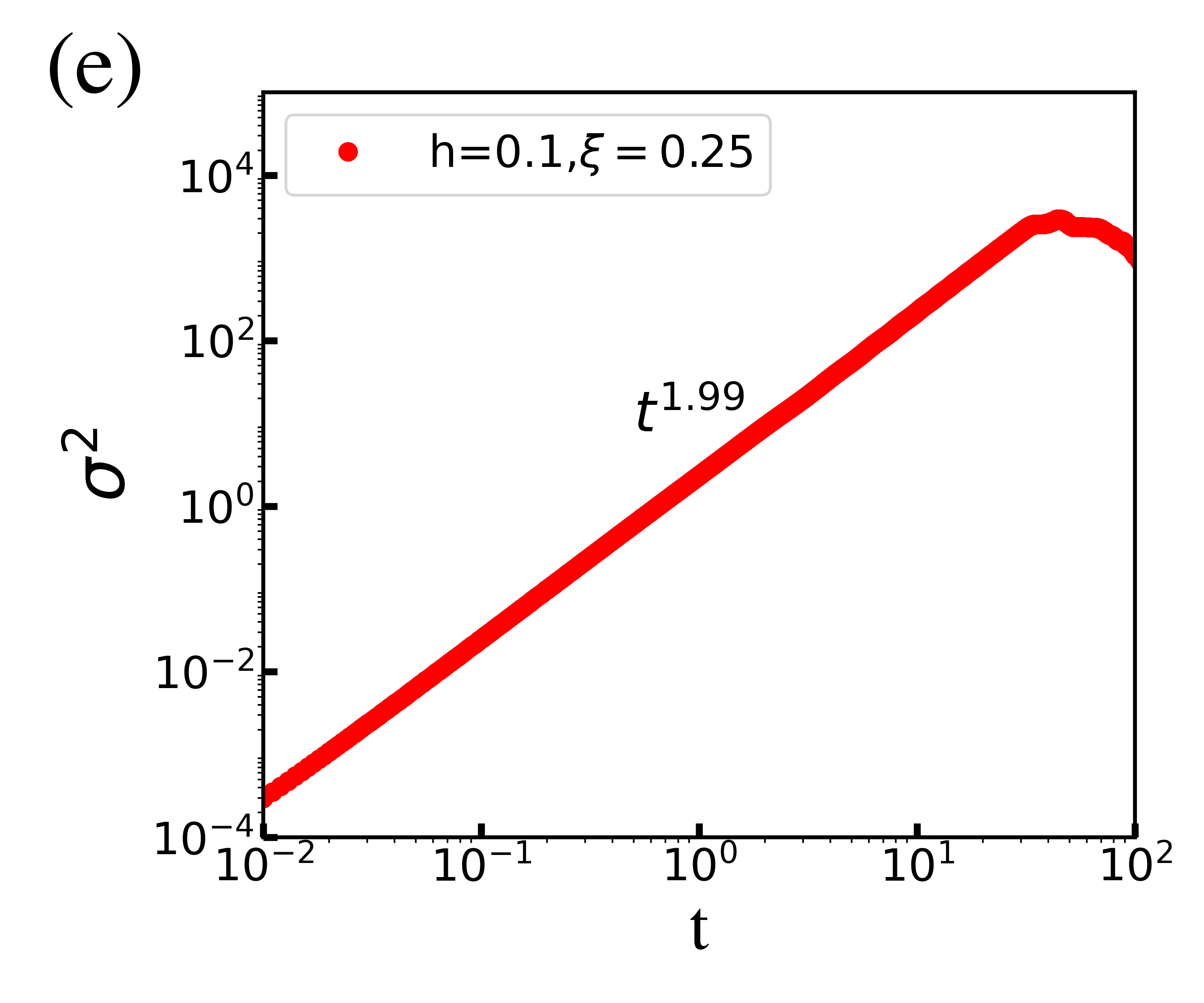}
			\includegraphics[width=0.232\textwidth,height=0.22\textwidth]{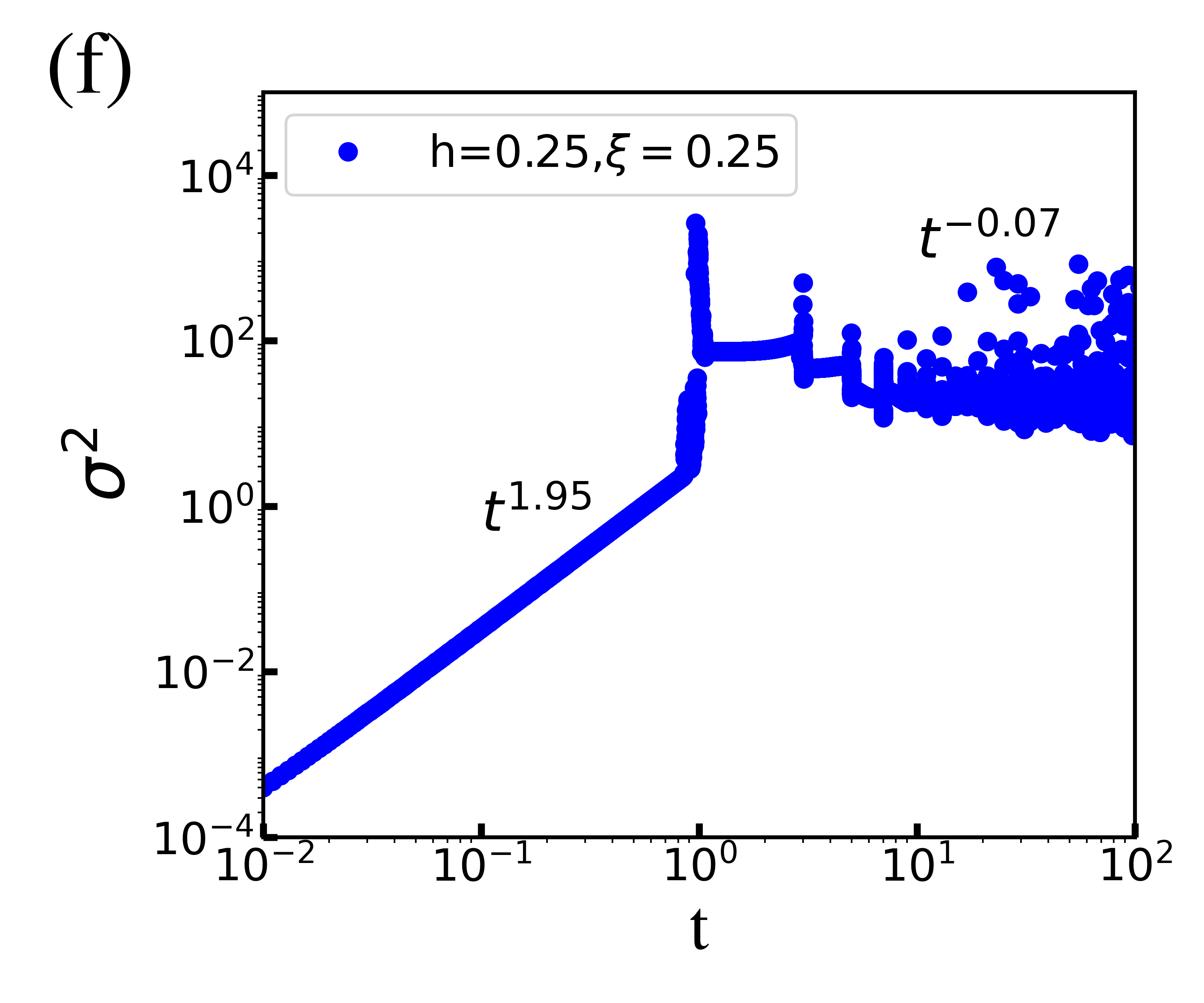}
			\includegraphics[width=0.232\textwidth,height=0.22\textwidth]{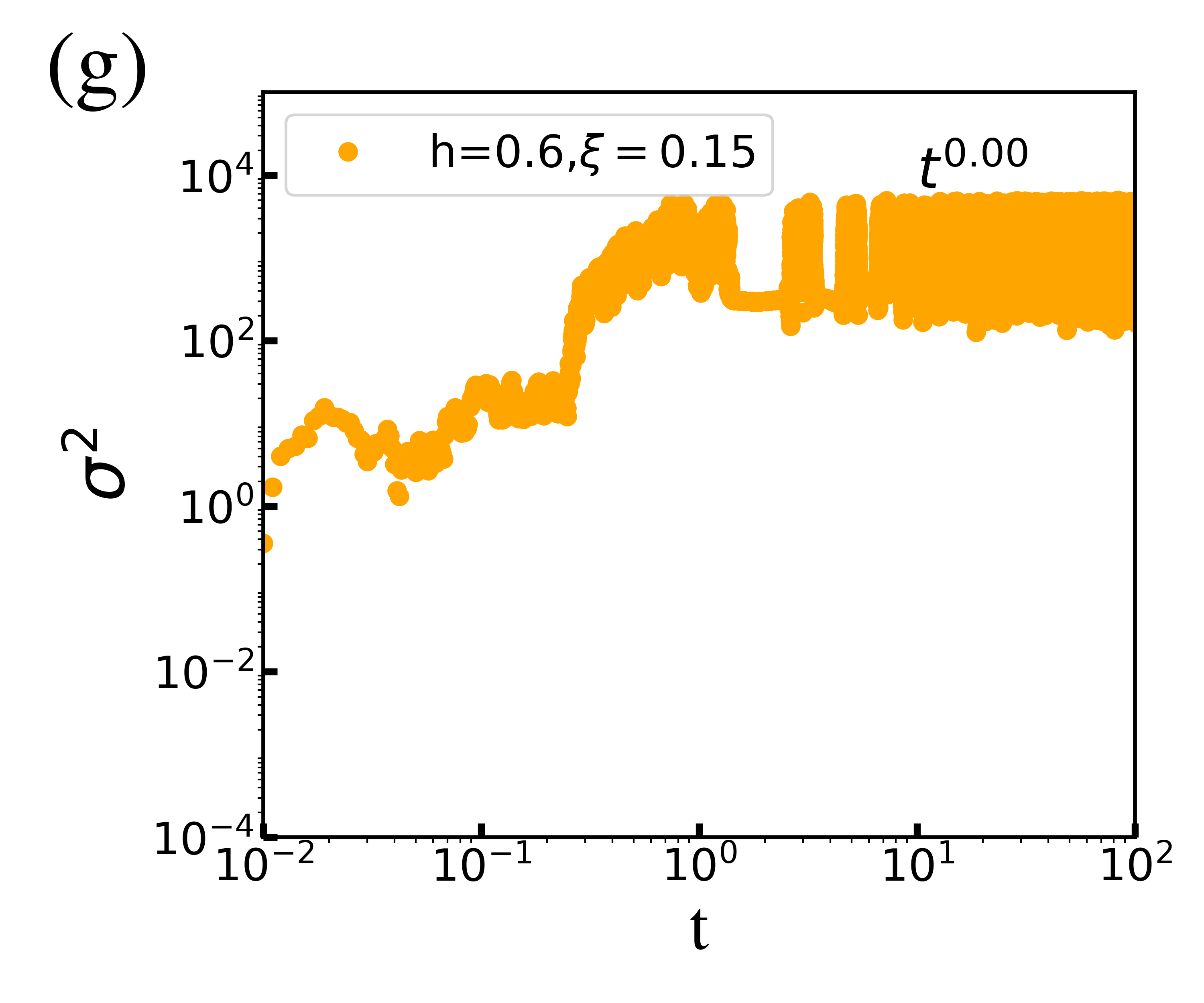}
			\includegraphics[width=0.232\textwidth,height=0.22\textwidth]{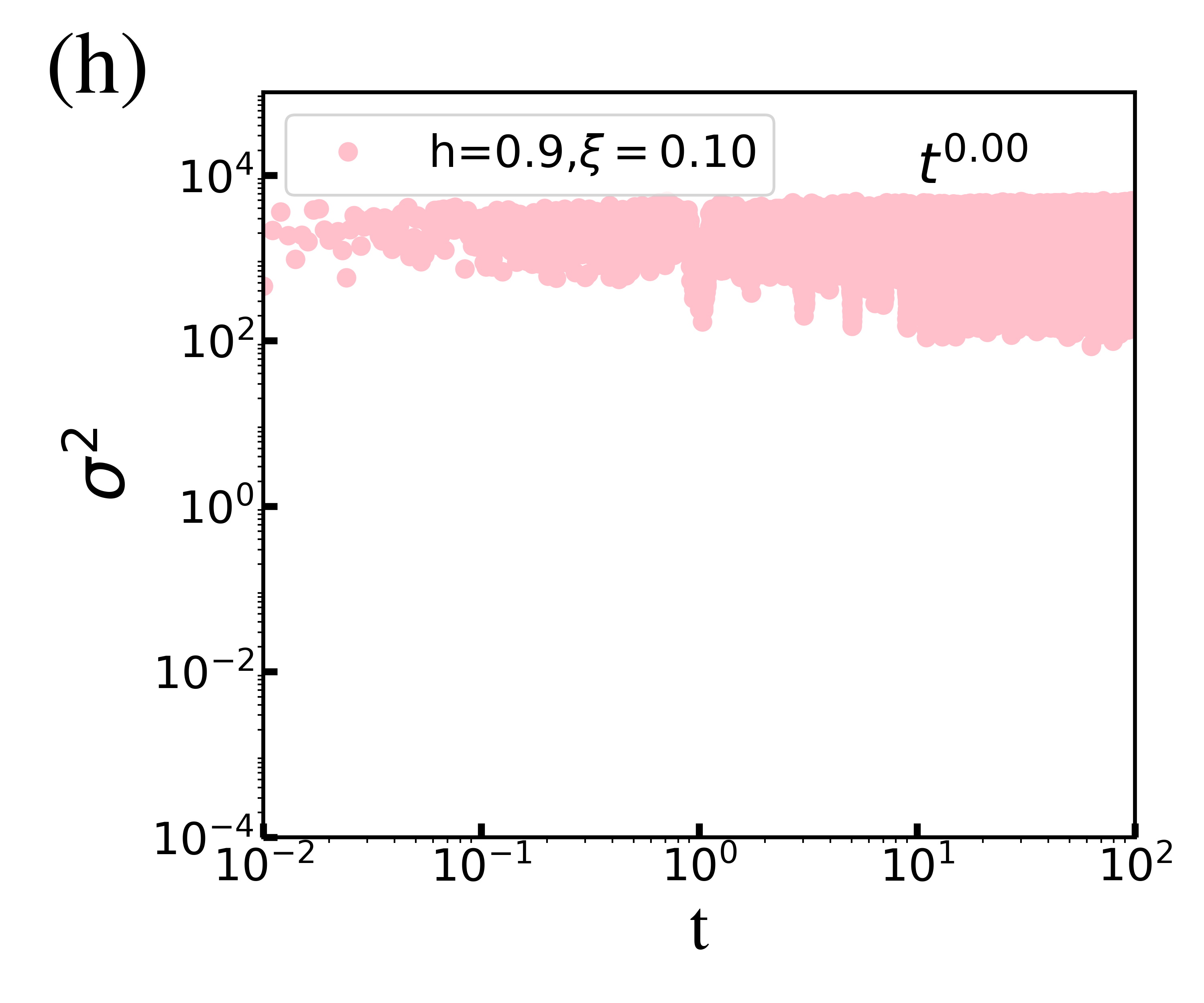}
			\caption{The evolution of the probability $|\psi|^2$ at different lattice sites ($n$) for a long time ($t$=100 secs) at $\omega=\pi/2$ under OBC for the different phases illustrated in Fig.~\ref{Fig:Fig_8}, i.e., at (a) ergodic phase without SE, (b) multifractal I phase without SE, (b)  multifractal II states without SE, and (d) multifractal states with SE. The inset shows the same distribution after the first stroboscopic period (4 secs). In the main plots, the wave function probability is shown at $t=10,30,50,80,100$ secs in dark blue, violet, pink, orange and yellow colors respectively, whereas in the inset $t=1,2,4$ secs. Figs.~(e)-(h) demonstrate the behavior of MSD in these phases along with the values of $\delta$ obtained on fitting.}
			\label{Fig:Fig_11}
	\end{figure*}

	\subsection{Dynamics: The Mean Square Displacement (MSD) and diffusivity}\label{Sec:Dynamics}
	In this section, we investigate the behavior of long-time dynamics to comprehend the nature of the delocalized, intermediate, and localized states under PBC and the multifractal nature of skin states under the OBC. To achieve this, we employ the investigation of the Mean Square Displacement (MSD) of an initial excitation at the centre
	of the lattice ($n_0$) as carried out in several works \cite{Dai, Longhi_2021, Chakrabarty_old}. The evolution of the time-dependent Hamiltonian at an instance $t$ is then given as,
	\begin{eqnarray}
		\psi(t)=U(t)\psi(t=0),
		\label{Eq:Micromotion}
	\end{eqnarray}
    where
    \begin{eqnarray}
    	U(t)=\mathcal{T}e^{-(i/\hbar)\int_{0}^{t}\mathcal{H}(t)dt},
    	\label{Eq:Micromotion_U}
    \end{eqnarray}
	and $\psi(t=0$) is the delta-type wavefunction released at the centre of the lattice as already mentioned.\\
	\indent
	The MSD is defined from the temporal spreading of the wavepacket obtained as,
	\begin{eqnarray}
		\sigma^2(t)=\frac{\sum_{n} (n-n_0)^2|\psi_n(t)|^2}{\sum_{n} |\psi_n(t)|^2}.
		\label{Eq:Mean_Square_Displacement}
	\end{eqnarray}
	In addition, the diffusion exponent ($\delta$) can be estimated from the MSD as follows:
	\begin{eqnarray}
		\sigma^2(t)=t^{2\delta}.
		\label{Eq:MSD_Diffusion_exponent_relation}
	\end{eqnarray}
	The diffusion exponent characterizes the long-time dynamics of the system. $\delta=1 (0)$ for pure ballistic/delocalized (localized) regime, whereas at the critical regime $\delta=0.5$ and the transport is termed as diffusive. Furthermore, the excitation transport is subdiffusive if $\delta<0.5$ and superdiffusive when $0.5<\delta<1$.\\
	\indent
	Figs.~\ref{Fig:Fig_9}(a-c) shows the dynamical evolution of the excitation in the delocalized, intermediate and localized regimes respectively when the frequency of the drive is low ($\omega=\pi/2$). The transport behavior for the fully delocalized and localized states are illustrated in Figs.~\ref{Fig:Fig_9}(a) and (c), identical to the observation in the regime of large driving frequency, i.e., at $\omega=4\pi$, as illustrated in Appendix \ref{App:High_frequency_transport}.
	Fig.~\ref{Fig:Fig_9}(a) demonstrates that the wave-packet dynamically delocalizes from its initial site of release, due to its ballistic nature.
	On the other hand, the excitation remains pinned at $n_0$ when the excitation is released in the localized regime as demonstrated in Fig.~\ref{Fig:Fig_9}(c).
	Fig.~\ref{Fig:Fig_9}(b) clearly indicates that a fraction of the states remain localized with time at the initial site of the release, whereas some states exhibit delocalized behavior in the intermediate regime (labelled as $I_1$). We have verified same qualitative observation in the other intermediate regime ($I_2$) as well. Furthermore, From Fig.~\ref{Fig:Fig_9}(d), it is clear that the diffusion exponents in the intermediate regimes $I_1$ and $I_2$ are $0.68$ and $0.54$ respectively suggestive of superdiffusive and nearly-diffusive behaviors, distinct from the delocalized and localized phases.
	An identical behavior in the transport of the initial excitation is also observed from Figs.~\ref{Fig:Fig_10}(a-d) when the frequency of the drive is $\omega=\pi$, where the intermediate regime $I_1$ has a superdiffusive behavior with $\delta=0.8$.
	In addition, in the presence of an external electric field, one may expect the occurence of Bloch oscillations with WS localization, as recently demonstrated in Ref.~\cite{Song_2024,Song_2024_arXiv}.
	Concommitant to our findings in the previous sections, we find no evidence of Bloch oscillations in all regimes of the driving frequency.
	It is thus established that the absence of WS localization and Bloch oscillations are interlinked.\\
	\indent To obtain a complete qualitative and quantitative picture of the multifractal skin states under OBC as shown in Figs.~\ref{Fig:Fig_8}(c-d), we analyze the time evolution of the initially localized wave-packet at a low driving frequency ($\omega=\pi/2$) in details in Fig.~\ref{Fig:Fig_11} over all the regimes in the phase diagram. In the ergodic phase (given in mauve regime of the multifractal phase diagram in Fig.~\ref{Fig:Fig_8}(c)), the wave-packet evolves with time and is spread over all the lattice sites at a long time (Fig.~\ref{Fig:Fig_11}(a)). This spreading is ballistic as given by the diffusion exponent in Fig.~\ref{Fig:Fig_11}(e). In the other two multifractal phases in yellow and green (labelled as multifractal I, no SE and multifractal II, no SE) of Fig.~\ref{Fig:Fig_8}(c)), the wave-packet spreads with time, but does not cover all the lattice sites, which is a typical nature of multifractal states and is illustrated in Figs.~\ref{Fig:Fig_11}(b) and (c). Initially the wave-packet spreads and thereafter the diffusion stops beyond a certain time as demonstrated in Figs.~\ref{Fig:Fig_11}(f) and (g).
	In the regime where SE exists (labelled as multifractal, SE in Fig.~\ref{Fig:Fig_8}(c)), the wave-packet has a tendency of moving in one direction due to the skin-effect (Fig.~\ref{Fig:Fig_8}(d)), right from the start of the evolution, which is also evident from Fig.~\ref{Fig:Fig_11}(h) without any significant diffusion. The MSD value is fixed at its maximum at all times, since the wave-packet is far away from the initial site of excitation, i.e., $n_0$. The insets correspond to the evolution after a complete stroboscopic period, where we verify that the anticipated behavior of states are concurrent to those obtained in Fig.~\ref{Fig:Fig_8}(a,c). Identical qualitative results are also obtained when the drive in the electric field is large enough, corresponding to Fig.~\ref{Fig:Fig_8}(b,d), as demonstrated in Figs.~\ref{Fig:Fig_S5} and \ref{Fig:Fig_S6} in Appendix ~\ref{App:Dynamics_large_frequency}.
		
	\section{Conclusions}\label{Sec:Conclusions}
	In summary, this work explores different quantum phases and provides a coherent and consistent finding of several intriguing phenomena in the non-Hermitian quasicrystals with a drive in the electric field.
	Unlike the time-independent setting, in such time-periodic quasicrystals, we have observed that the drive in the electric field retains the delocalized nature of the eigenstates upto a large strength of the electric field, as reported earlier.
	However, interestingly, we find the existence of multiple intermediate regimes, which turned out to be mobility edges, in the systems where the electric field is driven slowly.
	Such mobility edges are manifested due to the time-periodic drive, irrespective of the degree of non-Hermiticity, and are typically absent in the generic undriven AAH counterparts.
	In addition, the DL transition approaches the result of the static non-Hermitian AAH Hamiltonian with a rescaled hopping amplitude when the frequency becomes sufficiently large.
	Therefore, our result paves way to obtain novel phases in systems using a drive which had only two electronic phases to begin with.
	Furthermore, we find that the equally spaced energy ladders which form the backbone of WS localization are not manifested in systems with a drive, wherein the energy level statistics differ widely from the static counterpart and approach the Poisson distribution in the localized regime.
	In stark contrast to the undriven systems with an electric field that destroys the SE under OBC, we find the existence of SE in certain regime of the $\xi-h$ parameter space. Surprisingly, the skin states are identified to be multifractal in nature, which is also verified from the long-time dynamics of the system under OBC.
	This work also enriches the understanding of the interconnection between the existence of WS localization and the emergence of Bloch oscillations under the PBC, both of which remain absent in the time-dependent setting.
	We propose that the system will be beneficial in controlling the material
	properties and tailor new phases in quasiperiodic systems under the application of an externally tunable electric field in 1D optical lattices as briefly discussed in Sec.~\ref{Sec:Model}.
	
	\section{Acknowledgements}\label{Sec:Acknowledgements}
	A.C. would like to thank  the Council of Scientific $\&$ Industrial Research (CSIR)-HRDG, India, for the financial support received via File No.- 09/983(0047)/2020-EMR-I.
	The numerical findings in this work were achieved using the computational facilities availed from SERB (DST), India (Grant No. EMR/2015/001227) and
	the High Performance Computing (HPC) facility of the National Institute of Technology (Rourkela).
	
	\appendix
	  
	\begin{figure}[t]
		\begin{tabular}{p{\linewidth}c}
			\includegraphics[width=0.2450\textwidth,height=0.228\textwidth]  
			{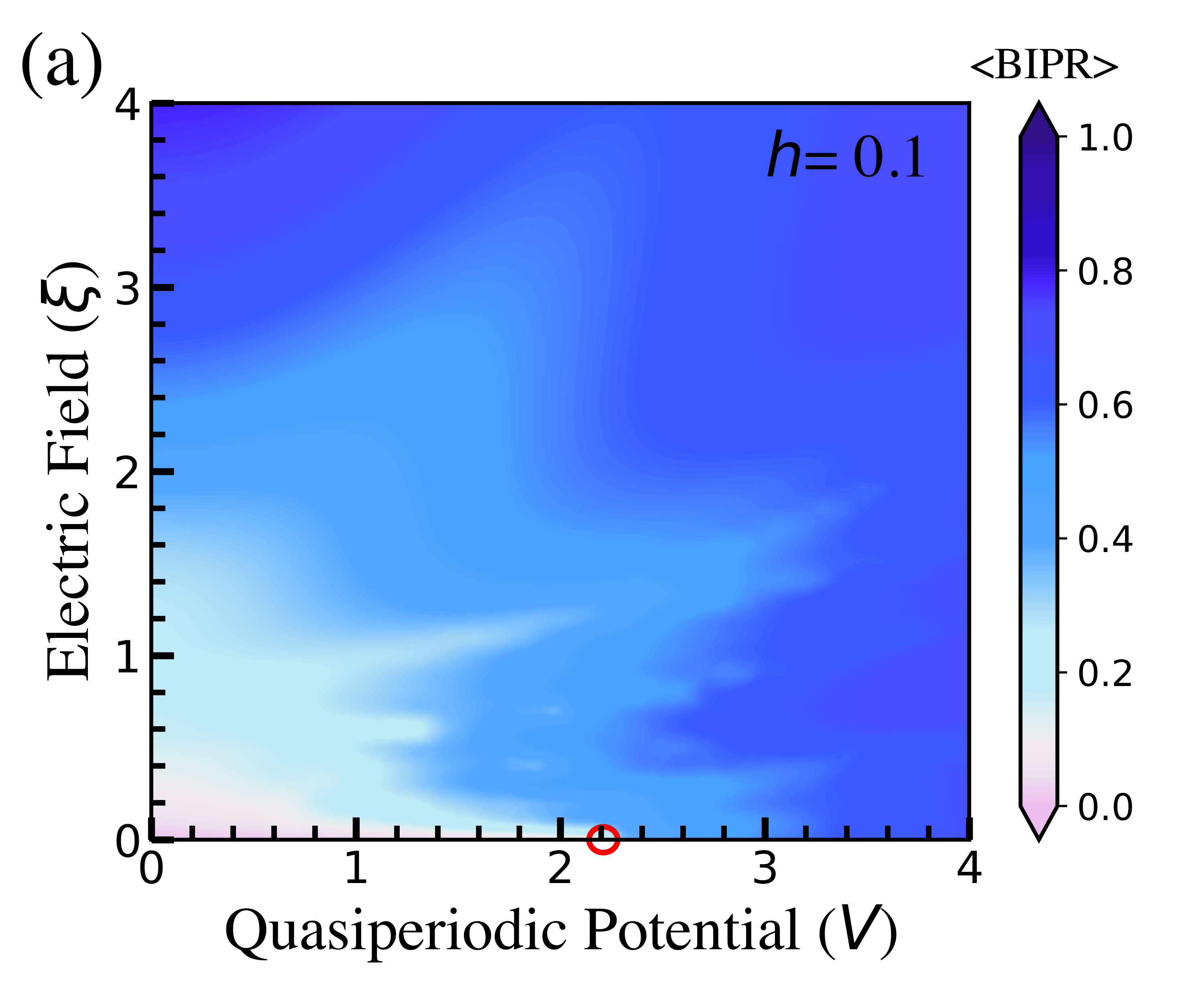}\hspace{-0.1cm}
			\includegraphics[width=0.2450\textwidth,height=0.225\textwidth]{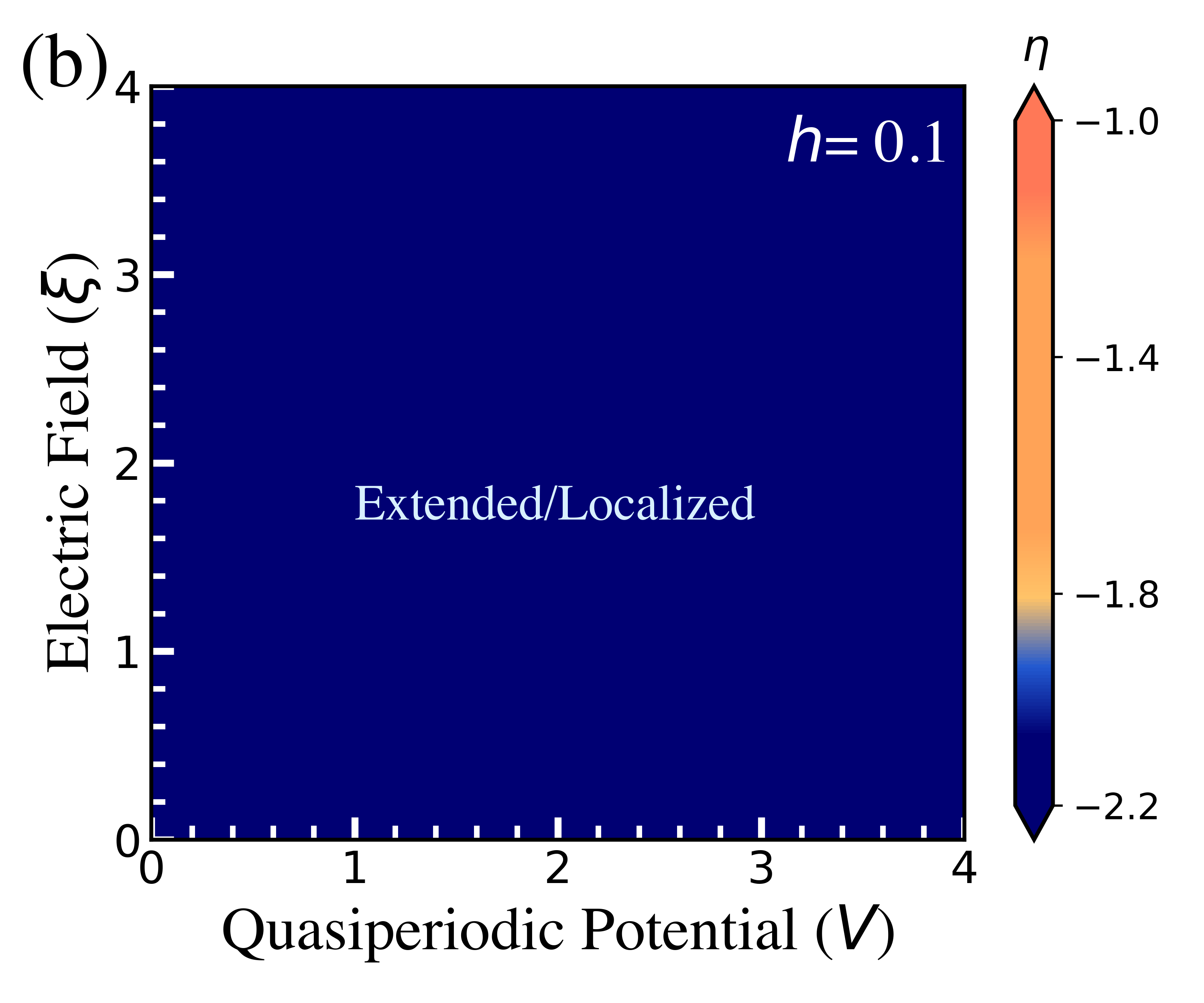}\\
			\hspace{-0.0cm}\includegraphics[width=0.234\textwidth,height=0.228\textwidth]{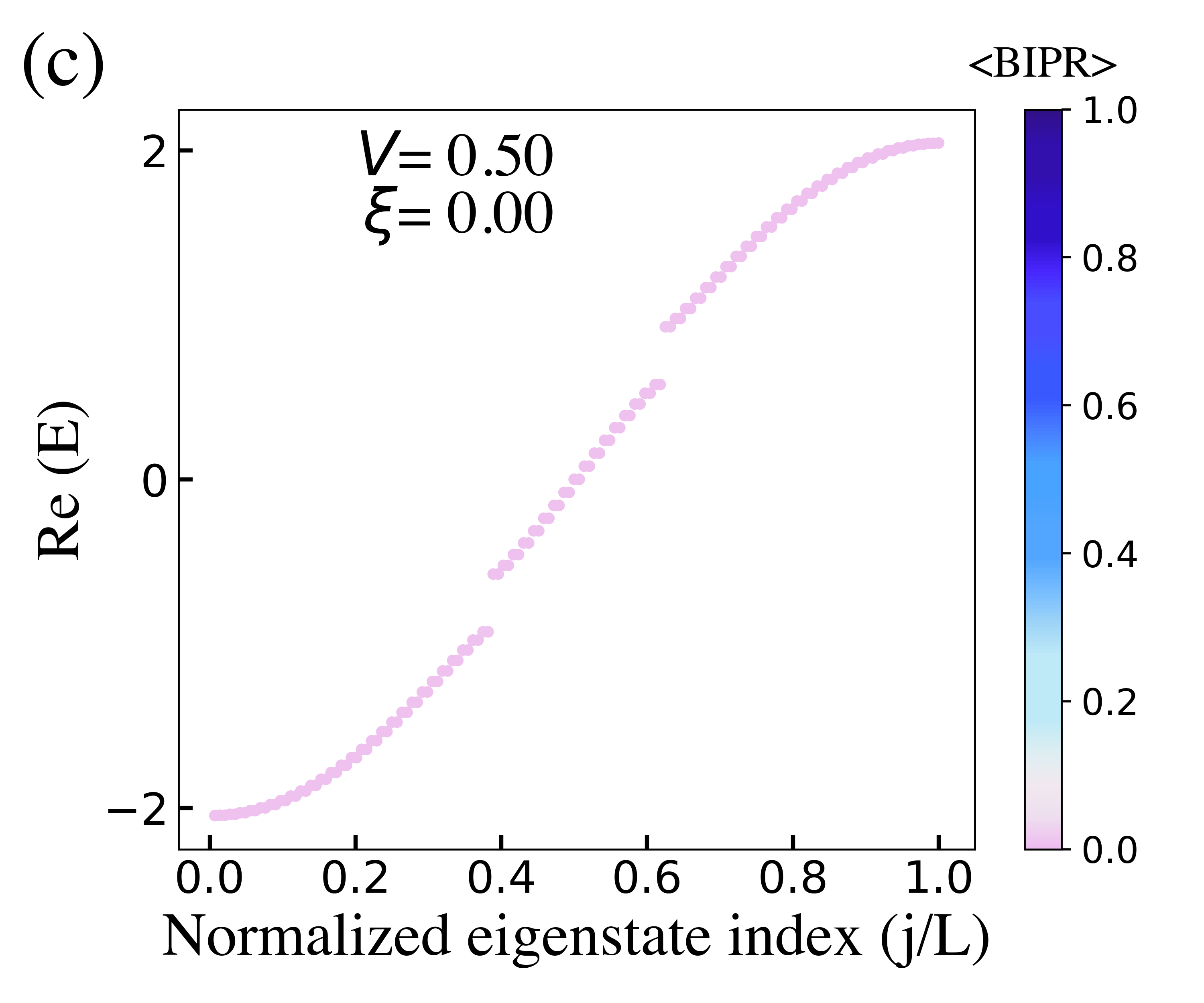}\hspace{-0.1cm}
			\includegraphics[width=0.250\textwidth,height=0.232\textwidth]{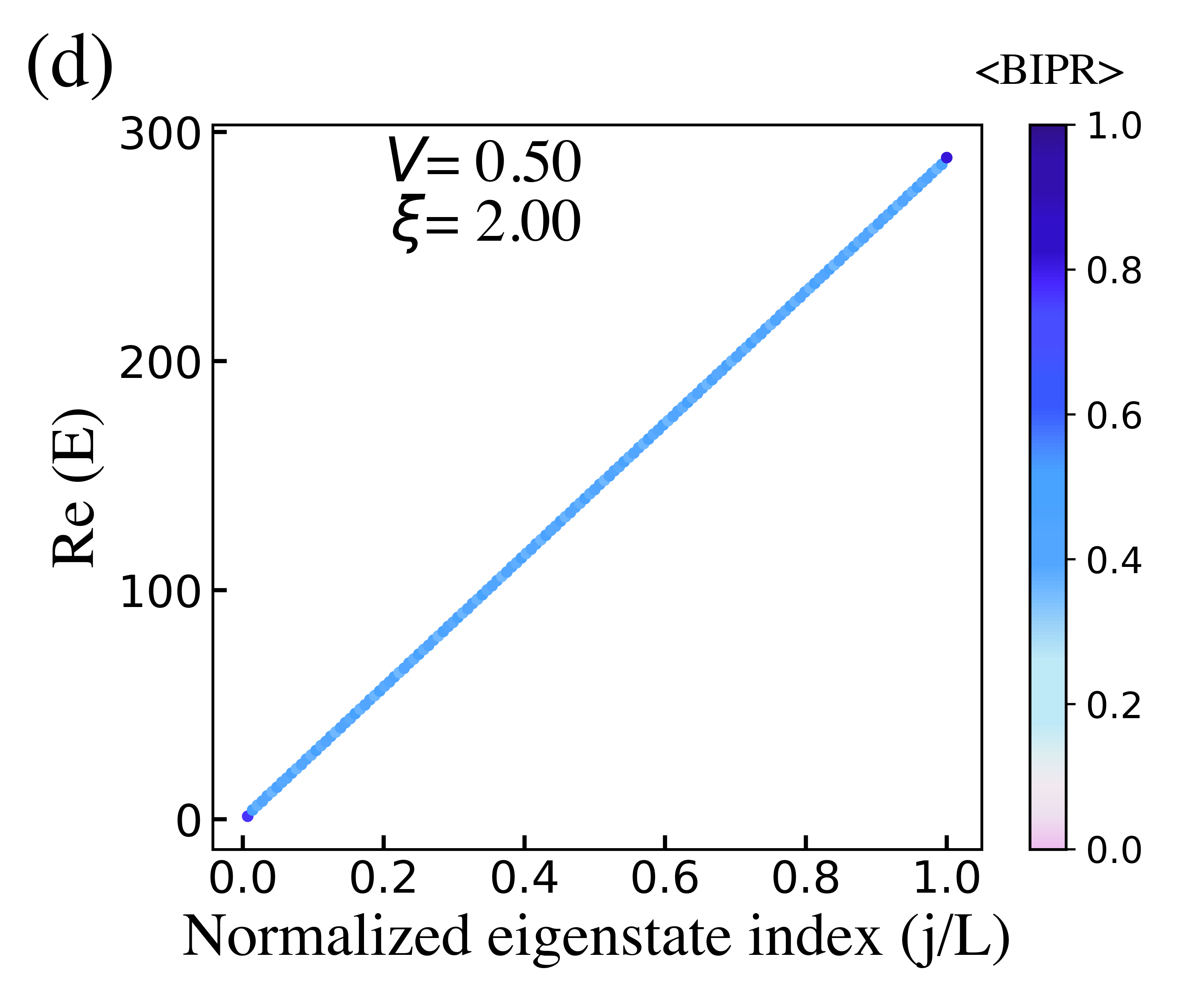}
			\caption{The phase diagram of (a) $\left\langle \text{BIPR} \right\rangle$ and (b) $\eta$, for different strengths of the electric field ($\xi$) and quasiperiodic potential ($V$) in the time-independent non-Hermitian system with $h=0.1$. The lower panels illustrate the real part of the energy spectrum of the undriven system at (c) $V=0.5 ~\text{and}~ \xi=0.0$ (delocalized regime), and (d) $V=0.5 ~\text{and}~ \xi=2.0$ (localized regime).
			The red marker shows the value of $V_c$ at $\xi=0.0$.}
			\label{Fig:Fig_S1}
		\end{tabular}
	\end{figure}

	\begin{figure}[]
		\begin{tabular}{p{\linewidth}c}
			\centering
			\includegraphics[width=0.245\textwidth,height=0.215\textwidth]{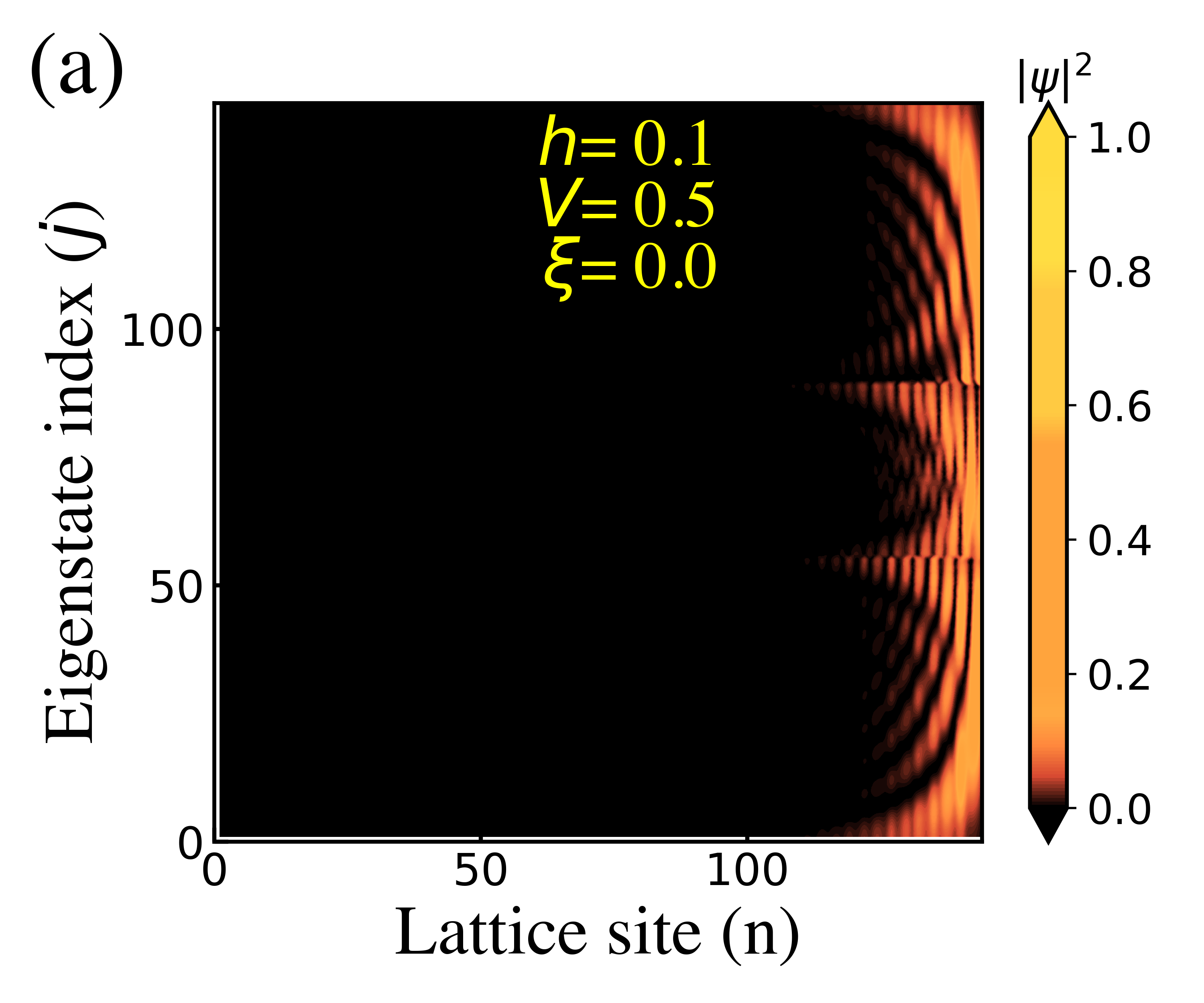}\hspace{-0.25cm}
			\includegraphics[width=0.245\textwidth,height=0.215\textwidth]{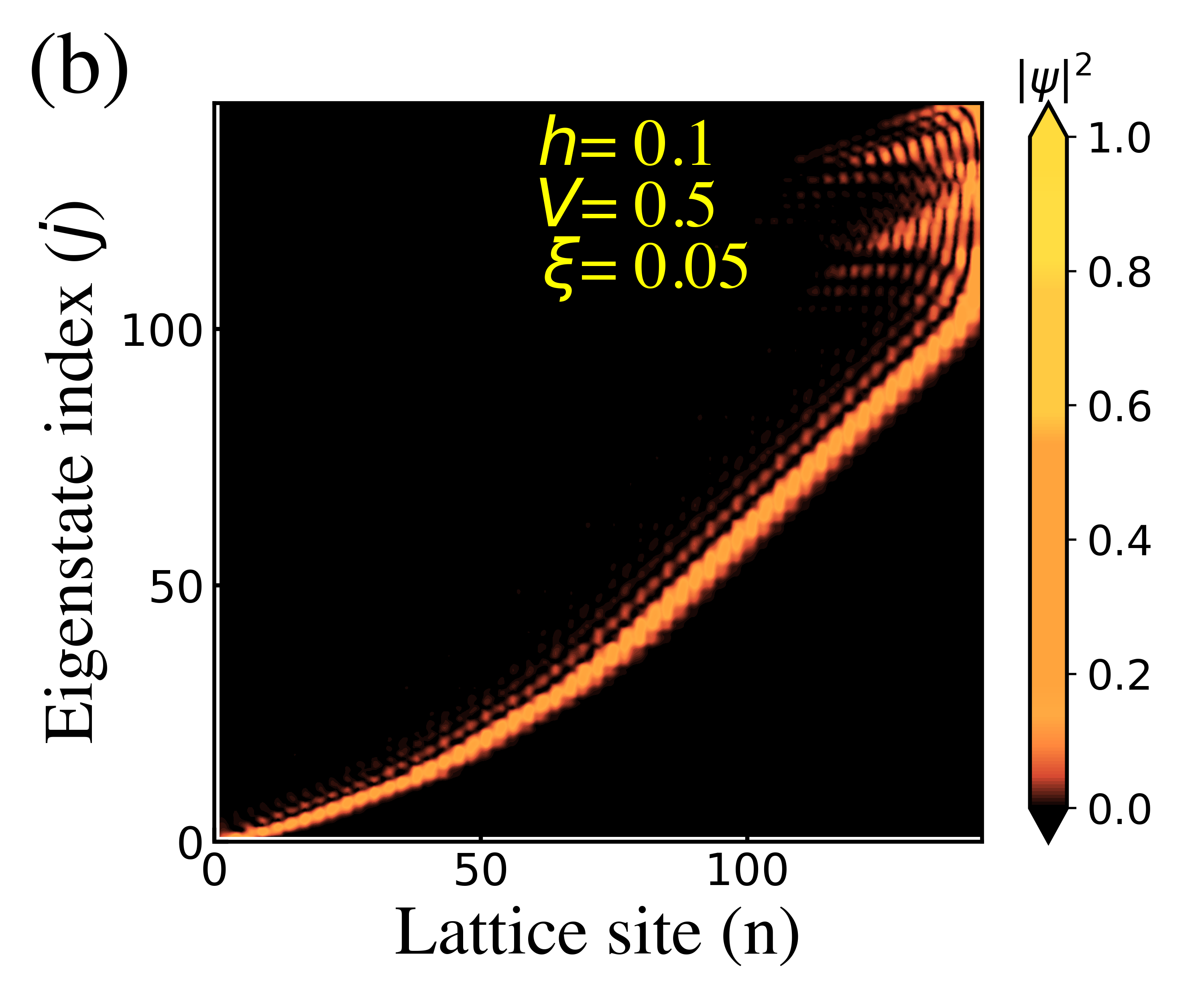}\\
			\includegraphics[width=0.245\textwidth,height=0.215\textwidth]{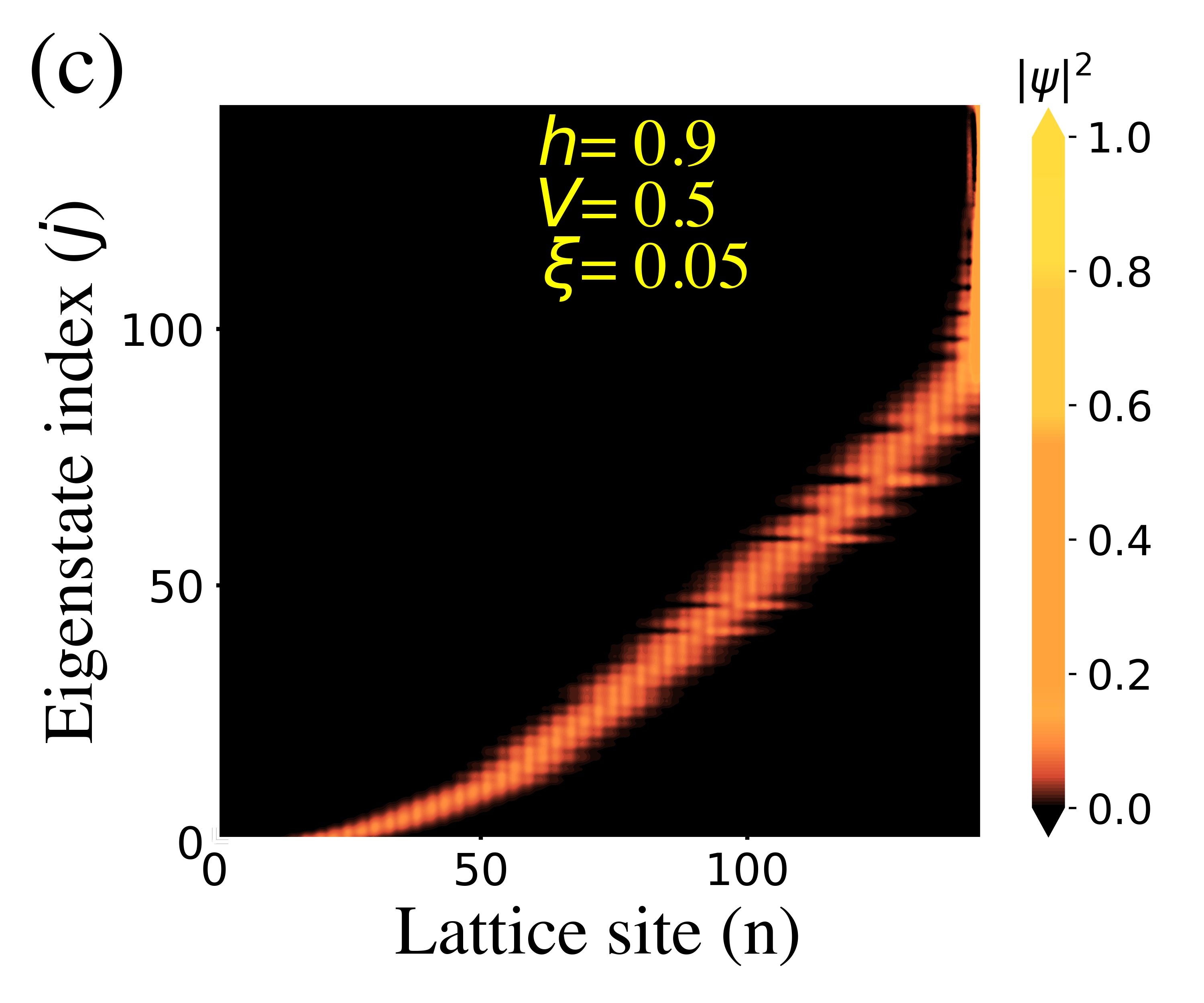}\hspace{-0.25cm}
			\includegraphics[width=0.245\textwidth,height=0.215\textwidth]{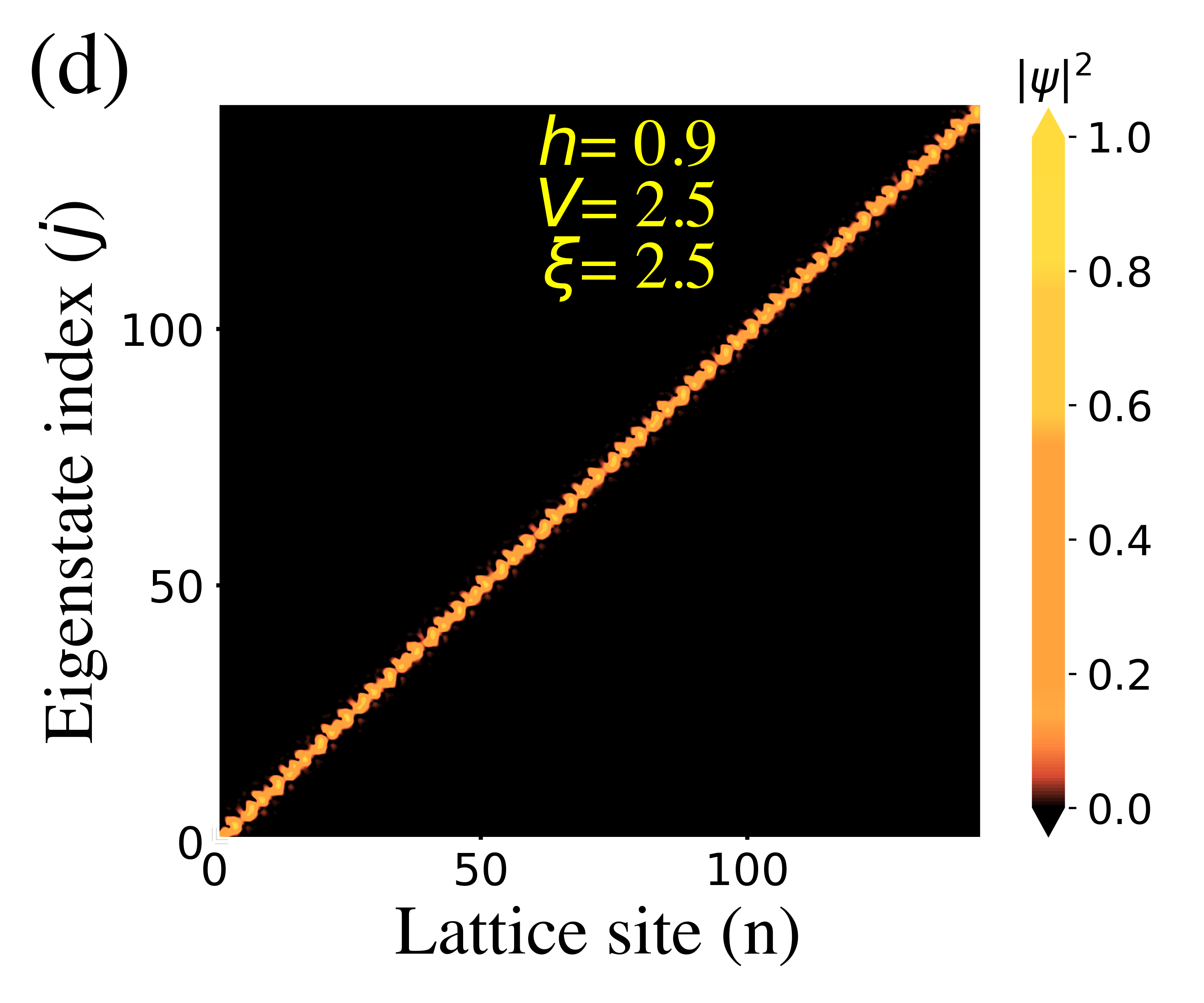}
			\caption{$|\psi|^2$ as a function of $n$ for all the eigenstates demonstrating the skin-effect under the open boundary condition (OBC) in undriven non-Hermitian system demonstrated at various parameters (a) $h=0.1, V=0.5, \xi=0.0$, (b) $h=0.1,V=0.5, \xi=0.05$, (c) $h=0.9, V=0.5, \xi=0.05$, and (d) $h=0.9, V=2.5, \xi=2.5$.}
			\label{Fig:Fig_S2}
		\end{tabular}
	\end{figure}

	\begin{figure}[]
		\begin{tabular}{p{\linewidth}c}
			\centering
			\includegraphics[width=0.245\textwidth,height=0.215\textwidth]{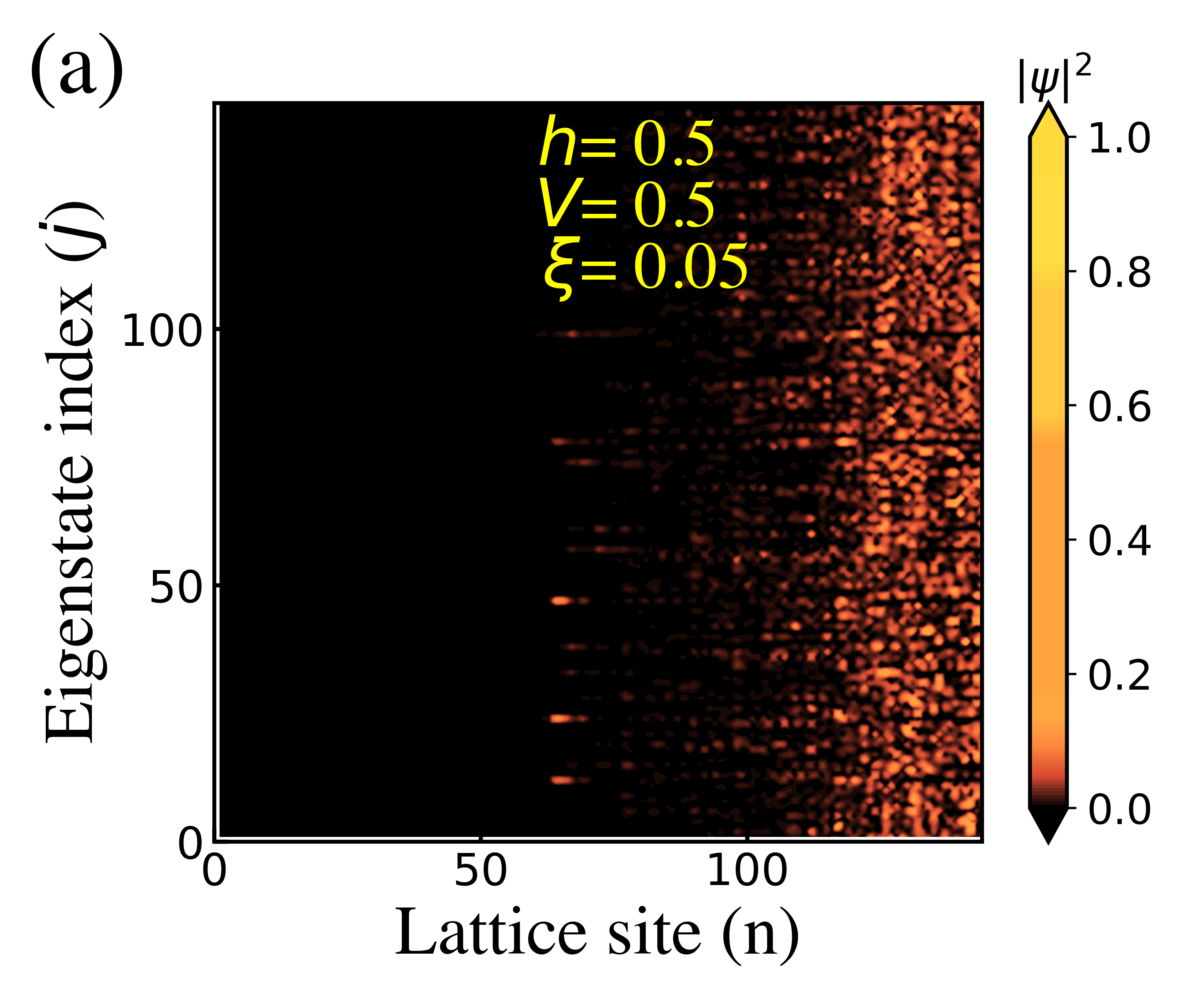}\hspace{-0.25cm}
			\includegraphics[width=0.245\textwidth,height=0.215\textwidth]{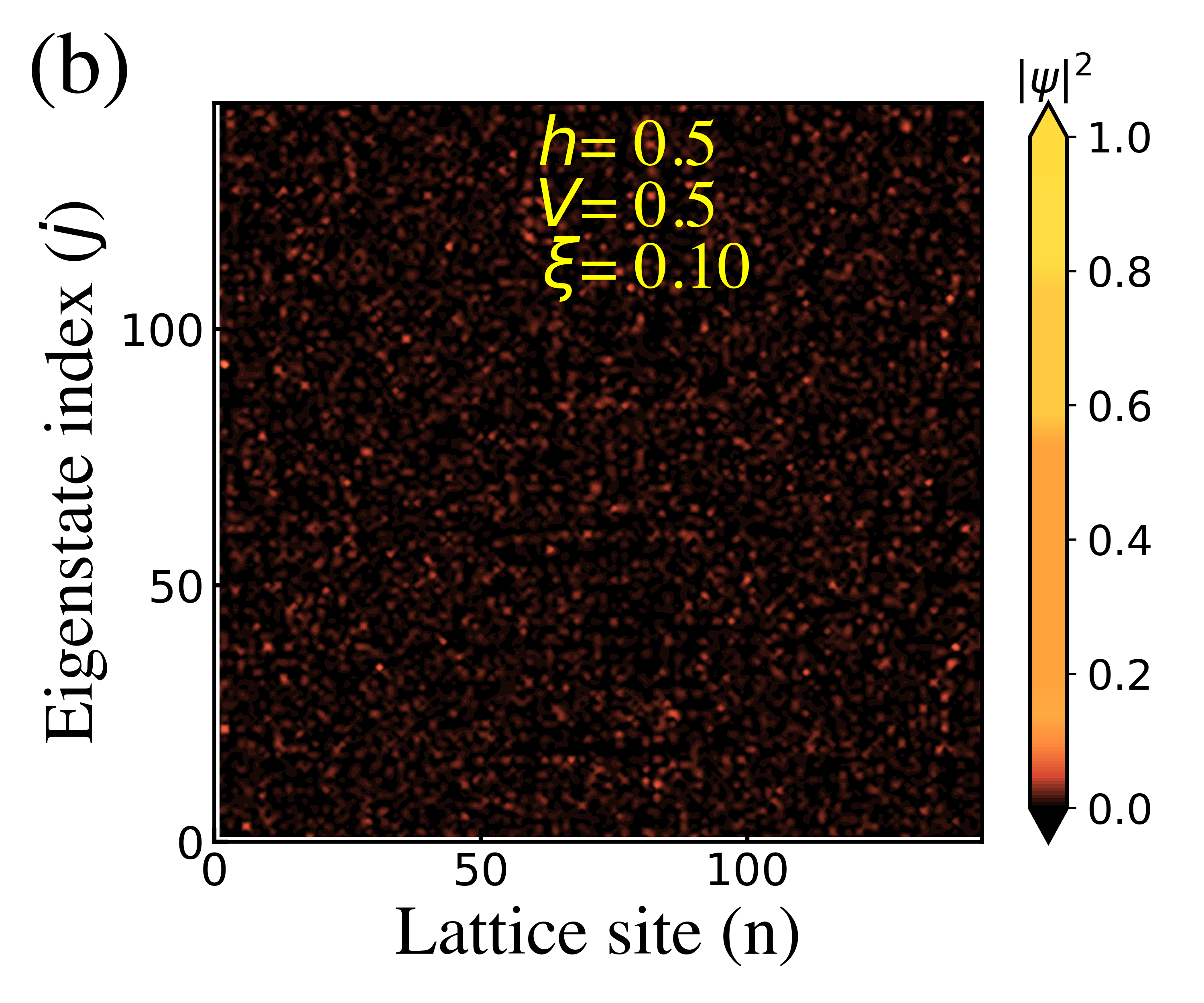}
			\caption{$|\psi|^2$ as a function of $n$ for the non-Hermitian driven system under OBC at $h=0.5$, $V=0.5$ (delocalized regime) where the electric field strength is (a) $\xi=0.05$, (b) $\xi=0.10$ in the regimes where the SE exists and vanishes respectively. $\omega=4\pi$ in both the cases.}
			\label{Fig:Fig_S3}
		\end{tabular}
	\end{figure}
	
	\begin{figure}[]
		\begin{tabular}{p{\linewidth}c}
			\centering
			\includegraphics[width=0.245\textwidth,height=0.215\textwidth]{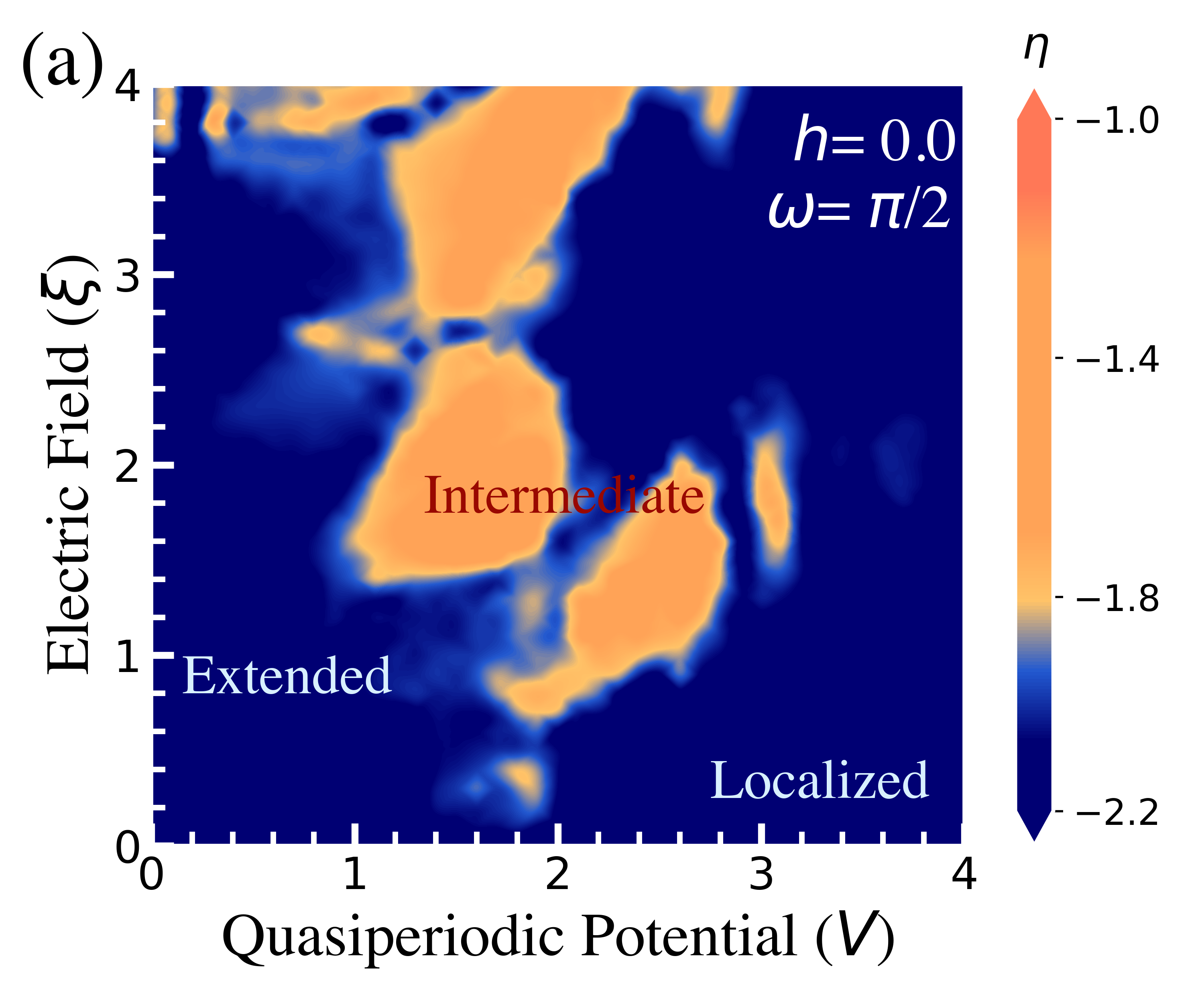}\hspace{-0.25cm}
			\includegraphics[width=0.245\textwidth,height=0.215\textwidth]{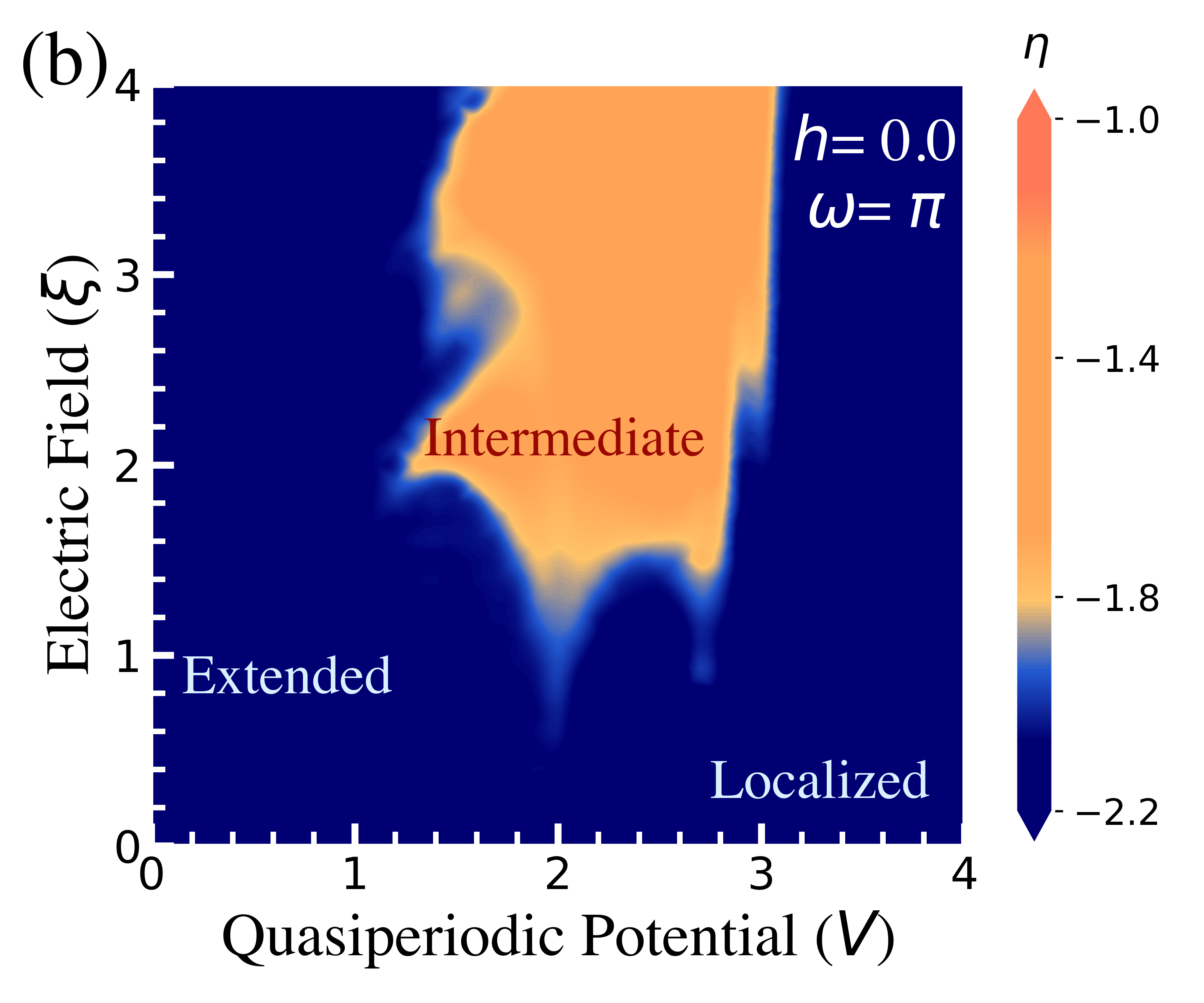}
			\caption{The existence of intermediate phases demonstrated in the $\xi-V$ parameter space in a Hermitian ($h=0.0$) driven system at driving frequencies (a) $\omega=\pi/2$, and (b) $\omega=4\pi$.}
			\label{Fig:Fig_N1}
		\end{tabular}
	\end{figure}
	
	\begin{figure}[]
		\begin{tabular}{p{\linewidth}c}
			\centering
			\includegraphics[width=0.245\textwidth,height=0.215\textwidth]{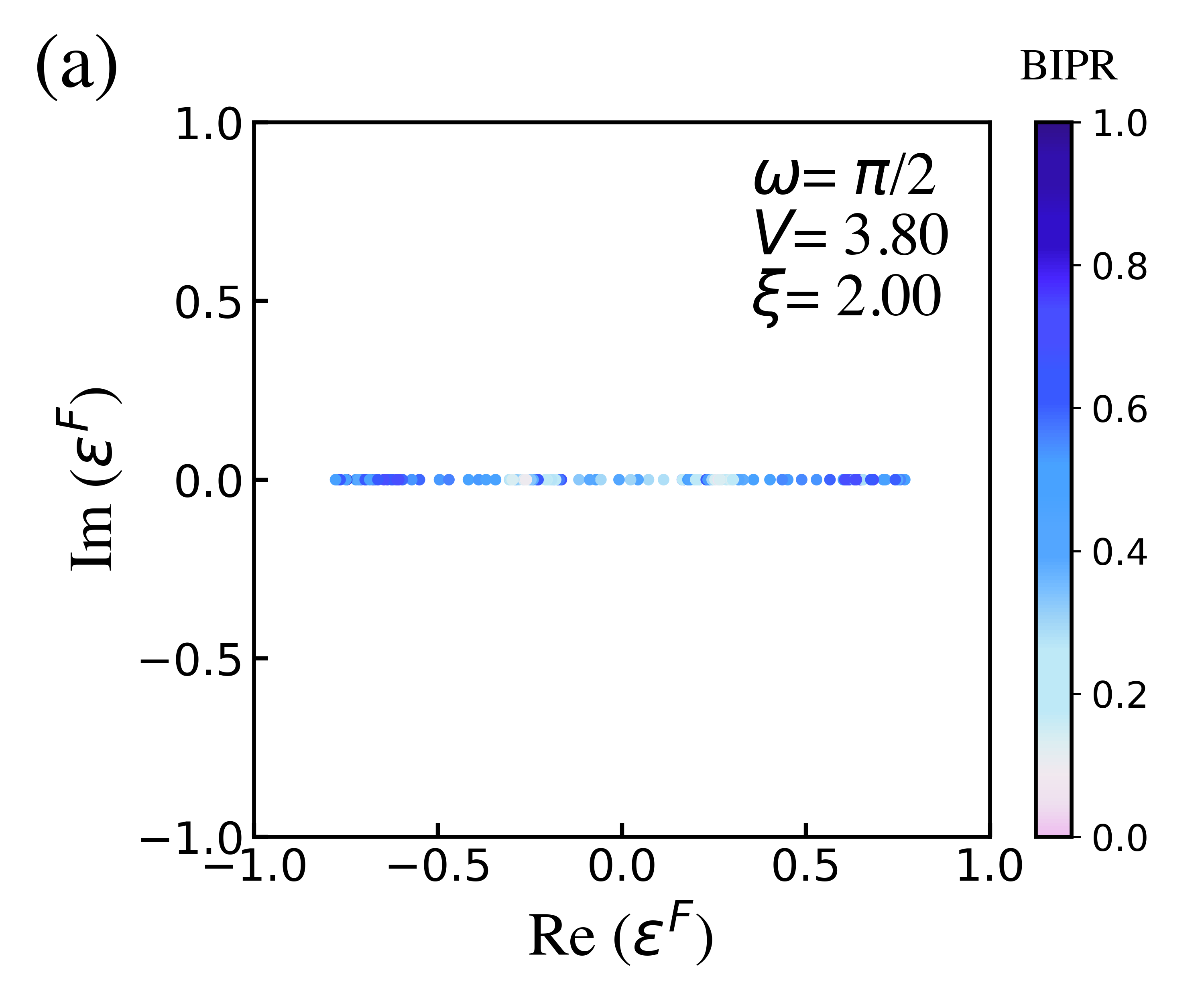}\hspace{-0.25cm}
			\includegraphics[width=0.245\textwidth,height=0.215\textwidth]{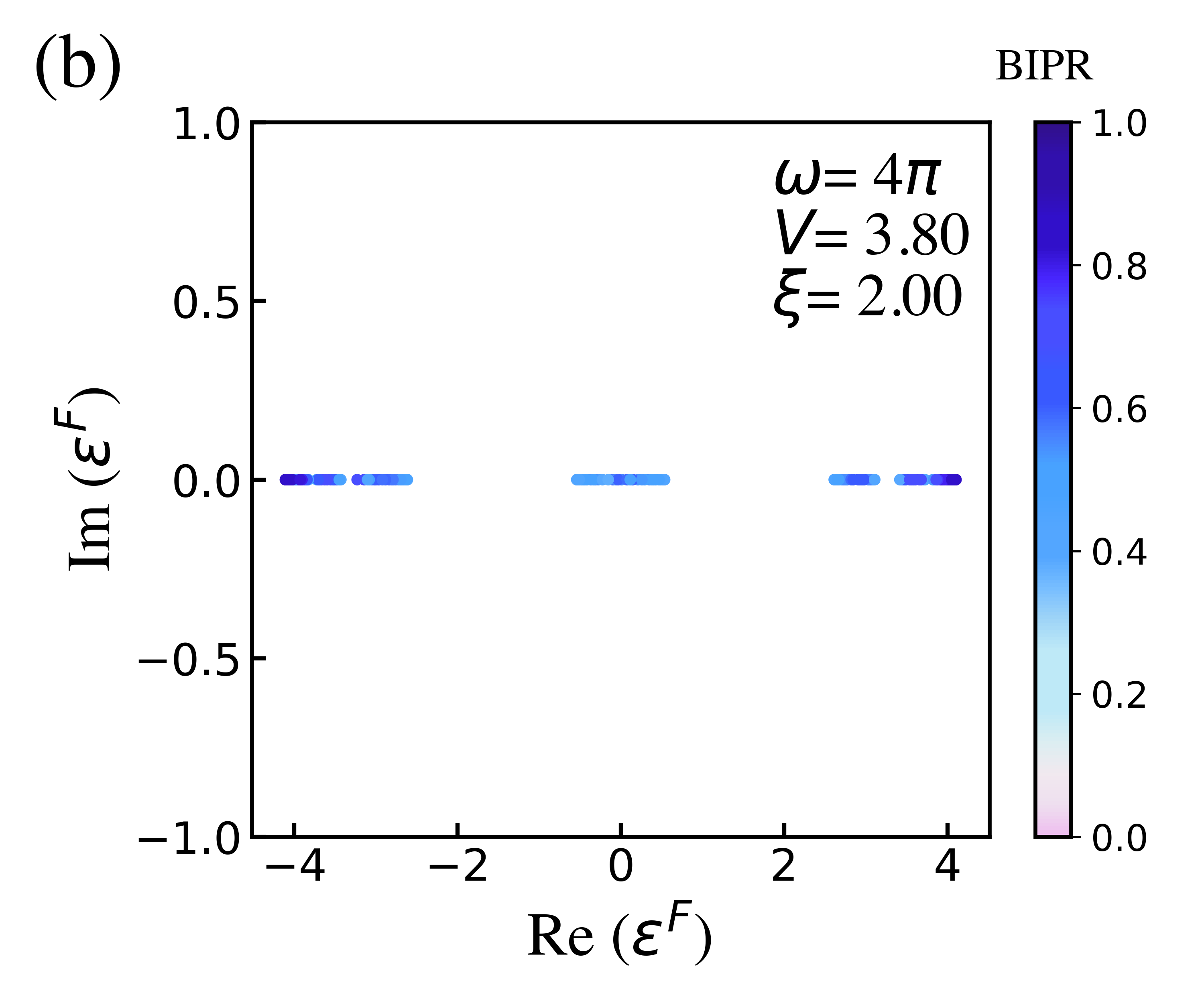}
			\caption{The Floquet quasienergy spectrum in the non-Hermitian ($h=0.1$) driven system in the localized phase at (a) $\omega=\pi/2$, and (b) $\omega=4\pi$, corresponding to Figs.~\ref{Fig:Fig_6}(a,b) in the main text.}
			\label{Fig:Fig_N2}
		\end{tabular}
	\end{figure}
	
	\begin{figure}[]
		\begin{tabular}{p{\linewidth}c}
			\vspace{0.5cm}
			\includegraphics[width=0.2450\textwidth,height=0.228\textwidth]  
			{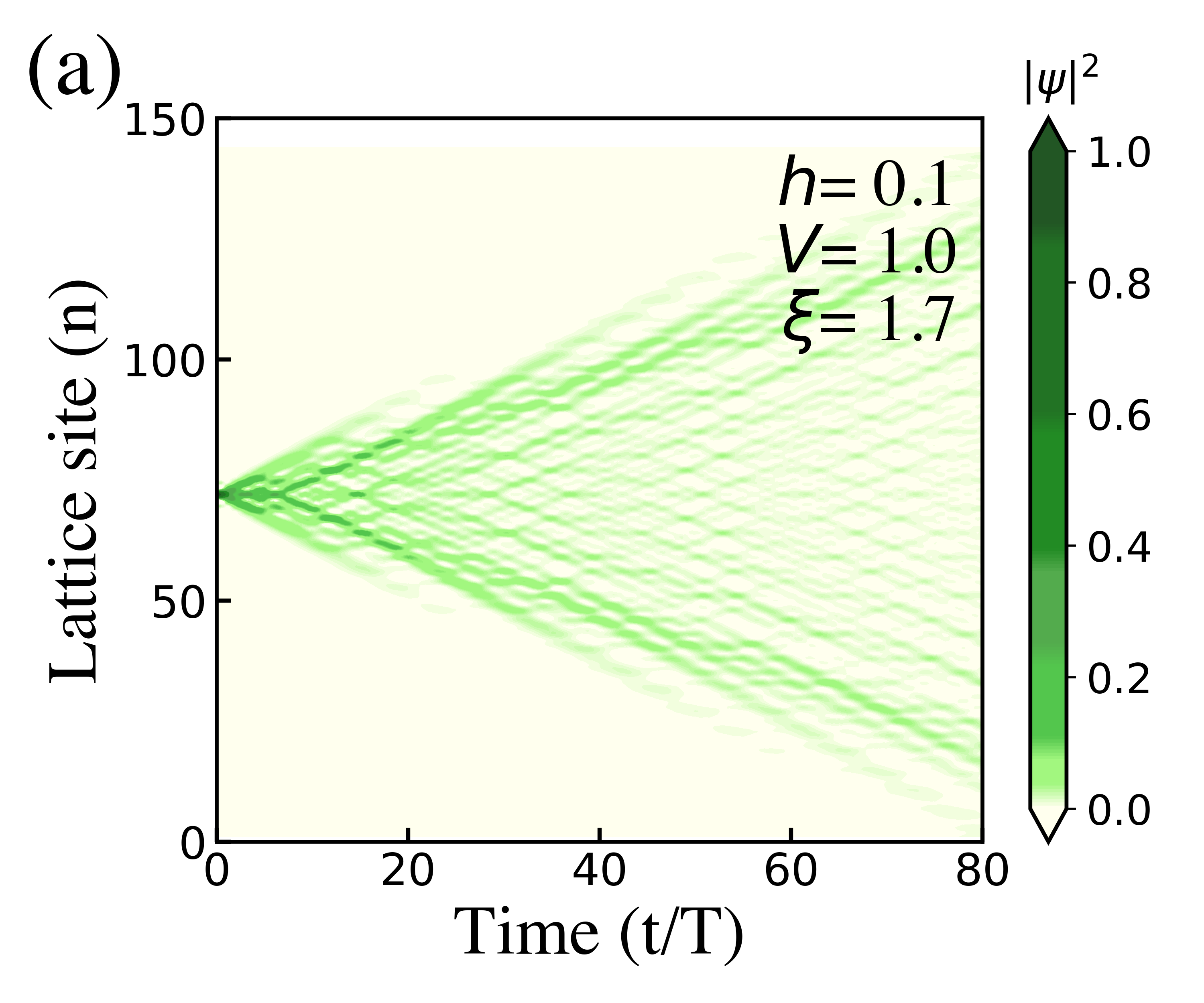}\hspace{-0.1cm}
			\includegraphics[width=0.2450\textwidth,height=0.228\textwidth]{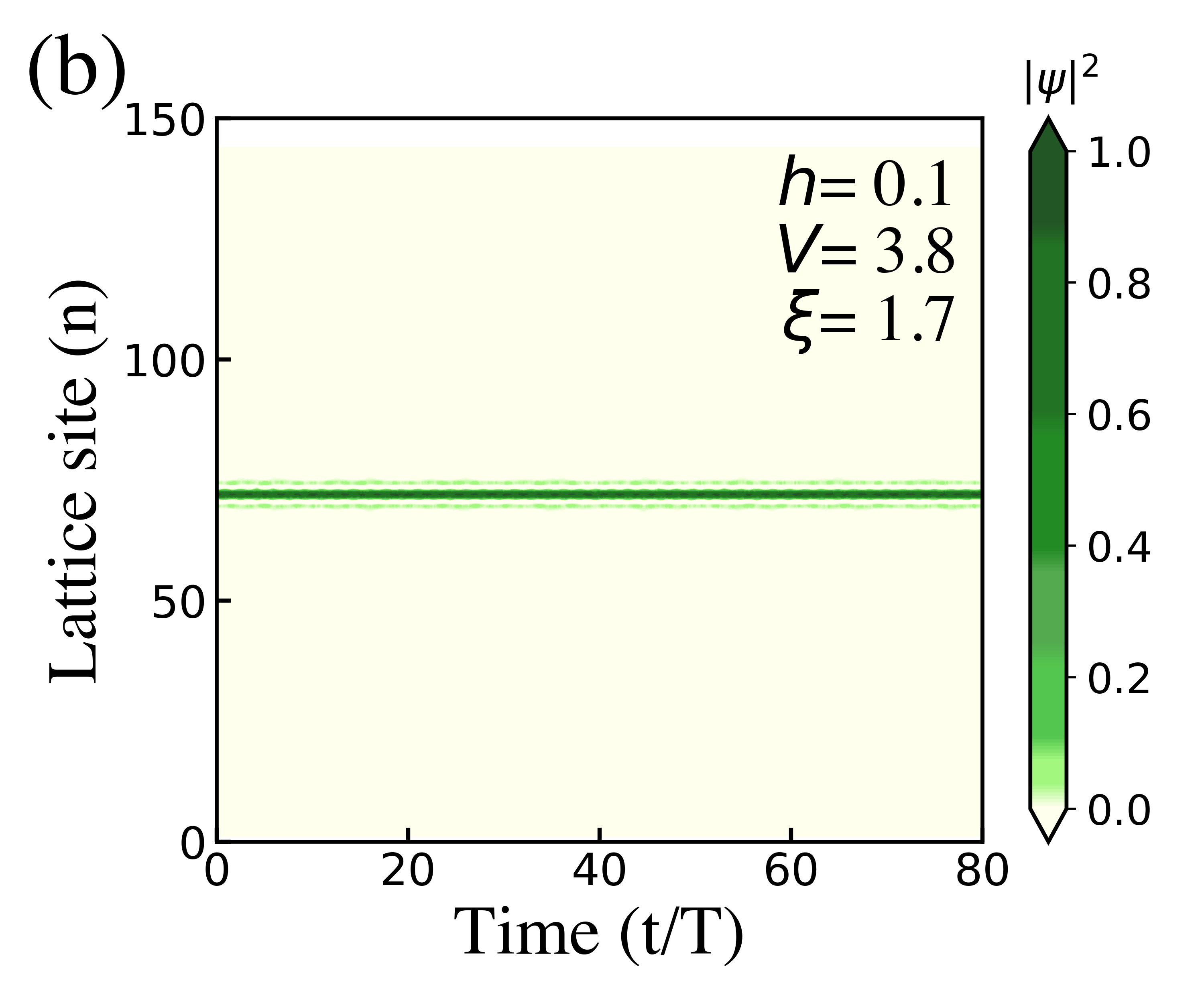}\\
			\hspace{1.75cm}\includegraphics[width=0.2450\textwidth,height=0.228\textwidth]{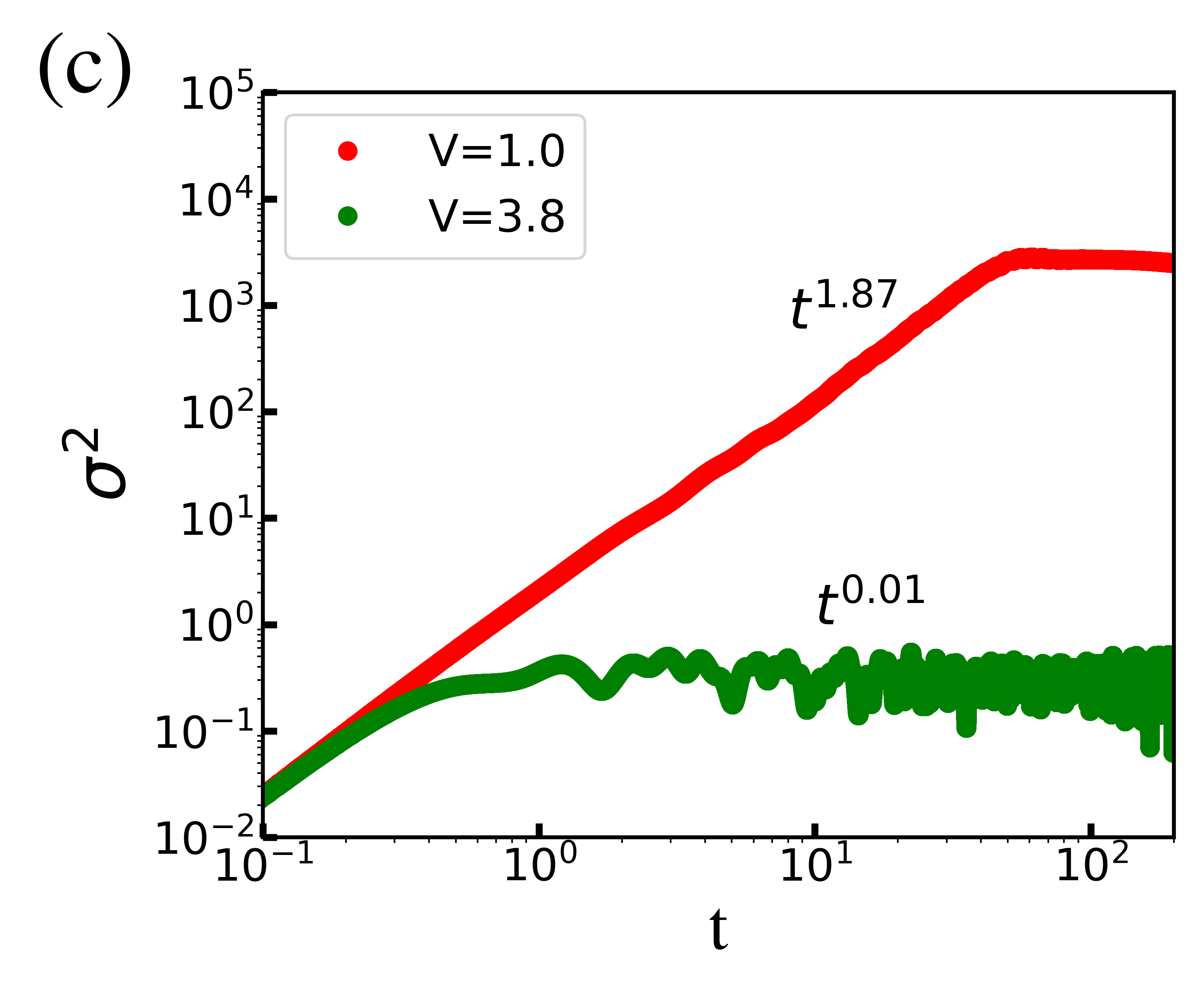}
			\caption{The time-dependence of the amplitude of the wave-packet similar to Figs.~\ref{Fig:Fig_9} and \ref{Fig:Fig_10} at $\omega=4\pi$ where no intermediate regime exists. (a) $V=1.0$ (delocalized regime) and (b) $V=3.8$ (localized regime). (c) MSD $vs.$ time in the completely delocalized and localized regimes at a high driving frequency.}
			\label{Fig:Fig_S4}
		\end{tabular}
	\end{figure}
	
	 \section{Non-Hermitian systems without a drive}\label{App:Undriven}
	
	\indent To recognize and appreciate the unconventional phases in a periodically driven system, we discuss the static analogue considered in this work which is an extension of the non-Hermitian AAH Hamiltonian in the presence of an electric field, expressed as,
	\begin{align}
		\mathcal{H}= \displaystyle\sum_{n} (Je^{h} c^\dag_{n+1} c_{n}+Je^{-h} c^\dag_{n} c_{n+1})~~~~\nonumber \\
		\tag{S-1}
		+\displaystyle\sum_{n} V \text{cos}(2\pi\alpha n) c^\dag_{n} c_{n}+ \displaystyle\sum_{n} n \xi \ c^\dag_{n} c_{n}.~~ \nonumber
		\label{Eq:Static_Hamiltonian}
	\end{align}
	All the individual terms of this Hamiltonian have been described in the main text.
	The phase diagram comprising of the $\left\langle \text{BIPR} \right\rangle$ estimated for a non-Hermitian ($h=0.1$) system at different combinations of $V$ and $\xi$ is demonstrated in Fig.~\ref{Fig:Fig_S1}(a).
	The light pink region indicates the delocalized regime, and as the color approaches blue, the states become more localized.
	The phase diagrams clearly indicates that the addition of a very tiny strength of electric field localizes all the eigenstates in the system, in contrary to the results in the absence of $\xi$, which manifests a clear DL transition at a critical strength of the quasiperiodic potential $V_c=2~\text{max}\{Je^h,Je^{-h}\}$ \cite{Jiang}.
	Furthermore, from Fig.~\ref{Fig:Fig_S1}(b), it is evident that all the states are either completely delocalized or localized determined from the value of $\eta$ as discussed in the main text.\\
	\indent In addition, it is evident that the real part of the eigenenergies in the delocalized phases of such systems possess gaps in the spectra (Fig.~\ref{Fig:Fig_S1}(c)) at values of $\alpha^z$ or $\alpha^z$+$\alpha^{z'}$, where $z,z'\in \mathbb{Z}$, since the energy spectrum belongs to a Cantor set \cite{Nilanjan}. The localized phase exhibits the usual WS localization with equally spaced energy ladders (Fig.~\ref{Fig:Fig_S1}(d)).
	In addition, in the non-Hermitian system ($h\neq0$), in the absence of the electric field ($\xi=0$) the states localize at one boundary when the boundaries of the system are open, as is visible from Fig.~\ref{Fig:Fig_S2}(a).
	Besides, the number of skin-modes diminish as we increase the strength of the electric field, as depicted in Figs.~\ref{Fig:Fig_S2}(b).
	This absence of SE in undriven systems is due to the presence of electric field and also occurs when $h$ is much large, i.e, say $h=0.9$.
	The skin-modes finally vanish, as expected in the localized regime, where the states become localized in the bulk of the lattice, demonstrating WS localization with equally spaced energy levels (Fig.~\ref{Fig:Fig_S2}(d)).\\
	
	\begin{figure*}[]
		\includegraphics[width=0.250\textwidth,height=0.25\textwidth]{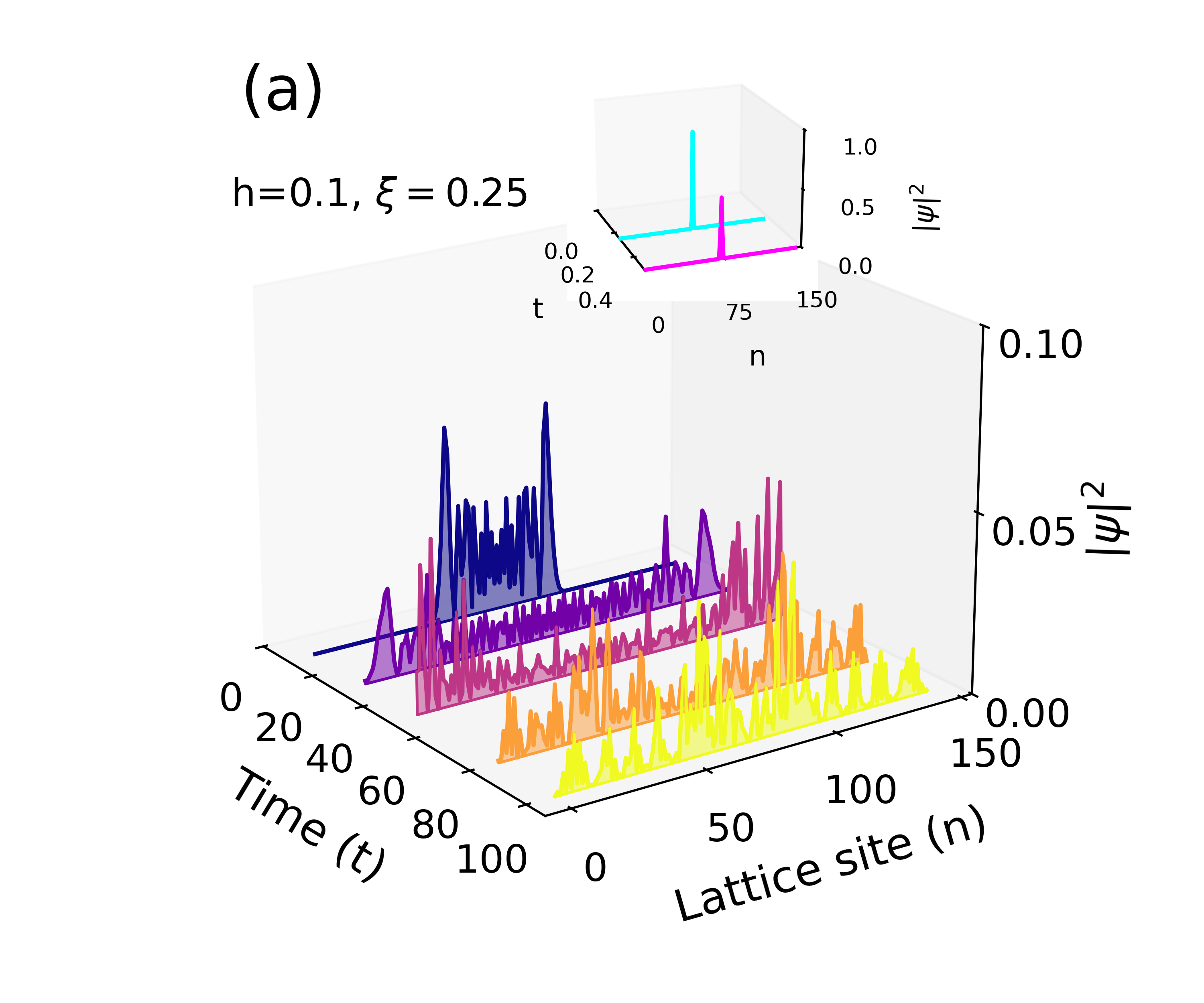}\hspace{-0.1cm}
		\includegraphics[width=0.250\textwidth,height=0.25\textwidth]{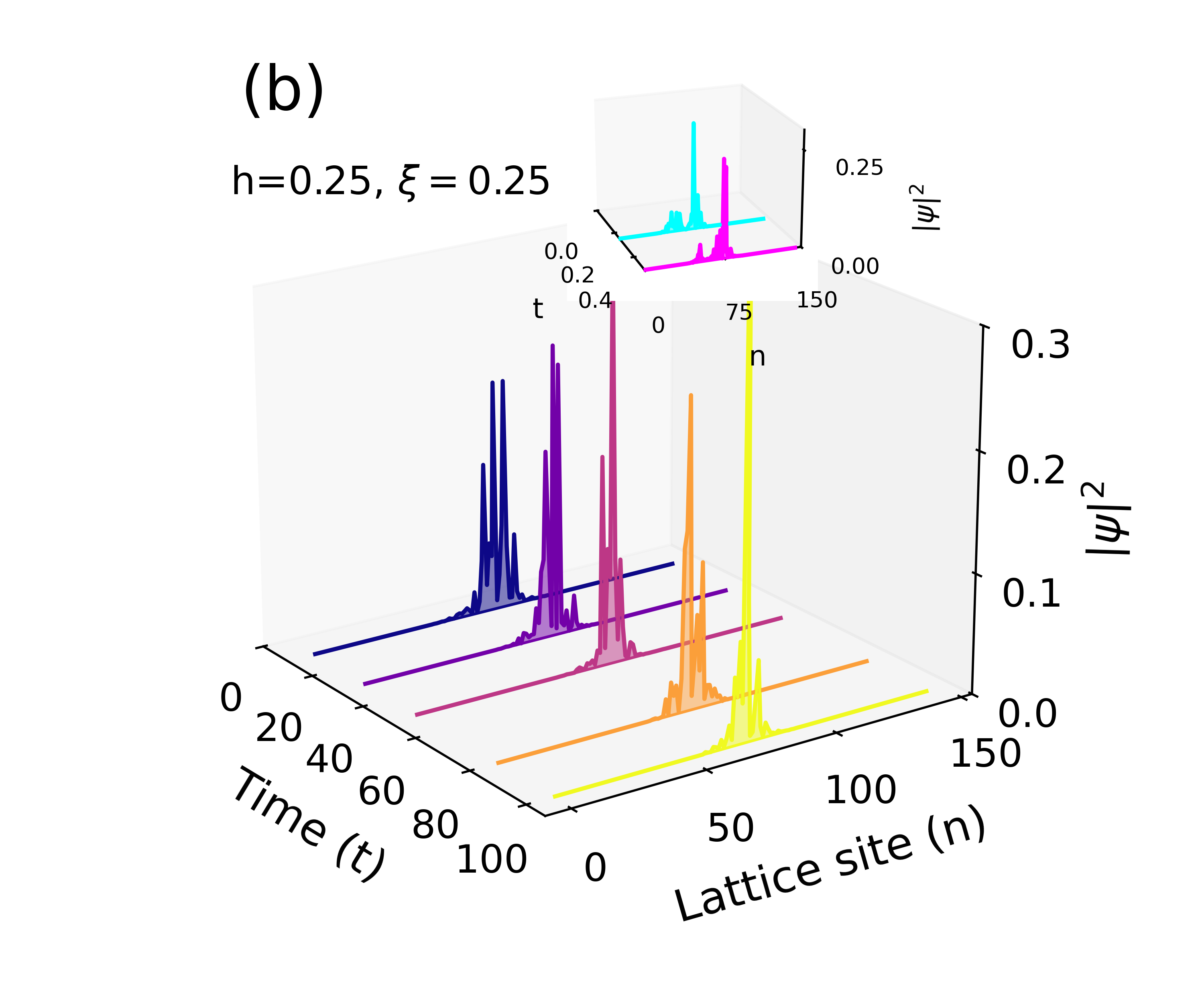}\hspace{-0.1cm}
		\includegraphics[width=0.25\textwidth,height=0.25\textwidth]{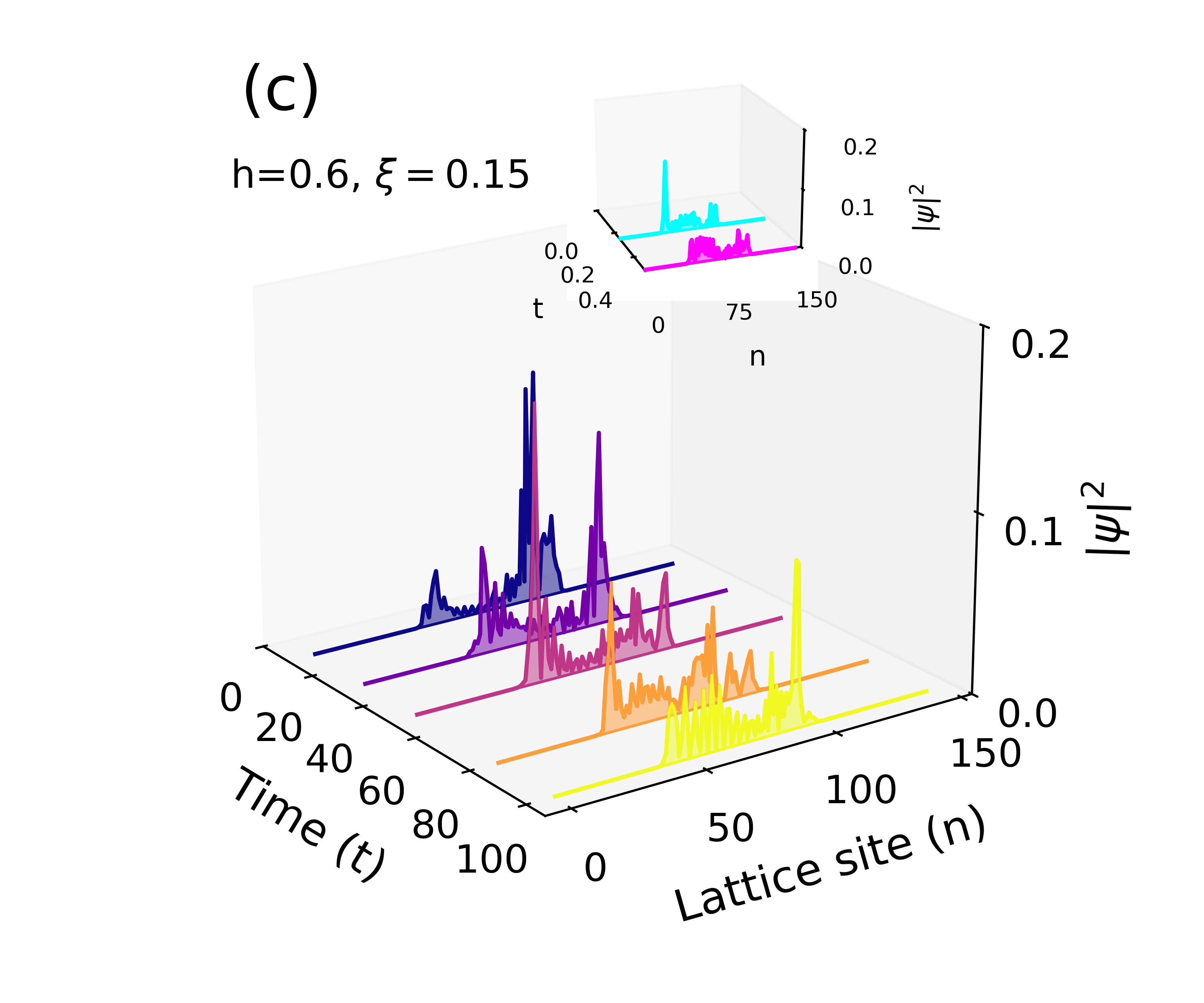}\hspace{-0.1cm}
		\includegraphics[width=0.25\textwidth,height=0.25\textwidth]{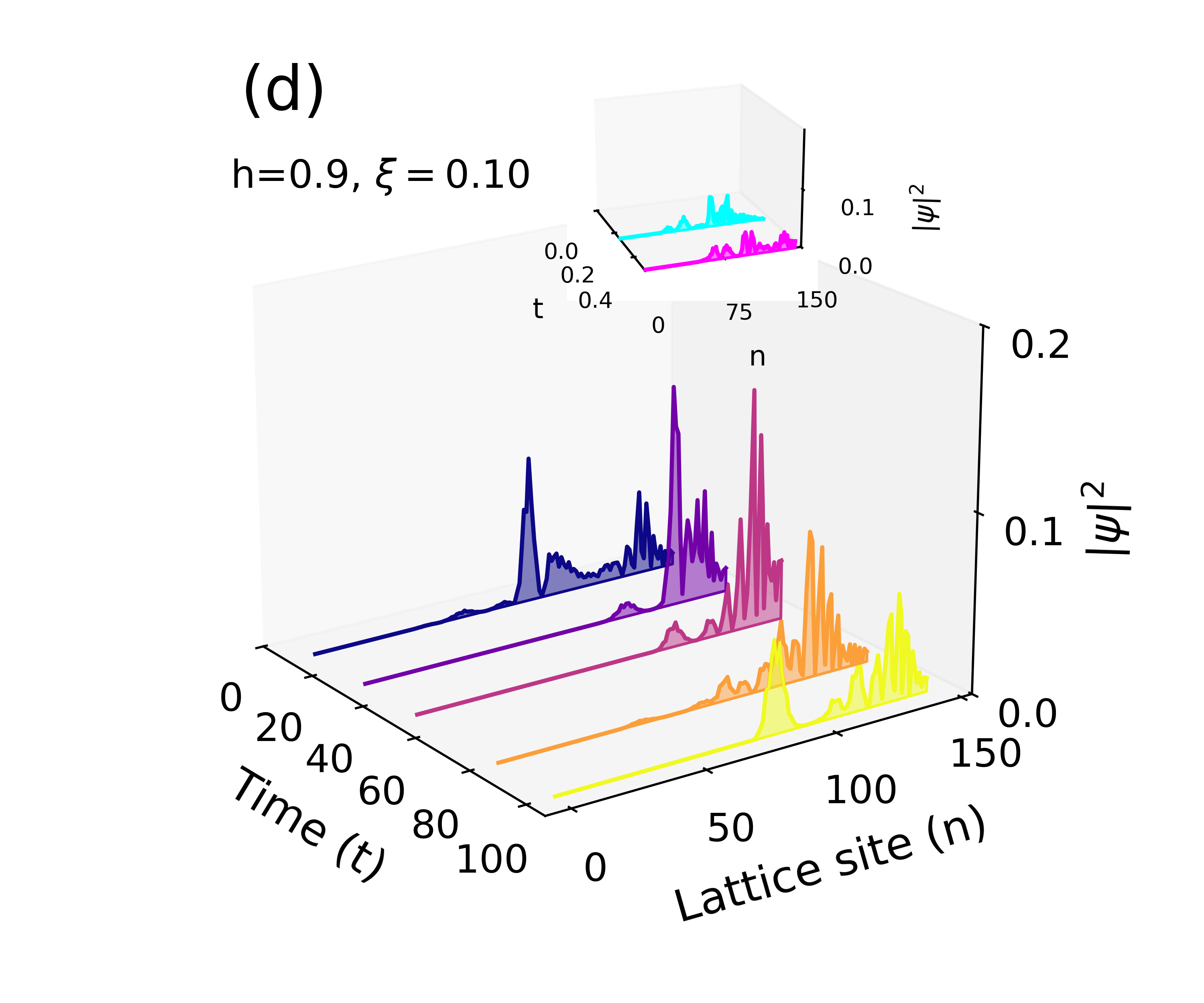}
		\caption{(a-d) Evolution of $|\psi|^2$ for $t$=100 secs as a function of all lattice sites at $\omega=4\pi$ under OBC, similar to Figs.~\ref{Fig:Fig_11}(a-d). The inset shows the probability distribution after a first stroboscopic period (0.5 secs).}
		\label{Fig:Fig_S5}
	\end{figure*}
	
	\begin{figure}[]
		\begin{tabular}{p{\linewidth}c}
			\centering
			\includegraphics[width=0.2450\textwidth,height=0.228\textwidth]  
			{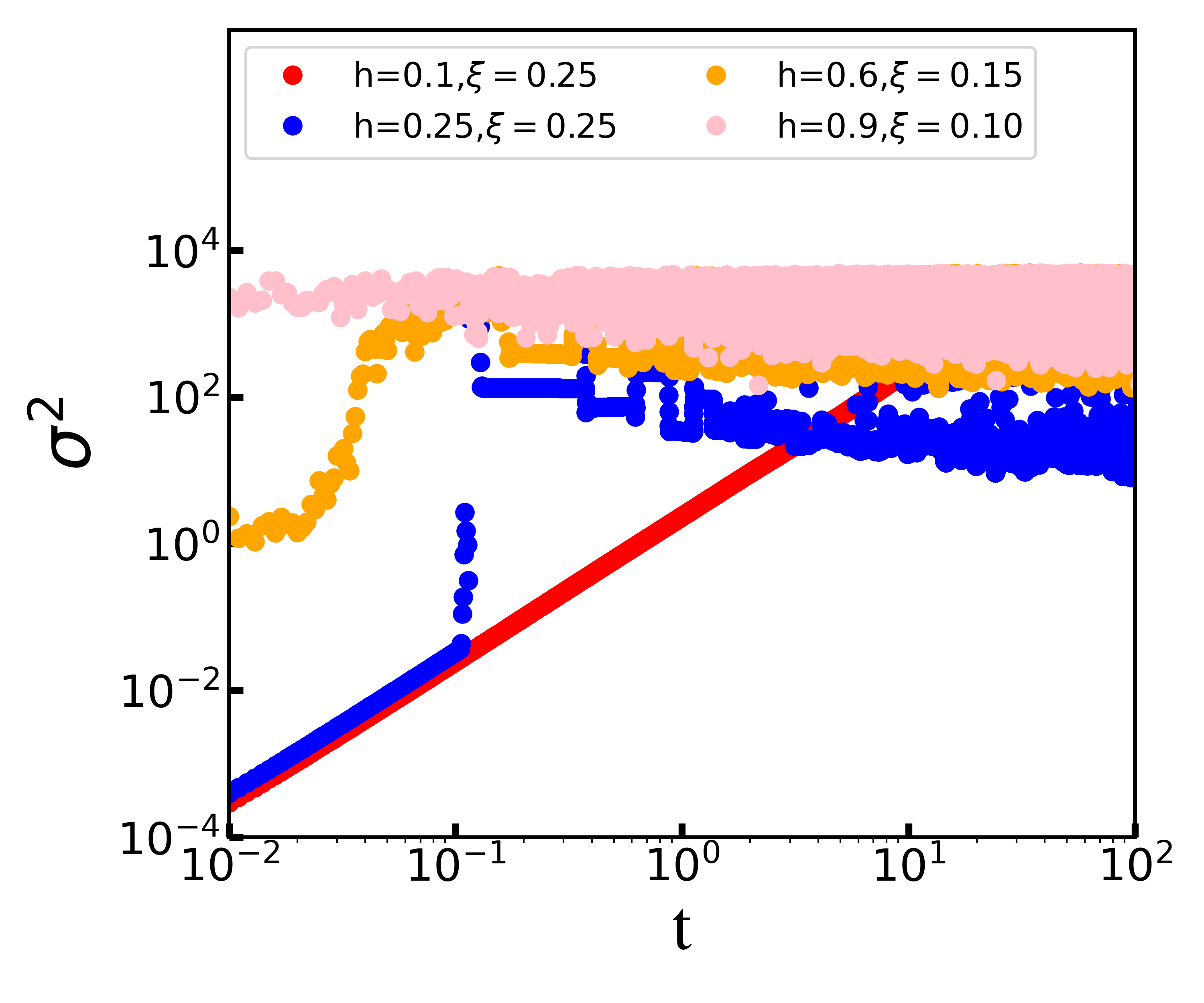}
			\caption{MSD $vs.$ time for the different phases at high frequency as demonstrated in Fig.~\ref{Fig:Fig_S5}.}
			\label{Fig:Fig_S6}
		\end{tabular}
	\end{figure}
	
	\section{Intermediate phases in the Hermitian counterpart with drive}\label{App:Hermitian}
	In Fig.~\ref{Fig:Fig_N1}(a,b), we demonstrate the appearance of intermediate phases at the driving frequencies $\omega=\pi/2$ and $\omega=4\pi$, similar to the ones that has been discussed in the main text for a non-Hermitian system ($h=0.1$). Therefore, it is clear that the intermediate phases arise solely because of the time-periodic nature of the drive, and appears irrespective of the degree of non-Hermiticity.
	
	\section{Real spectrum of the Floquet quasienergies in the localized regime}\label{App:Floquet_energies}
To verify the nature of the Floquet quasienergies under PBC in the localized limit ($V=3.8, \xi=2.0$), we plot them in the complex plane in Figs.~\ref{Fig:Fig_N2}(a,b) at $\omega=\pi/2$ and $\omega=4\pi$ respectively. It is evident that the Floquet quasienergy spectrum is real, such that the measure of $\left\langle r \right\rangle$ can be used as discussed in Sec.~\ref{Sec:WS_localization}.
	
	\section{Skin effect in the driven non-Hermitian system}\label{App:SE_driven}
	It is clear from the type of the Hamiltonian considered in our work that there exists a counter force to the unidirectionality imposed by the asymmetricity in the Hamiltonian ($h$) from the driven electric field.
	We verify that the measure ($\rho_{im}$) gives a proper understanding for the existence/absence of SE. For a particular value of $h=0.5$, in Fig.~\ref{Fig:Fig_S3}(a-b), we clearly illustrate that the SE appears at $\xi=0.05$ but disappears when the electric field $\xi=0.10$, which correspond to the green and magenta regimes in the phase diagram shown in Fig.~\ref{Fig:Fig_8} and occurs when $\rho_{im}=1$ and $\rho_{im}=0$ respectively.
	The entire phase diagram spanning $\xi$ and $h$ can therefore be chalked out using the measure $\rho_{im}$.
	
	\section{Dynamics in the pure localized and delocalized phases at a large driving frequency}\label{App:High_frequency_transport}
	To distinguish the non-trivial transport behavior of the intermediate regimes obtained when the driving frequency is not sufficiently high, we demonstrate the behavior of the completely localized and localized states of the time-dependent Hamiltonian at a high driving frequency ($\omega=4\pi$) as shown in Figs.~\ref{Fig:Fig_S4}(a-c). It is clear that the initial excitation has a ballistic nature ($\delta \simeq 0.94$) in the delocalized phase, while the transport ceases completely ($\delta \simeq 0.01$) when the excitation is released in the localized phase.
	
	\section{Dynamics under the OBC}\label{App:Dynamics_large_frequency}
	In Figs.~\ref{Fig:Fig_S5}(a-d), we analyze the long-time dynamics of the initial wave-packet excitation at a large driving frequency ($\omega=4\pi$) in the different phases as shown in Fig.~\ref{Fig:Fig_8}(d). Similar to the evolution as illustrated in Figs.~\ref{Fig:Fig_11}(a-d) when the driving frequency is low, in this case as well, one obtains the multifractal behavior of the eigenstates exhibiting SE.
	The MSD $vs.$ time in all the four regimes of the phase diagram is also presented in Fig.~\ref{Fig:Fig_S6} for clarity.
	
	\bibliography{ref.bib}
\end{document}